\documentclass[a4paper]{article}
\pdfoutput=1
\usepackage[T1]{fontenc}
\usepackage{graphicx}
\usepackage{amsmath}
\usepackage{xspace}
\usepackage[numbers,sort&compress]{natbib}
\usepackage{subfig}
\usepackage[shortlabels]{enumitem}

\usepackage{pstricks}
\usepackage{graphicx}
\usepackage{xspace}
\usepackage{array}

\usepackage{jheppub}
\hypersetup{
  pdfauthor={S. Kallweit, J. M. Lindert, S. Pozzorini, M. Schoenherr},
  pdftitle={NLO QCD+EW predictions for off-shell vector boson pair-production}
}

\newcommand{\GF}{{G_\mu}}

\newcommand{\LO}{\text{LO}\xspace}
\newcommand{\NLO}{\text{NLO}\xspace}
\newcommand{\NNLO}{\text{NNLO}\xspace}

\newcommand{\NNLL}{\text{NNLL}\xspace}
\newcommand{\QED}{\text{QED}\xspace}
\newcommand{\QCD}{\text{QCD}\xspace}

\newcommand{\YFS}{\text{YFS}\xspace}
\newcommand{\CSS}{\text{CSS}\xspace}
\newcommand{\EW}{\text{EW}\xspace}

\newcommand{\EWvirt}{\ensuremath{\EW_\text{VI}}\xspace}

\newcommand{\QCDpEW}{\text{\QCD{}+\EW}\xspace}

\newcommand{\QCDtEW}{\ensuremath{\text{\QCD{}\ensuremath{\times}\EW}}\xspace}
\newcommand{\QCDotEW}{\ensuremath{\text{\QCD{}\ensuremath{\otimes}\EW}}\xspace}
\newcommand{\QCDtEWvirt}{\ensuremath{\text{\QCD{}\ensuremath{\times}\EW{}$_\text{VI}$}}\xspace}
\newcommand{\MSbar}{\ensuremath{\overline{\text{MS}}}\xspace}
\newcommand{\NLOQCDtEWYFS}{\ensuremath{\text{\NLO \QCDtEWvirt}\otimes\text{\YFS{}}}\xspace}
\newcommand{\NLOQCDtEWCSS}{\ensuremath{\text{\NLO \QCDtEWvirt}\otimes\text{\CSS{}}}\xspace}

\newcommand{\Collier}{{\rmfamily\scshape Collier}\xspace}
\newcommand{\Munich}{{\rmfamily \scshape Munich}\xspace}
\newcommand{\Sherpa}{{\rmfamily\scshape Sherpa}\xspace}
\newcommand{\SherpaOpenLoops}{{\rmfamily\scshape Sherpa+OpenLoops}\xspace}
\newcommand{\OpenLoops}{{\rmfamily\scshape OpenLoops}\xspace}
\newcommand{\Recola}{{\rmfamily\scshape Recola}\xspace}
\newcommand{\MunichOpenLoops}{{\rmfamily \scshape Munich+OpenLoops}\xspace}
\newcommand{\MunichSherpaOpenLoops}{{\rmfamily \scshape Munich/Sherpa+OpenLoops}\xspace}
\newcommand{\Rivet}{{\rmfamily\scshape Rivet}\xspace}
\newcommand{\LHAPDF}{{\rmfamily\scshape Lhapdf}\xspace}
\newcommand{\CSShower}{{\rmfamily\scshape Csshower}\xspace}

\newcommand{\llnn}{\ensuremath{2\ell 2\nu}\xspace}

\newcommand{\Pa}{\ensuremath{\gamma}\xspace}
\newcommand{\Pt}{\ensuremath{t}\xspace}
\newcommand{\Pb}{\ensuremath{b}\xspace}
\newcommand{\Pe}{\ensuremath{e}\xspace}
\newcommand{\Pmu}{\ensuremath{\mu}\xspace}
\newcommand{\Ptau}{\ensuremath{\tau}\xspace}
\newcommand{\PV}{\ensuremath{V}\xspace}
\newcommand{\PW}{\ensuremath{W}\xspace}

\newcommand{\PZ}{\ensuremath{Z}\xspace}
\newcommand{\PH}{\ensuremath{H}\xspace}

\newcommand{\Pq}{\ensuremath{q}\xspace}
\newcommand{\Pqbar}{\ensuremath{\bar{\Pq}}\xspace}

\newcommand{\wpwm}{\ensuremath{\PW^+\PW^-}\xspace}
\newcommand{\wpmz}{\ensuremath{\PW^\pm\PZ}\xspace}
\newcommand{\ww}{\ensuremath{\PW\PW}\xspace}
\newcommand{\zz}{\ensuremath{\PZ\PZ}\xspace}
\renewcommand{\aa}{\ensuremath{\Pa\Pa}\xspace}  

\newcommand{\jet}{\ensuremath{j}\xspace}
\newcommand{\shortequal}{\ensuremath{\!\!\!=\!\!\!}}

\newcommand{\ri}{\mathrm{i}}
\newcommand{\rF}{\mathrm{F}}
\newcommand{\rR}{\mathrm{R}}

\newcommand{\rT}{\mathrm{T}}
\newcommand{\rd}{\mathrm{d}}

\newcommand{\rS}{\mathrm{S}}

\newcommand{\MW}{M_\PW}
\newcommand{\MZ}{M_\PZ}
\newcommand{\MH}{M_\PH}
\newcommand{\Mt}{m_{\Pt}}
\newcommand{\Mb}{m_{\Pb}}
\newcommand{\Me}{m_{\Pe}}
\newcommand{\Mmu}{m_{\Pmu}}
\newcommand{\GW}{\Gamma_{\PW}}
\newcommand{\GZ}{\Gamma_{\PZ}}
\newcommand{\GH}{\Gamma_{\PH}}
\newcommand{\Gb}{\Gamma_{\Pb}}
\newcommand{\Gt}{\Gamma_{\Pt}}

\newcommand{\CA}{{\ensuremath{C_A}}}
\newcommand{\CF}{{\ensuremath{C_F}}}
\newcommand{\TR}{{\ensuremath{T_R}}}
\newcommand{\NC}{{\ensuremath{N_C}}}
\newcommand{\mur}{\mu_{\rR}}
\newcommand{\muf}{\mu_{\rF}}
\newcommand{\murf}{\mu_{\rR,\rF}}

\newcommand{\keV}{\text{keV}\xspace}
\newcommand{\MeV}{\text{MeV}\xspace}
\newcommand{\GeV}{\text{GeV}\xspace}
\newcommand{\TeV}{\text{TeV}\xspace}

\newcommand{\alphaS}{\alpha_{\rS}}
\newcommand{\ord}{\mathcal{O}}

\newcommand{\HTl}{H_{\mathrm{T}}^{\mathrm{lep}}}
\newcommand{\HTjet}{H_{\mathrm{T}}^{\mathrm{jet}}}

\newcommand{\pT}{\ensuremath{p_{\mathrm{T}}}\xspace}

\newcommand{\missingET}{\ensuremath{\displaystyle{\not}E_{\rT}}\xspace}

\newcommand{\sw}{s_{\mathrm{w}}}
\newcommand{\cw}{c_{\mathrm{w}}}

\newcommand{\beqar}{\begin{eqnarray}}
\newcommand{\eeqar}{\end{eqnarray}}
\newcommand{\beq}{\begin{equation}}
\newcommand{\eeq}{\end{equation}}
\newcommand{\bsp}{\protect\begin{split}}
\newcommand{\esp}{\protect\end{split}}
\newcommand{\bit}{\begin{itemize}}
\newcommand{\eit}{\end{itemize}}

\newcommand{\xs}[3]{\ensuremath{{#1}_{-#2}^{+#3}}}

\newcommand{\vp}{\ensuremath{\vphantom{\int}}}
\newcommand{\vP}{\ensuremath{\vphantom{\int_a^b}}}

\newlength{\unitcharwidth}
\newcolumntype{C}{>{\centering\arraybackslash}p{0.165\textwidth}}

\newcommand{\refeq}[1]{\mbox{\eqref{#1}}}
\newcommand{\refeqs}[2]{\mbox{\eqref{#1}--\eqref{#2}}}
\newcommand{\reffi}[1]{\mbox{Fig.~\ref{#1}}}
\newcommand{\reffis}[2]{\mbox{Figs.~\ref{#1}--\ref{#2}}}
\newcommand{\refta}[1]{\mbox{Table~\ref{#1}}}
\newcommand{\reftas}[2]{\mbox{Tables~\ref{#1}--\ref{#2}}}
\newcommand{\refse}[1]{\mbox{Section~\ref{#1}}}
\newcommand{\refses}[2]{\mbox{Sections~\ref{#1}--\ref{#2}}}
\newcommand{\refapp}[1]{\mbox{Appendix~\ref{#1}}}

\newcommand{\ie}{i.e.\ }
\newcommand{\eg}{e.g.\ }

\newcommand{\aPDF}{\ensuremath{\Pa{}\text{PDF}}\xspace}
\newcommand{\boldaPDF}{\ensuremath{\boldsymbol{\Pa{}}\text{PDF}}\xspace}
\newcommand{\CTQED}{CT14qed\xspace}
\newcommand{\LUXQED}{LUXqed\xspace}
\newcommand{\NNPDFQED}{NNPDF3.0qed\xspace}

\newcommand{\numutau}{\nu_{\Pmu/\tau}}
\newcommand{\nuetau}{\nu_{\Pe/\tau}}
\newcommand{\barnumutau}{\bar\nu_{\Pmu/\tau}}
\newcommand{\barnuetau}{\bar\nu_{\Pe/\tau}}

\newcommand{\llmetdf}{\Pe^+\Pmu^-+\missingET}
\newcommand{\llmetsf}{\Pe^+\Pe^-+\missingET}
\newcommand{\llnndf}{\Pe^+\Pmu^-\nu_{\Pe}\bar\nu_{\Pmu}}
\newcommand{\llnndfflip}{\Pmu^+\Pe^-\nu_{\Pmu}\bar\nu_{\Pe}}
\newcommand{\llnnsf}{\Pe^+\Pe^-\nu\bar\nu}
\newcommand{\eenn}{\Pe^+\Pe^-\nu\bar\nu}
\newcommand{\llnnsfwwzz}{\Pe^+\Pe^-\nu_{\Pe}\bar\nu_{\Pe}}
\newcommand{\llnnsfwwzzflip}{\Pmu^+\Pmu^-\nu_{\Pmu}\bar\nu_{\Pmu}}
\newcommand{\llnnsfzz}{\Pe^+\Pe^-\numutau\barnumutau}
\newcommand{\llnnsfzzflip}{\Pmu^+\Pmu^-\nuetau\barnuetau}
\newcommand{\llnnsfzzone}{\Pe^+\Pe^-\nu_{\Pmu}\bar\nu_{\Pmu}}

\newcommand{\DFWW}{\ensuremath{\mathrm{DF_{\ww}}}}
\newcommand{\SFZZ}{\ensuremath{\mathrm{SF_{\zz}}}}
\newcommand{\SFWWZZ}{\ensuremath{\mathrm{SF_{\ww{}/\zz}}}}

\newcommand{\relplotwidth}{0.47}

\newcommand{\Zgammavirt}{\delta Z_{\Pa,\mathrm{virt}}}
\newcommand{\ceps}{C_{\epsilon}}
\newcommand{\lograt}[2]{\ln\left(\frac{#1}{#2}\right)}
\newcommand{\mud}{\mu_{\mathrm{D}}}
\newcommand{\ncqsf}{N_{\mathrm{C},f}\, Q^2_f}

\newcommand{\OS}{\mathrm{OS}}
\newcommand{\PDF}{\mathrm{PDF}}
\newcommand{\ep}{\epsilon}

\newcommand{\bom}[1]{{\mbox{\boldmath $#1$}}}

\newcommand{\qtilde}[2]{Q_{#1,#2}}
\newcommand{\Sin}{\mathcal{S}_{\mathrm{in}}}
\newcommand{\Sout}{\mathcal{S}_{\mathrm{out}}}

\newcommand{\Sinout}{\mathcal{S}_{\mathrm{in}+\mathrm{out}}}
\newcommand{\Shatin}{\mathcal{S}_{\mathrm{in}}}
\newcommand{\Shatout}{\mathcal{S}_{\mathrm{out}}}
\newcommand{\Shatinout}{\mathcal{S}_{\mathrm{in}+\mathrm{out}}}
\newcommand{\bornIJ}{\rd\sigma^{B}|_{IJ\to\IJtilde}}

\newcommand{\gammai}{\Pa_i}
\newcommand{\epsFSgammai}{\epsilon_{\mathrm{FS},\gammai}}
\newcommand{\DeltaFSgamma}{\Delta_{\mathrm{FS},\Pa}}

\newcommand{\ngammain}{n_{\Pa}^{(\mathrm{in})}}

\newcommand{\ngammaouteps}{n_{\Pa,\epsilon}^{(\mathrm{out})}}
\newcommand{\IJtilde}{\widetilde{IJ}}
\newcommand{\wtffbar}{\widetilde{\!f\bar f\;}\!}
\newcommand{\light}{\mathrm{light}}
\newcommand{\heavy}{\mathrm{heavy}}

\newcommand{\Rrec}{R_{\mathrm{rec}}}
\newcommand{\nmassless}[1]{N_{0, #1}}
\newcommand{\thetaw}{\theta_{\mathrm{w}}}
\newcommand{\sinw}{\sin\thetaw}
\newcommand{\cosw}{\cos\thetaw}
\newcommand{\mferm}{F_{\mathrm{m}}}

\newcommand{\epsqq}{\epsilon_{q\bar q}}
\newcommand{\epsaa}{\epsilon_{\gamma\gamma}}

\preprint{
  \begin{flushright}
    IPPP/17/32 \\ MCNET--17--05 \\ CERN-TH-2017-097 \\ ZU-TH 09/17
  \end{flushright}
}

\author[a]{S.~Kallweit,}
\author[b]{J.~M.~Lindert,}
\author[c]{S.~Pozzorini,}
\author[c]{and M.~Sch{\"o}nherr}

\affiliation[a]{TH Division, 
                Physics Department, 
                CERN, 
                CH-1211 Geneva 23, 
                Switzerland}
\affiliation[b]{Institute for Particle Physics Phenomenology, 
		Durham University, 
		Durham DH1 3LE, 
		UK}
\affiliation[c]{Physik-Institut, Universit\"at Z\"urich,
		Winterthurerstrasse 190, 
		CH-8057 Z\"urich,
		Switzerland }

\emailAdd{stefan.kallweit@cern.ch}
\emailAdd{lindert@physik.uzh.ch}
\emailAdd{pozzorin@physik.uzh.ch}
\emailAdd{marek.schoenherr@physik.uzh.ch}

\title{NLO \protect\QCDpEW predictions for $\mathbf{2\ell2\nu}$ diboson signatures
at the LHC}

\abstract{
We present next-to-leading order~(\NLO)  calculations including \QCD and 
electroweak (\EW) corrections for  $2\ell2\nu$ diboson signatures with
two opposite-charge leptons and two neutrinos.
Specifically, we study the processes 
$pp\to\llnndf$ and  $pp\to\eenn$,
including all relevant off-shell diboson channels,
\wpwm, \zz, $\gamma\PZ$, as well as non-resonant contributions.
Photon-induced processes are computed at \NLO \EW, and we discuss subtle
differences related to the definition and the renormalisation of 
the coupling $\alpha$ for processes with initial- and final-state photons.
All calculations are performed within the automated 
\MunichSherpaOpenLoops frameworks, and we provide 
numerical predictions for the LHC at 13\,\TeV.
The behaviour of the corrections is investigated with emphasis on the high-energy
regime, where \NLO \EW effects can amount to tens of percent due to large
Sudakov logarithms.
The interplay between \ww and \zz contributions to the 
same-flavour channel, $pp\to\eenn$, is discussed in detail, and
a quantitative analysis of photon-induced contributions is presented.
Finally, we consider approximations that account for all 
sources of large logarithms, at high and low energy, 
by combining virtual \EW corrections with a \YFS soft-photon resummation or a \QED parton shower.
}


\begin{document}

\maketitle
\flushbottom

\section{Introduction}
\label{sec:intro}

The production of vector-boson pairs, \wpwm, \wpmz and \zz,
plays an important role in various areas of the LHC physics programme.
Experimental studies of this family of processes 
permit to test key aspects of the Standard Model (SM) 
at energies that range from the \EW scale up to the 
\TeV regime.
In particular, due to the high sensitivity to anomalous 
trilinear couplings, differential measurements at high 
transverse  momentum  allow one to test the gauge symmetry structure of \EW 
interactions and to search for indirect effects of 
physics Beyond the Standard Model (BSM).
Diboson final states are widely studied
also in the context of direct BSM searches. 
Moreover, they play the role of nontrivial backgrounds 
in a broad range of measurements and searches. 
Most notably, they represent the irreducible 
background to Higgs-boson analyses 
in the $\PH\to \wpwm$ and $\PH\to\zz$ decay modes.  
These motivations, together with the increasing level of 
accuracy of experimental measurements, call for continuous improvements 
in the theoretical description of diboson production at the LHC.

Leptonically decaying vector-boson pairs yield clean experimental signatures
with charged leptons and neutrinos.
In this paper we focus on final states with two opposite-charge leptons and two
neutrinos, generically denoted as $2\ell2\nu$.  Their production is 
dominated by \wpwm resonances, resulting in the highest cross sections among the
various channels with dibosons decaying into charged leptons and neutrinos.  
The resonant structure of $pp\to 2\ell2\nu$
depends on the lepton-flavour configuration, and 
we consider both the case of different 
and same 
charged-lepton flavours.
In the different-flavour case, $\ell^+_i\ell^-_j\nu_i\bar\nu_j$ with 
$\ell_i \neq \ell_j$, only \wpwm
resonances contribute, whereas same-flavour final states, 
$\ell^+_i\ell^-_i\nu_k\bar\nu_k$,
can arise both through \wpwm and \zz resonances.
While $2\ell 2\nu$ production is dominated by resonant contributions,
off-shell effects and non-resonant topologies play an important role for
various phenomenological studies, 
for instance in $H\to\PV\PV$ studies, where
selection cuts or kinematic discriminants can 
force diboson backgrounds into the off-shell regime.

Theoretical predictions for \wpwm and \zz  production and decays are
available up to next-to-next-to-leading order~(\NNLO) in
\QCD~\cite{Gehrmann:2014fva,Grazzini:2016ctr,Cascioli:2014yka,Grazzini:2015hta}. 
More precisely, \NNLO \QCD predictions for $2\ell 2\nu$ production 
have been published only 
in the \wpwm mediated channel $pp\to \llnndf$~\cite{Grazzini:2016ctr},
while \NNLO \QCD calculations for \zz
mediated processes exist only for the $pp\to 4\ell$ channel to date.  At
higher orders in \QCD, both processes receive sizeable contributions from the
opening of gluon-induced channels, and the important impact of \QCD radiation
results in a pronounced sensitivity to jet vetoes.  Also
loop-induced contributions from gluon fusion, known up to
$\ord(\alphaS^3)$~\cite{Caola:2015psa,Caola:2015rqy}, play an important
role.

In order to reach the level of precision required by present and future 
experimental analyses,
higher-order \QCD predictions need to be supplemented 
by  \EW correction effects.
In general, the dominant \EW corrections 
are due to \QED radiation effects in the distributions of final-state leptons,
and large Sudakov logarithms that arise at scattering energies
$Q^2\gg \MW^2$~\cite{Denner:2000jv}. 
The importance of \EW Sudakov logarithms for $pp\to\wpwm/\zz$ at the LHC 
was demonstrated in~\cite{Accomando:2004de} and confirmed by full \NLO \EW calculations
for on-shell vector-boson production~\cite{Bierweiler:2012kw,Baglio:2013toa,Gieseke:2014gka}.
At the \TeV scale, due to the large SU(2) charges of $\PW$ and $\PZ$ bosons,
\EW Sudakov corrections can reach the level of 50\% at $\ord(\alpha)$, 
and also higher-order Sudakov \EW effects become significant.
For the case of \wpwm production, corresponding results are available
up to $\ord(\alpha^2)$ to \NNLL accuracy~\cite{Kuhn:2011mh}.

A first calculation that includes diboson production and decays at \NLO \EW
was performed for the different-flavour process $pp\to\wpwm\to \llnndf$ using
a spin-correlated double-pole approximation (DPA)~\cite{Billoni:2013aba}. 
More recently, full \NLO \EW predictions for the \zz and \wpwm mediated
processes $pp\to 4\ell$~\cite{Biedermann:2016yvs,Biedermann:2016lvg} and
$pp\to \llnndf$~\cite{Biedermann:2016guo} became available.  Here, at
variance with the DPA, off-shell effects are fully included, and also
non-resonant topologies are taken into account.  This is crucial for
analyses targeted at off-shell phase-space regions, such as $H\to\PV\PV$
measurements, but also for lepton-$\pT$ distributions and other
observables~\cite{Biedermann:2016guo}.

Besides the dominant $q\bar q$ annihilation channel, 
also the \aa channel
enters $pp\to \wpwm$ at leading order~(\LO), contributing twice as much as the
$c\bar{c}$-channel.  The \aa channel raises the inclusive cross section by
about $+1\%$ at \LO, and, due to the comparably large photon PDF at high $x$,
it contributes significantly more at large transverse momenta or invariant
masses.  In the literature, photon-induced contributions to $pp\to\PV\PV$ are
typically included at \LO, and the corresponding \NLO \EW corrections have
been studied only for $\aa\to \wpwm$ at a \aa 
collider~\cite{Denner:1995jv,Jikia:1996uu, Bredenstein:2005zk}
and for the production of stable vector
bosons at the LHC~\cite{Baglio:2013toa}.
The quantitative impact of photon-induced diboson production and the related
uncertainty strongly depend on the photon distribution function (\aPDF)
supplied by the different PDF groups
\cite{Martin:2004dh,Ball:2013hta,Schmidt:2015zda,Manohar:2016nzj}.

In this paper we present new \NLO calculations of $pp\to 2\ell2\nu$ that
extend previous results in various directions. 
First, we include both \NLO \QCD and \EW corrections and
address also the issue of their combination, which is of particular
relevance in phase-space regions where both types of corrections are large,
e.g.\ in the tails of transverse-momentum distributions. 
Second, besides revisiting the  different-flavour $\llnndf$ channel, for
the first time we also study the same-flavour $\llnnsf$ channel at \NLO
\EW, including all relevant off-shell and non-resonant effects,
as well as interferences and spin correlations. 
In the same-flavour channel, we investigate the relative importance of 
\wpwm and \zz resonances and of their interference.
In particular, while \zz resonances are generally 
subdominant, we point out that for certain 
distributions they can play a significant role.
Third, at variance with previous studies, we treat $q\bar q$- and
$\aa$-induced channels on the same footing, including \NLO \EW corrections
throughout, and not only for the $q\bar q$ channel.  
In this respect, we note that the \EW corrections to 
the $q\bar q$ channel involve $q\Pa$-induced processes that 
are related---via cancellations of collinear singularities---to the
\EW corrections to the \aa channel. Thus, the \EW corrections 
to the \aa channel are mandatory for a 
fully consistent treatment of $pp\to
2\ell2\nu$ at \NLO \EW.
Fourth, we assess the importance of photon-induced contributions 
and related uncertainties based on various state-of-the-art 
PDFs and their comparison.
Fifth, we study a convenient approximation of the \EW corrections
amenable to a simplified form of matching to parton showers and 
multi-jet merging at \NLO \QCDpEW ~\cite{Kallweit:2015dum}. Specifically, we 
consider IR regularised virtual \EW corrections supplemented with 
\QED radiation as described by \YFS soft-photon resummation or, alternatively, 
by a \QED parton shower. 

Finally, motivated by subtleties that arise from photon-induced 
processes at \NLO \EW, we present a complete $\ord(\alpha)$ analysis 
of the interplay between the definition of the electromagnetic coupling  
and the renormalisation of the photon wave function and of the \aPDF in 
processes with external photons. In particular, we demonstrate that,
in order to avoid large logarithms of the light-quark and lepton masses
associated with $\Delta \alpha(\MZ^2)$, the coupling of initial-state 
photons should be defined at at the scale $\muf^2$ or at the \EW scale,
using, for instance, the $G_\mu$ scheme or $\alpha(M_Z)$ scheme. 
This was first pointed out in~\cite{Harland-Lang:2016lhw}, based on 
considerations related to the evolution of the \aPDF at \LO.
In contrast, as is well known, for final-state photons $\alpha(0)$ should be used. 

The calculations presented in this paper have been performed with the
fully automated \NLO \QCDpEW framework~\cite{Kallweit:2014xda,Kallweit:2015dum}
provided by the \OpenLoops matrix-element generator~\cite{Cascioli:2011va,hepforge} 
in combination with the Monte Carlo programs \Munich~\cite{munich} and
\Sherpa~\cite{Gleisberg:2008ta,Krauss:2001iv,Gleisberg:2007md,Schonherr:2017qcj}.  

This paper is organised as follows: In \refse{se:anatomy} we introduce 
general features and ingredients of $pp\to 2\ell 2\nu$, 
while technical aspects of the calculations are detailed 
in \refse{se:setup}.  Numerical predictions for the 13\,\TeV LHC are 
presented in \refse{se:results}, with emphasis on the 
behaviour of \QCD and \EW corrections, and our findings are summarised in 
\refse{se:conclusions}.  In \refapp{app:subren} we
document the implementation of Catani--Seymour subtraction at $\ord(\alpha)$
in \Sherpa and \Munich, and we discuss the issue of the definition and
renormalisation of $\alpha$ for processes with external photons.  Technical
details related to the separation of single-top contamination at \NLO \QCD
are addressed in \refapp {se:conversion}. 
\refapp{app:splitcorr} details a breakdown of the electroweak corrections 
presented in \refse{se:results} by flavour channels.
Finally, in \refapp{app:xsecs} we
present benchmark cross sections for $pp\to 2\ell2\nu$ in various fiducial
regions.

\section{Anatomy of  hadronic \texorpdfstring{$\boldsymbol{2\ell2\nu}$}{2l2n} 
         production at \NLO \texorpdfstring{$\boldsymbol{\QCD\!+\!\EW}$}{QCD+EW}}
\label{se:anatomy}

\subsection[Categorisation of \texorpdfstring{$2\ell2\nu$}{2l2n} final states]
           {Categorisation of $\boldsymbol{2\ell2\nu}$ final states}
\label{se:categorisation}
In the Standard Model, the signature of two opposite-charged leptons and missing energy 
is dominantly produced through $pp\to \wpwm/\zz\to 2\ell2\nu$, \ie with 
two types of diboson resonances that decay into two leptons and two neutrinos.
Such signatures can be categorized according to the flavour of the two charged leptons 
into a different-flavour~(DF) mode and a same-flavour~(SF) 
mode, with different implications
on the underlying production mechanisms.
We restrict our discussion to final states with electrons and muons, 
and we focus on $pp\to 2\ell2\nu$  processes with 
DF and SF final states corresponding, respectively, to 
 $\llmetdf$ and $\llmetsf$. Note that such processes are invariant with respect to
$\Pe\leftrightarrow\Pmu$ interchange. More precisely, taking into account appropriate momentum mappings, we have
\beq
  \label{eq:emusymmetry}
  \begin{split}
    \rd\sigma(pp\to \llnndfflip)
    \;=\;&\;
    \rd\sigma(pp\to \llnndf)\;,
    \\
    \rd\sigma(pp\to \llnnsfwwzzflip)
    \;=\;&\;
    \rd\sigma(pp\to\llnnsfwwzz )\;,
    \\
    \rd\sigma(pp\to \llnnsfzzflip)
    \;=\;&\;
    \rd\sigma(pp\to \llnnsfzz)\;.
  \end{split}
\eeq

In our calculation we do not apply any resonance approximation, but include the full set of 
Feynman diagrams that contribute to $pp\to 2\ell2\nu$
at each perturbative order,
thereby including all sub-dominant contributions with single- and non-resonant diagrams besides the
dominant double-resonant ones. All off-shell effects, interferences and spin correlations are consistently
taken into account, treating resonances in the complex-mass scheme~\cite{Denner:2005fg} throughout.

\begin{figure}
\begin{center}
\begin{tabular}{ccccc}
\includegraphics[width=.20\textwidth]{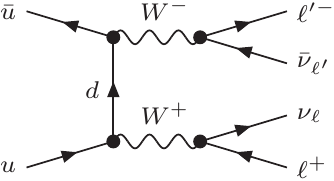} & &
\includegraphics[width=.20\textwidth]{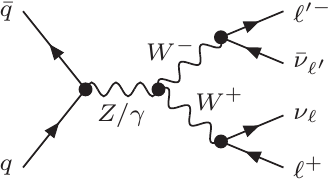} & &
\includegraphics[width=.20\textwidth]{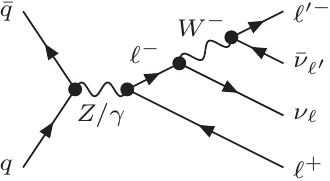} \\[0ex]
(a) & & (b) & & (c)
\end{tabular}
\end{center}
\caption[]{
  \label{fig:BorndiagramsWW}
  Sample of Born diagrams contributing to $2\ell2\nu$ 
  production in the different-flavour case ($\ell\neq \ell^\prime$) 
  and in the same-flavour case ($\ell=\ell^\prime$). 
  Both double-resonant (a,b) and single-resonant (c) diagrams are shown.
}
\vspace*{.5ex}
\begin{center}
\begin{tabular}{ccc}
\includegraphics[width=.20\textwidth]{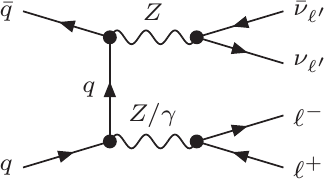} & &
\includegraphics[width=.20\textwidth]{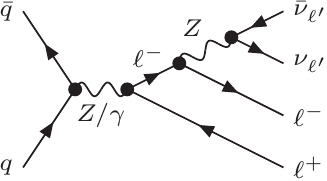} \\[0ex]
(a) & & (b)
\end{tabular}
\end{center}
\caption[]{
  \label{fig:BorndiagramsZZ}
  Sample of Born diagrams contributing to $2\ell2\nu$ 
  final states only in the case of same lepton flavour (neutrinos can have 
  flavour $\ell'=\ell$ or $\ell'\neq\ell$).  Both double-resonant (a) and 
  single-resonant (b) diagrams are shown.
}
\end{figure}

At \LO, the DF process $pp\to \llnndf$, 
is dominated by resonant $\wpwm$ production
in the $q\bar q$ channel and subsequent decays. 
The full set of Feynman diagrams contributing to $pp\to \llnndf$
will be referred to as \DFWW{} channel. 
Representative tree-level diagrams both for double-resonant
and sub-leading contributions are shown in \reffi{fig:BorndiagramsWW}.

The situation in the SF case is 
more involved since its signature can be produced by different partonic
processes, $pp\to \llnnsfzz$ and 
$pp\to \llnnsfwwzz$. Their final states are indistinguishable on an event-wise level, as the produced 
neutrinos can only be detected as missing transverse energy and their flavours cannot be resolved.
Consequently, predictions for $\llmetsf$ production originate as the incoherent sum over all three possible 
neutrino-flavour contributions.

The SF process $pp\to \llnnsfzz$ is 
dominated by resonant \zz production in $q\bar q$ annihilation
and subsequent $\PZ\to \Pe^+\Pe^-$ and $\PZ\to \nu\bar \nu$ decays.
Such double-resonant contributions are accompanied by all allowed
topologies with sub-leading resonance structures, including diagrams with 
$\Pa^*\to \Pe^+\Pe^-$ subtopologies, as well as other single- and non-resonant topologies. 
The full set of Feynman diagrams contributing to $pp\to \llnnsfzz$
will be referred to as \SFZZ{} channel.
Sample tree-level diagrams are 
depicted in \reffi{fig:BorndiagramsZZ}.

Finally, the SF process $pp\to \llnnsfwwzz$ proceeds both via 
\wpwm and \zz diboson resonances. The corresponding amplitudes are built 
by coherently summing over all
diagrams entering the two previously discussed \DFWW{} and \SFZZ{} channels. 
Consequently, this channel is referred to
as \SFWWZZ{} channel, and all diagrams shown in \reffis{fig:BorndiagramsWW}{fig:BorndiagramsZZ}
are representatives of the tree-level diagrams contributing here. 

Due to the fact that the phase-space regions with resonant intermediate \wpwm
and \zz states are typically distinct, the assumption is justified that the
\SFWWZZ{} cross section is dominated by the incoherent sum of double-resonant contributions 
of one and the
other type, while the effect of quantum interferences is small.  It is,
however, not obvious if this assumption still holds in phase-space regions
away from such double-resonant topologies.  Interference effects 
are studied in detail in
\refse{se:sfresults} by comparing exact predictions in the \SFWWZZ{} channel against the
incoherent sum of the \wpwm and \zz channels.

\subsection{Photon-induced production}
\begin{figure}
\begin{center}
\begin{tabular}{ccccccc}
\includegraphics[width=.20\textwidth]{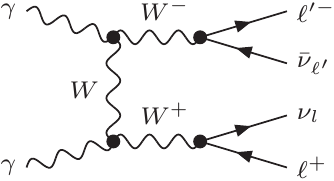} & &
\includegraphics[width=.20\textwidth]{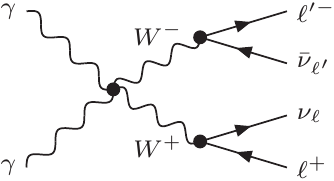} & &
\includegraphics[width=.20\textwidth]{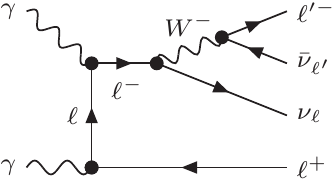} & &
\includegraphics[width=.20\textwidth]{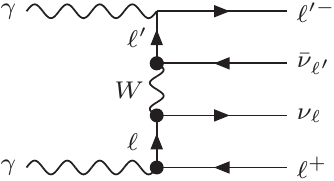} \\[0ex]
(a) & & (b) & & (c) & & (d)
\end{tabular}
\end{center}
\caption[]{
  \label{fig:AABorndiagramsWW}
  Sample of photon-induced Born diagrams contributing to $2\ell2\nu$ 
  production in the different-flavour case ($\ell\neq \ell^\prime$) 
  and in the same-flavour case ($\ell=\ell^\prime$). 
  Double-resonant~(a,b), single-resonant~(c) and non-resonant~(d) 
  diagrams are shown.
}
\vspace*{.5ex}
\begin{center}
\begin{tabular}{ccc}
\includegraphics[width=.20\textwidth]{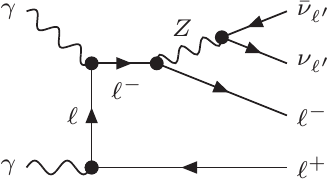} & \hspace{10mm} &
\includegraphics[width=.20\textwidth]{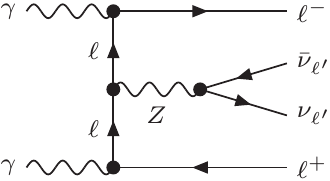} \\[0ex]
(a) & & (b)
\end{tabular}
\end{center}
\caption[]{
  \label{fig:AABorndiagramsZZ}
  Sample of photon-induced Born diagrams contributing to $2\ell2\nu$ 
  final states only in the same lepton-flavour case, both for $\ell'=\ell$ 
  or $\ell'\neq\ell$. Only single-resonant diagrams contribute.
}
\end{figure}

Besides the dominant $q\bar q$ production mode,  $2\ell2\nu$ final states can
also be produced in photon--photon scattering. As we do not count the 
photon PDF as an $\ord(\alpha)$ suppressed quantity, such $\aa\to 2\ell2\nu$ processes
contribute already at the \LO, \ie at  $\ord(\alpha^4)$.
Their quantitative relevance varies significantly between the channels.
Photon-induced contributions to the DF channel are dominated 
by $\aa\to\wpwm\to\llnndf$ topologies, which are accompanied
by single-resonant topologies involving $t$-channel lepton-pair 
production with an emission of a \PW boson off
one of the produced leptons, and non-resonant diagrams with multiperipheral topologies. 
Sample tree diagrams for the described DF topologies 
are collected in \reffi{fig:AABorndiagramsWW}. 
Due to a $t$-channel pole, regulated by the \PW mass, the 
contribution of the double-resonant diagram depicted 
in \reffi{fig:AABorndiagramsWW}(a) is  enhanced for large invariant masses of the 
intermediate \wpwm pair~\cite{Bierweiler:2012kw,Baglio:2013toa}. 
In fact, for on-shell \wpwm pair production the contribution of the \aa channel was found to 
increase beyond $10\%$ of the \LO $q\bar q$ annihilation 
mode for $m_{\ww}>800\,\GeV$~\cite{Bierweiler:2012kw}. 
In this paper we investigate  the significance of the \Pa-induced production mode 
using state-of-the-art PDFs and taking into account \NLO \EW corrections, as well as realistic 
selection cuts on the ${2\ell2\nu}$ final state.

The DF channel $\aa\to \llnnsfzz$ does not involve any double-resonant 
topology due the lack of triple and quartic gauge couplings among neutral \EW bosons. 
Similarly, non-resonant multiperipheral topologies do not exist due to lepton-flavour 
conservation. 
Thus, lepton-pair production in $t$-channel topologies with subsequent 
emission of a \PZ boson with $\PZ\to\nu\bar\nu$ 
is the only photon-induced production mechanism at \LO, as shown 
in the sample diagrams of \reffi{fig:AABorndiagramsZZ}.
Consequently, the invariant mass of the charged-lepton
pair does not 
show a Breit--Wigner peak around $\MZ$.

Similarly as for quark--antiquark annihilation, the $\aa\to\llnnsfwwzz$ channel
is build from the coherent sum of all diagrams entering
$\aa\to\llnndf$ and $\aa\to\llnnsfzz$.

\subsection{Ingredients of \QCD and \EW corrections}
\label{se:ingredients}

\begin{figure}
\begin{center}
\begin{tabular}{ccccccc}
\includegraphics[width=.20\textwidth]{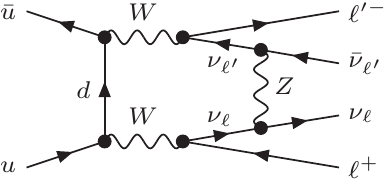} & &
\includegraphics[width=.20\textwidth]{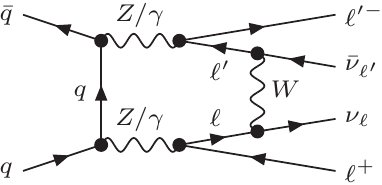} & &
\includegraphics[width=.20\textwidth]{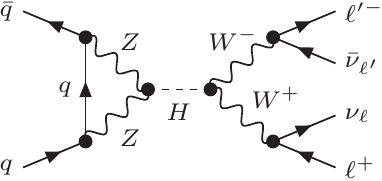} & &
\includegraphics[width=.20\textwidth]{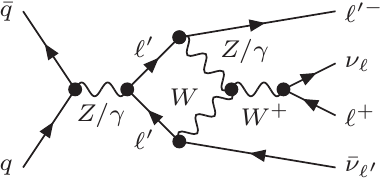} \\[0ex]
(a) & & (b) & & (c) & & (d) \\[2ex]
\includegraphics[width=.20\textwidth]{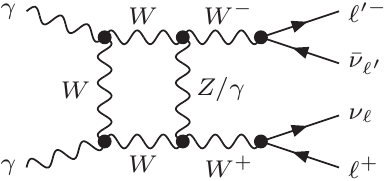} & &
\includegraphics[width=.20\textwidth]{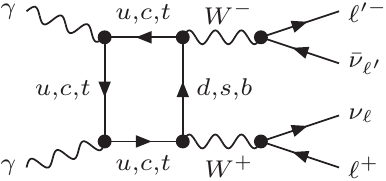} & &
\includegraphics[width=.20\textwidth]{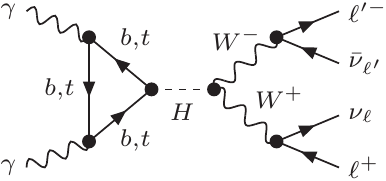} & &
\includegraphics[width=.20\textwidth]{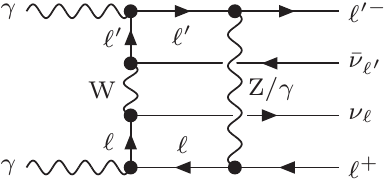} \\[0ex]
(e) & & (f) & & (g) & & (h)
\end{tabular}
\end{center}
\caption[]{
  \label{fig:VirtdiagramsWW}
  Sample of one loop diagrams contributing to $2\ell2\nu$ production in 
  the different-flavour case ($\ell\neq \ell^\prime$) and in the 
  same-flavour case ($\ell=\ell^\prime$) in the quark-induced (a-d) 
  and photon-induced (e-h) channels.}
\vspace*{.5ex}
\begin{center}
\begin{tabular}{ccccccc}
\includegraphics[width=.20\textwidth]{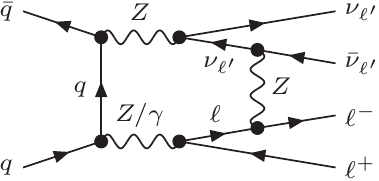} & &
\includegraphics[width=.20\textwidth]{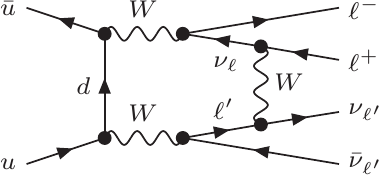} & &
\includegraphics[width=.20\textwidth]{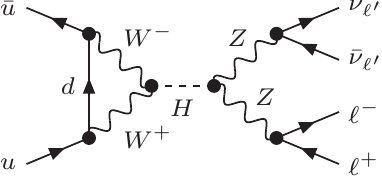} & &
\includegraphics[width=.20\textwidth]{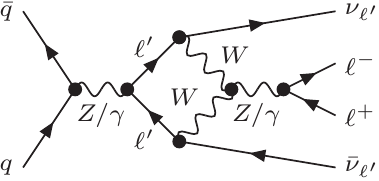} \\[0ex]
(a) & & (b) & & (c) & & (d) \\[2ex]
\includegraphics[width=.20\textwidth]{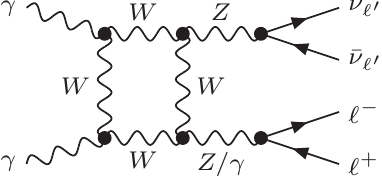} & &
\includegraphics[width=.20\textwidth]{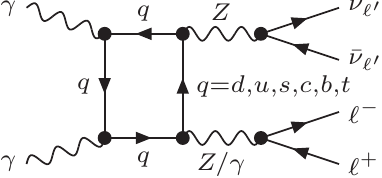} & &
\includegraphics[width=.20\textwidth]{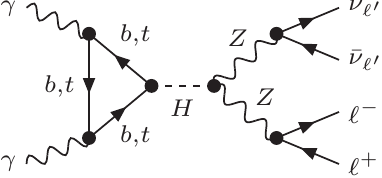} & &
\includegraphics[width=.20\textwidth]{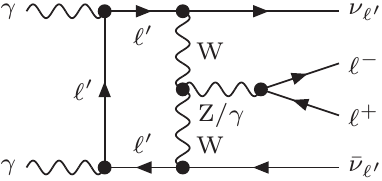} \\[0ex]
(e) & & (f) & & (g) & & (h)
\end{tabular}
\end{center}
\caption[]{
  \label{fig:VirtdiagramsZZ}
  Sample of one-loop diagrams contributing to $2\ell2\nu$ final states 
  only in the same-flavour (wrt.\ the charged leptons) case in the 
  quark-induced (a-d) and photon-induced (e-h) channels.}
\end{figure}

\begin{figure}
\begin{center}
\begin{tabular}{ccccccc}
\includegraphics[width=.20\textwidth]{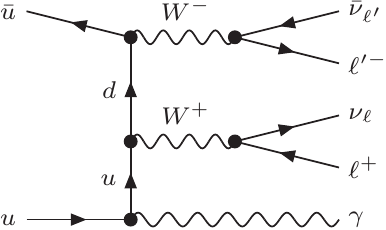} & &
\includegraphics[width=.20\textwidth]{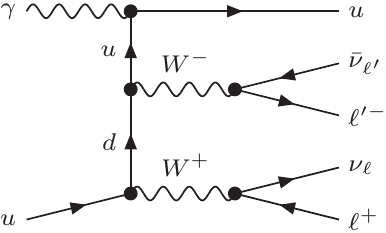} & &
\includegraphics[width=.20\textwidth]{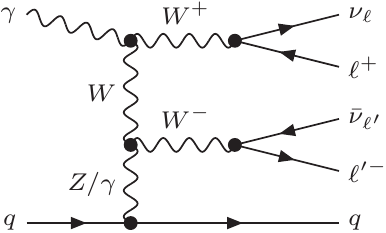} & &
\includegraphics[width=.20\textwidth]{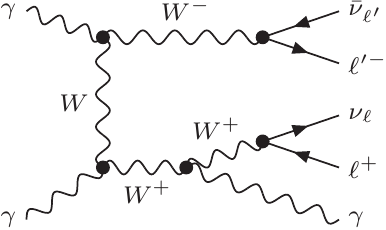} \\[0ex]
(a) & & (b) & & (c) & & (d)
\end{tabular}
\end{center}
\caption[]{
  \label{fig:RealdiagramsWW}
  Sample of real emission diagrams contributing to $2\ell2\nu$ production in 
  the different-flavour case ($\ell\neq \ell^\prime$) and in the 
  same-flavour case ($\ell=\ell^\prime$), in the 
  quark--antiquark channel (a), the (anti-)quark--photon channel (b,c) and 
  the photon--photon channel (d).}
\vspace*{.5ex}
\begin{center}
\begin{tabular}{ccccccc}
\includegraphics[width=.20\textwidth]{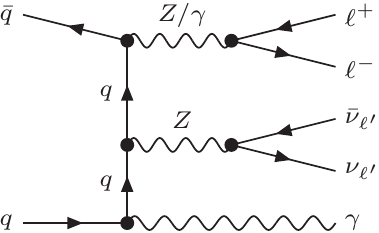} & &
\includegraphics[width=.20\textwidth]{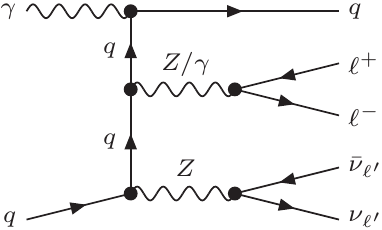} & &
\includegraphics[width=.20\textwidth]{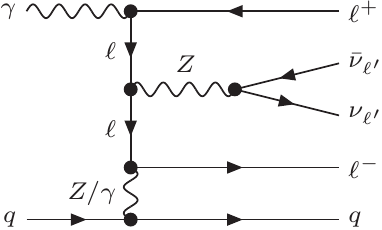} & &
\includegraphics[width=.20\textwidth]{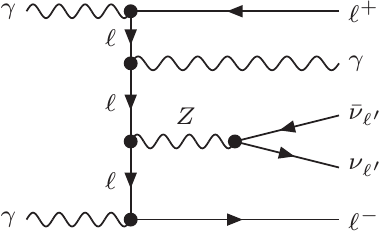} \\[0ex]
(a) & & (b) & & (c) & & (d)
\end{tabular}
\end{center}
\caption[]{
  \label{fig:RealdiagramsZZ}
  Sample of real emission diagrams contributing to $2\ell2\nu$ final states 
  only in the same-flavour (wrt.\ the charged leptons) case in the 
  quark--antiquark channel (a), the (anti-)quark--photon channel (b,c) and 
  the photon--photon channel (d).}
\end{figure}

At \NLO \QCD all $\ord(\alpha_s\alpha^4)$ contributions
to $pp\to 2\ell2\nu$ are taken into account.
In the $q\bar q$  channel, the only \QCD loop corrections arise 
from virtual-gluon exchange, while the real corrections 
result from real-gluon emission and crossed topologies describing 
(anti-)quark--gluon channels. 
The infrared divergences separately arising in these two contributions 
are mediated by the standard dipole-subtraction approach 
\cite{Catani:1996vz,Catani:2002hc}.
We note that the \aa channels do not receive \QCD corrections 
at \NLO, due to the absence of any \QCD partons in all tree-level diagrams.

At \NLO \EW we include the full set of $\ord(\alpha^5)$ contributions to $pp\to 2\ell2\nu$.
At this order  both the $q\bar q$ and \aa channels receive 
corrections from virtual \EW bosons
and from closed fermion loops, cf.\ 
\reffis{fig:VirtdiagramsWW}{fig:VirtdiagramsZZ}. 
These corrections include Higgs resonances with decay into four fermions 
coupled to weak bosons (in the $q\bar q$ channel) or 
coupled to a heavy-fermion loop (in the $\aa$ channel).
The real corrections in the $q\bar q$ channel can be split into 
real-photon emission channels and $\Pa q\to 2\ell2\nu q$ channels\footnote{Corresponding
$\Pa\bar q$-induced channels are implicitly understood here and in the following.} 
with initial-state $\Pa\to \Pq\Pqbar$ splittings.
The $\aa$ channel also receives real corrections from photon 
bremsstrahlung, and also from $\Pa q\to 2\ell2\nu q$ channels
with initial-state $q\to q\Pa$ splittings, cf.\ 
\reffis{fig:RealdiagramsWW}{fig:RealdiagramsZZ}.
While the separation into $q\bar q$ and \aa channels 
can still be preserved for virtual and photon-bremsstrahlung contributions, 
such separation is no longer meaningful for the 
\mbox{$q\Pa$-initiated} channels due to their singularity 
structure: both above-mentioned splittings result in 
infrared-divergent configurations, and these $q\Pa$ channels 
simultaneously cancel infrared poles arising in $q\bar q$ and 
$\aa$ channels.
This situation demands the inclusion of the full \NLO \EW corrections to 
the $q \bar q$ and \aa Born processes to guarantee infrared safety and consistency.
To deal with the mediation of these 
divergences between virtual and real corrections
the \QED extension of the dipole-subtraction 
method~\cite{Dittmaier:1999mb,Dittmaier:2008md,Gehrmann:2010ry} 
is applied (see \refapp{app:subren}).  

Instead of a separation of \NLO contributions into $q\bar q$ and \aa channels, 
we quantify the impact of photon-induced processes by 
considering contributions involving at least one photon 
PDF factor and all other contributions that are also present under the 
assumption of vanishing photon PDFs. At \LO this distinction 
coincides with the splitting according to production modes, while at \NLO \EW
it combines \aa and $\Pa q$ channels 
in spite of the fact that the latter 
involves $q\bar q$-related contributions.

\section{Technical ingredients and setup of the simulations}
\label{se:setup}

\subsection{Tools}
\label{se:setup:tools}

The calculations presented in this paper have been performed with the 
automated frameworks \MunichOpenLoops and \SherpaOpenLoops. They automate 
the full chain of 
all
operations---from process definition to collider 
observables---that enter \NLO \QCDpEW simulations at parton level. 
The recently achieved automation of \EW corrections~\cite{Kallweit:2014xda,Kallweit:2015dum} is based on the well established 
\QCD implementations and allows for \NLO \QCDpEW simulations for a vast 
range of SM processes, up to high particle multiplicities, 
at current and future colliders. 

In these frameworks virtual amplitudes are provided by the \OpenLoops
program~\cite{hepforge}, which is based on the open-loops algorithm~\cite{Cascioli:2011va} --
a fast numerical recursion for the evaluation of one-loop scattering 
amplitudes.
Combined with the \Collier tensor reduction library
\cite{Denner:2014gla}, which implements the Denner--Dittmaier reduction
techniques~\cite{Denner:2002ii,Denner:2005nn} and the scalar integrals
of~\cite{Denner:2010tr}, or with {\sc
CutTools}~\cite{Ossola:2007ax}, which 
implements the OPP method~\cite{Ossola:2006us}, together with the {\sc OneLOop} 
library~\cite{vanHameren:2010cp}, the employed recursion permits to achieve very 
high CPU performance and a high degree of numerical stability.  
We validated phase-space point wise the contributing tree amplitudes between \Sherpa and \OpenLoops,
and the one-loop amplitudes between \OpenLoops and an in-house algebraic amplitude
generator and also against \Recola~\cite{Actis:2016mpe}. 
All remaining tasks, \ie the bookkeeping of
partonic subprocesses, phase-space integration, and the subtraction of \QCD
and \QED bremsstrahlung, are supported by the two independent and fully
automated Monte Carlo generators, \Munich~\cite{munich} and
\Sherpa~\cite{Gleisberg:2008ta,Krauss:2001iv,Gleisberg:2007md,Schonherr:2017qcj}.  
These two tools have been validated extensively against each other. As a
further validation of the Monte Carlo integration employed for the results presented here,
we want to note the perfect agreement between \Sherpa and the results 
of \cite{Biedermann:2016guo} for the related process $pp\to 4\ell$ presented in \cite{Biedermann:2017yoi}.

\subsection{YFS soft-photon resummation and QED parton shower}
\label{se:setup:yfs}

As discussed in \refse{sec:intro}, the \NLO \EW corrections to $pp\to
2\ell2\nu$ are dominated by \EW Sudakov logarithms of virtual origin and \QED
logarithms stemming from photon radiation off leptons. 
In~\cite{Kallweit:2015dum} it was shown that, for observables that are
sufficiently inclusive with respect to photon radiation, full \NLO \EW results
can be reproduced with good accuracy by an approximation 
consisting only of
virtual \EW corrections upon appropriate subtraction of IR singularities. 
This approximation, which was dubbed \EWvirt, 
is defined through
\beq
  \label{eq:ewvi}
  \begin{split}
    \rd\sigma_\text{\NLO \EWvirt}
    \,=\;&
      \rd\sigma_\text{\LO}
      +\rd\sigma_{\EW}^\text{V}
      +\rd\sigma_{\EW}^\text{I}
    \,=\,
      \rd\sigma_{\LO}\left(1+\delta_{\EWvirt}\right)\;.
  \end{split}
\eeq
Therein, $\rd\sigma_{\LO}$ is the leading order differential 
cross section, while $\rd\sigma_{\EW}^\text{V}$ and 
$\rd\sigma_{\EW}^\text{I}$ are the NLO EW virtual correction 
and the endpoint part of the integrated Catani-Seymour subtraction 
terms, ensuring a finite result by construction. In practice, 
a logarithmic approximation over the real photon emission phase space 
is added to the virtual corrections. This approach 
captures all Sudakov effects at \NLO \EW \cite{Kuhn:2007cv} and is very suitable 
for a combination of \QCD and \EW higher-order effects through
a simplified multi-jet merging approach at 
\NLO \QCDpEW~\cite{Kallweit:2015dum}.
As a further possible step towards a fully consistent implementation of
matching and merging at \NLO \QCDpEW, in this paper we investigate the
possibility of supplementing the \EWvirt approximation with \QED radiation
effects by means of naive matching to \QED parton showers or \QED resummation.  Specifically,
we  consider a soft-photon resummation in the Yennie-Frautschi-Suura
(\YFS) scheme \cite{Yennie:1961ad} and, alternatively, the \CSShower \QED parton
shower~\cite{Schumann:2007mg,Hoeche:2009xc} based on Catani-Seymour
splitting kernels. Combined with the \EWvirt approximation and a 
differentially applied \NLO \QCD $K$-factor, the \NLOQCDtEWYFS and 
\NLOQCDtEWCSS approximations are defined. As in the fixed-order 
calculation, both $q\bar q$ and \aa channels are taken into account on 
the same footing.

The original \YFS scheme resums real and virtual soft-photon corrections 
to arbitrary scattering processes. The implementation in \Sherpa 
\cite{Schonherr:2008av} is specialised to correct decays of 
massive resonances, and for both 
cases  relevant in this paper, \ie for \PW and \PZ resonances,
the resummation of soft logarithms is matched to exact 
$\ord(\alpha)$ corrections.\footnote{
  To be precise, the virtual corrections used neglect terms of 
  $\ord\left(m_\ell^2/m_V^2\right)$ or higher, which are however 
  negligible.
}  
Its accuracy 
in charged- and neutral-current Drell-Yan processes has 
been validated in \cite{Badger:2016bpw,Alioli:2016fum}. 
As neither photon emissions off the initial states nor 
$\Pa\to f\bar f$ splittings are included,
it is worth 
noting that no $\Pa q$ channels, occurring in the real emission 
correction of the full calculation, are accounted for.

The \YFS implementation in \Sherpa includes a generic resonance 
identification, ensuring that collective multipole radiation off 
the charged lepton ensemble preserves all resonance structures 
present in the event. To this end, 
first the final state of a scattering process is analysed, 
and possible resonances decaying into leptons and neutrinos 
are identified on the basis of event kinematics 
and existing vertices in the model. 
For the process studied in this paper, $pp \to 2\ell 2\nu$,
multiple resonance structures are possible. 
They are disentangled on the basis of the distance measures\footnote{
  We choose to identify a resonance only if
  $\Delta_{\PZ,\PW}<\Delta_{\mathrm{cut}}=10$. Thus, 
  in the far off-shell regions no resonance is identified.
  We have checked that the results presented here are independent of 
  $\Delta_{\mathrm{cut}}$ if it is chosen not too small, 
  which would exclude higher-order corrections for a 
  significant resonant phase-space region.
} 
$\Delta_\PZ=|m_{\ell\ell}-\MZ|/\GZ$ and
$\Delta_\PW=|m_{\ell\nu}-\MW|/\GW$.
In $2\ell 2\nu$ production this leads to three distinct cases: 
(a) two pairs of leptons are identified to come from a specific resonance; 
(b) one pair of leptons is identified to come from a specific resonance, 
the other is classified as non-resonant;
(c) all leptons are classified as non-resonant. 
Subsequently, identified resonant-production subprocesses are separated 
from the rest of the event, 
and the emerging decay is dressed with photon radiation 
respecting the Breit--Wigner distribution of the resonance,
\ie preserving the original virtuality of the off-shell 
lepton/neutrino system.
Finally, all left-over non-resonantly produced leptons are grouped in a 
fictitious $X\to n\ell+m\nu$ process, with suitably adjusted charges and 
masses for $X$. In this case, resummed real and virtual radiative \QED corrections are 
applied in the soft limit only, including however hard 
collinear real-emission corrections through suitably subtracted 
Catani-Seymour dipole splitting functions~\cite{Schonherr:2008av}. 

In the \CSShower, the construction of the emitting dipoles 
follows the subtraction terms used in the fixed-order 
calculation. Owing to the unitary nature of all parton showers, 
dipoles whose splitting functions are negative, \ie all 
dipoles formed by partons with like-sign electric charges,
are inactive and do not contribute.\footnote{
  Radiation from negative-valued splitting functions could in 
  principle be taken into account using the algorithms of 
  \cite{Hoeche:2009xc,Hoeche:2011fd}, but are not implemented 
  in the general shower.
}
In the \QCD case this corresponds to the leading-colour limit, and
keeping  $\CF$ and $\CA$ at their $\NC=3$ values guarantees a full-colour treatment of 
the collinear limit, while the soft-limit remains at 
$\NC\to\infty$. No such limit is meaningful in \QED. Consequently, 
the absence of the like-signed dipoles has a degrading impact both 
on the description of the collinear and the soft limit.
Moreover, the \CSShower has no knowledge of the internal resonance 
structure of the Born process. Thus, dipoles of charged particles 
spanning across one or multiple resonances will inevitably distort 
their line shape through their recoil assignments.\footnote{ 
  Comparing various resonance blind recoil schemes 
  \cite{Carli:2010cg} and different evolution variables \cite{Hoeche:2014lxa}
  we found similar effects for all observables discussed in \refse{se:results}.
}
At the same time, however, all processes including photon 
radiation off the initial state quarks and $\Pa\to f\bar f$ splittings 
are present. Thus, every channel occurring in the fixed-order calculation 
is described in its respective soft-collinear limits.

\subsection{Input parameters, scale choices and variations}
\label{se:setup:params}

\begin{table}[tb]
  \begin{center}
    \begin{tabular}{rclrcl}
      $\GF$ & \shortequal & $1.1663787\cdot 10^{-5}~\GeV^2$ & \qquad\qquad & &\\
      $\MW$ & \shortequal & $80.385~\GeV$  & $\GW$ & \shortequal & $2.0897~\GeV$ \\
      $\MZ$ & \shortequal & $91.1876~\GeV$ & $\GZ$ & \shortequal & $2.4955~\GeV$ \\
      $\MH$ & \shortequal & $125~\GeV$     & $\GH$ & \shortequal & $4.07~\MeV$ \\
      $\Mb$ & \shortequal & $4.75~\GeV$    & $\Gb$ & \shortequal & $0$ \\
      $\Mt$ & \shortequal & $173.2~\GeV$   & $\Gt$ & \shortequal & $1.339~\GeV$ \\
      \\[-2ex] \hline\\[-2ex]
      $\Me$ & \shortequal & $511~\keV$     & $\Mmu$       & \shortequal & $105~\MeV$ \\
      $\alpha(0)$ & \shortequal & $1/137.03599976$ &&&
    \end{tabular}
  \end{center}
  \caption{
    Numerical values of all input parameters. The gauge boson masses are 
    taken from \cite{Agashe:2014kda}, while their widths are obtained from 
    state-of the art calculations. The Higgs mass and width are taken from 
    \cite{Heinemeyer:2013tqa}. The top quark mass is taken from 
    \cite{Agashe:2014kda} while its width has been calculated at \NLO \QCD.
    The electron and muon masses as well as the electromagnetic coupling 
    in the Thomson limit, $\alpha(0)$,  are only relevant for calculations involving \YFS 
    soft-photon resummation and the \CSShower.
    \label{tab:inputs}
  }
\end{table}

The input parameters for the 
\NLO \QCDpEW calculations of $pp\to2\ell2\nu$ presented in \refse{se:results}
are summarised in \refta{tab:inputs}.
All unstable particles are treated in the complex-mass scheme~\cite{Denner:2005fg},
where width effects are absorbed into the complex-valued renormalised squared masses
\beq\begin{split}\label{eq:complexmasses}
\mu^2_i\,=&\;\;M_i^2-\ri\Gamma_iM_i
\qquad\mbox{for}\;i=\PW,\PZ,\PH,\Pt\;.
\end{split}\eeq
As top-quark and Higgs-boson contributions enter only at loop level, 
the dependence of our results on $\Gamma_\Pt$ and $\Gamma_\PH$ is completely 
negligible. The CKM matrix is assumed to be diagonal.
In fact, due to the negligible mixing of the first two and the third quark 
generations and because all quarks of the first two quark generations are 
taken to be massless, the unitarity of the CKM matrix ensures the independence 
of all physical results from the values of its matrix elements.
The \EW couplings are derived from the gauge-boson masses 
and the Fermi constant using 
\beq\begin{split}\label{eq:defalpha}
\alpha\,=&\;\left|\frac{\sqrt{2}\;\sw^2\,\mu^2_\PW\,\GF}{\pi}\right|,
\end{split}\eeq
where the $\PW$-boson mass and the squared sine of the mixing angle, 
\beq\begin{split}\label{eq:defsintheta}
\sw^2\,=&\;\;1-\cw^2\,=\;1-\frac{\mu_\PW^2}{\mu_\PZ^2}\;,
\end{split}\eeq
are complex-valued.
The $G_\mu$-scheme guarantees an  optimal description of pure SU(2) interactions
at the \EW scale. It is used for all channels, including  
photon-induced ones. In this respect, while it is 
well known that the coupling of 
final-state photons should be parametrised in terms of 
$\alpha(0)$, in 
\refapp{app:subren} 
analysing
the interplay between the counterterms associated with the renormalisation 
of $\alpha$, the photon wave function, and the $\aPDF$, 
we demonstrate that the coupling of initial-state photons
cannot be parametrised in terms of $\alpha(0)$. Instead  
a high-energy definition of $\alpha$, for example in the $\alpha(M_Z)$- or the $G_{\mu}$-scheme, 
for the coupling of initial-state photons should be employed.

In all fixed-order results the renormalisation scale $\mur$ and 
factorisation scale $\muf$ are set to
\beq\begin{split}\label{eq:RFscales} 
\murf\,=&\;\xi_{\rR,\rF}\,\mu_0\;,
\quad\mbox{with}\quad 
\mu_0\,=\;\tfrac{1}{2}\,\HTl
\quad\mbox{and}\quad 
\tfrac{1}{2}\le \xi_{\rR},\xi_{\rF}\le 2\;.
\end{split}\eeq 
Therein, $\HTl$ is the scalar sum of the transverse momenta of all charged
final-state leptons plus the missing transverse momentum,
\beq\begin{split}\label{eq:scale} 
\HTl = \sum_{i\in \{\ell^\pm\}} p_{\rT,i} + \missingET\,, 
\end{split}\eeq
with $\missingET=\left|\vec p_{\rT,\nu} +\vec p_{\rT,\bar\nu}\right|$. 
In order to guarantee infrared safety at \NLO \EW, the scale of 
\refeq{eq:scale} must be insensitive to collinear photon emissions 
off charged leptons.  To this end, any charged leptons are dressed 
with collinear photons with $\Delta R_{\ell\Pa} < 0.1$. 
Our default scale choice corresponds to $\xi_{\rR}=\xi_{\rF}=1$,
and theoretical uncertainties are assessed by
applying the scale variations $(\xi_\rR,\xi_\rF)=(2,2)$,
$(2,1)$, $(1,2)$, $(1,1)$, $(1,\tfrac{1}{2})$, $(\tfrac{1}{2},1)$, 
$(\tfrac{1}{2},\tfrac{1}{2})$.
For all considered processes at the inclusive level the difference with 
respect to a fixed scale choice $\mu_0=\MW$ is below 2\% at \NLO \QCD, 
while inclusive \NLO \EW corrections agree at the level of one permille.

\subsection{PDFs}
\label{se:setup:PDFs}

For the calculation of hadron-level cross sections we employ the \CTQED
parton distributions~\cite{Schmidt:2015zda}, which include \NLO \QCD and 
\LO \QED effects,\footnote{
  To be precise we use the \texttt{CT14qed\_inc\_proton} set interfaced 
  through \LHAPDF 6.1.6 \cite{Buckley:2014ana}.
}
with the corresponding $\alphaS(\MZ)=0.118$. 
The \NLO PDF set is used for \LO computations as well as for 
\NLO \QCD and \NLO \EW predictions.  
In order to assess the potentially large uncertainties stemming from
photon-induced processes, two alternative sets based on different
determinations of the photon PDF are considered, namely the recently
calculated \LUXQED PDFs \cite{Manohar:2016nzj} and the data driven fit of
\NNPDFQED \cite{Ball:2014uwa,Ball:2013hta}.  Specifically, we replace the
photon PDF of the default set by the alternative parametrisations, while
using \CTQED quark and gluon  PDFs throughout.  This is justified by the
negligible dependence of the quark and gluon densities on the \aPDF.

The three considered sets implement different treatments of the 
photon PDF.
The \CTQED PDFs assume as initial condition for the \aPDF 
at $Q_0=1.295$\,\GeV an inelastic contribution that results from the 
convolution of primordial quark distributions with \QED splitting functions. 
This ansatz involves a free normalisation parameter, which is traded 
for the inelastic photon momentum fraction, 
$p_0^\Pa=\int_0^1\rd x \,x\, \Pa(x,Q_0)$, and fitted to DIS data 
with isolated photons.
For our default predictions we use a \CTQED set corresponding to the 
best fit value, $p_0^\Pa=0.05\%$.
The inelastic component, which describes processes where the 
proton breaks, is complemented by an elastic component,
corresponding to the case where the proton remains intact.
The latter is determined at the scale $Q_0$ using
the equivalent photon approximation (EPA)~\cite{Budnev:1974de}.
The sum of inelastic and elastic contributions at $Q_0$ 
is evolved as a single photon density%
\footnote{
  Note that, in contrast to ``inelastic photons'', which are 
  inherently off-shell, ``elastic'' photons as obtained form 
  the EPA at $Q_0^2$ are exactly on-shell, even when they enter 
  hard-scattering processes at $Q^2\gg m^2_p$.
  Nevertheless, also elastic photons can undergo 
  $\Pa\to q\bar q$ splittings at arbitrary $Q^2$. Thus, 
  elastic and inelastic photons contribute to the PDF evolution 
  towards high $Q^2$ on the same footing. In practice, the photon 
  PDF at high-$Q^2$ receives contributions form the elastic and 
  inelastic \aPDF at $Q_0^2$, both decreased due to 
  $\Pa\to q\bar q$ splittings, and positive contributions from 
  (anti)quark distributions via $q\to \Pa q$ splittings. It 
  turns out that, due to the much larger quark density, the latter 
  contributions dominate by far. Thus, the details of the evolution 
  of the elastic and inelastic \aPDF{}s play only a marginal 
  role~\cite{Yuan:2016xxx,Yuan:2016xxy}.
}
through coupled DGLAP equations for photons, quarks and gluons at 
\NLO \QCD{}\,+\,\LO \QED. 

In the \LUXQED  approach, the usual description of $\Pe p\to \Pe+X$ data,
where a virtual photon radiated from the electron beam probes quarks inside
the proton via $\Pa^*q$ scattering, is related to an alternative
interpretation, where the lepton beam probes the photon content of
the proton via $\ell\Pa$ scattering.
In this way, the photon density can be derived from 
proton structure functions in a model-independent way, and
building on available  global fits of \QCD PDFs, 
parametrisations of $\Pe p$ data at low $Q^2$, and elastic contributions, one arrives
at an accurate determination of the \aPDF.
Then, starting at $Q_0=10\,\GeV$, 
the photon density is evolved with all other \QCD partons 
through DGLAP equations including \QED 
corrections up to $\ord(\alphaS\alpha)$. 

The \NNPDFQED photon PDF is based on a much more general 
multiparameter neural-network parametrisation, which can naturally 
account for both the elastic and inelastic components.
Thus the \NNPDFQED photon density is much more receptive to the poor 
sensitivity of current data to photon--induced processes. This leads to 
much larger admissible photon densities combined with much bigger 
uncertainties as compared to the other PDF sets.
The resulting photon density is evolved at 
\NLO \QCD{}\,+\,\LO \QED. 

In order to avoid undesired contaminations from single-top contributions of type
$pp\to\PW\Pt\to \PW\PW\Pb\to 2\ell2\nu$
in the \NLO \QCD and \NLO \EW corrections, 
in our calculations we apply a full veto against final-state
\Pb{}-quarks.
Since such a veto would jeopardize IR cancellations for $\Mb=0$,
we consider the \Pb{}-quark to be massive, i.e.\ we assume the presence of 
only four light flavours. In order to reconcile this choice with the 
fact that the employed PDFs involve five active flavours, 
an appropriate PDF-scheme conversion~\cite{Cacciari:1998it} is applied. 
As discussed in \refapp{se:conversion}, this transformation 
is almost trivial for the process at hand. 
At LO,
$pp\to 2\ell2\nu$ 
comprises neither gluon channels 
nor $\alphaS$ terms. Thus, only the \aa channel requires
a correction related to the scheme dependence of the \aPDF.
Taking this into account,
we can safely perform our calculations using five-flavour PDFs,
omitting initial- and final-state \Pb{}-quarks,
and using $\Mb>0$ in the loops. 
Up to terms beyond NLO QCD+EW, this approach is 
perfectly consistent with a conventional calculation in the 4F scheme.

\section{Results}
\label{se:results}

In this section we present numerical predictions for 
the DF
and SF processes, $pp\to\llnndf$ and 
$pp\to\llnnsf$, at $\sqrt{s}=13$\,\TeV.
The impact of \NLO corrections is illustrated 
by comparing against \LO predictions, which 
include $q\bar q$ and $\aa$-induced processes at $\ord(\alpha^4)$.
For the combination of \QCD and \EW higher-order effects 
we consider both an additive and a multiplicative approach, defined, respectively, as
\beq
  \label{eq:EW+QCDfactorisation}
  \rd\sigma_\text{\NLO \QCDpEW}
  =
  \rd\sigma_\text{\LO}
  \left(1+\delta_\text{\QCD}+\delta_\text{\EW}\right)
\eeq
and
\beq
  \label{eq:EWxQCDfactorisation}
  \rd\sigma_\text{\NLO \QCDtEW}
  =
  \rd\sigma_\text{\LO}
  \left(1+\delta_\text{\QCD}\right)
  \left(1+\delta_\text{\EW}\right)\;.
\eeq
Therein, the relative \QCD and \EW corrections are defined as
\beq
  \label{eq:delta_QCD_EW}
  \delta_\text{\QCD}
  =
  \frac{\rd\sigma_{(1,4)}}{\rd\sigma_{(0,4)}}
  \qquad\text{and}\qquad
  \delta_\text{\EW}
  =
  \frac{\rd\sigma_{(0,5)}}{\rd\sigma_{(0,4)}}\;,
\eeq
where the $\rd\sigma_{(i,j)}$ are the cross section contributions of 
$\ord(\alphaS^i\alpha^j)$, thus $\rd\sigma_{(0,4)}\equiv\rd\sigma_\text{\LO}$. 
In order to illustrate the interplay of the various partonic channels
in the multiplicative \QCDtEW combination,
we write each $\rd\sigma_{(i,j)}$ as a sum over contributions
$\rd\sigma_{(i,j)}^{ab}$ where $a$ and $b$ are the proton constituents 
initiating the subprocess at the given order.
At LO, for the decomposition into 
 $q\bar q$ and $\gamma\gamma$
channels and their relative weights we write
\beq
 \rd\sigma_\text{\LO} = 
\rd\sigma_\text{(0,4)}^{\Pq\Pqbar}
		+\rd\sigma_\text{(0,4)}^{\aa}
\eeq
and
\beq
\epsqq=\frac{\rd\sigma_\text{(0,4)}^{\Pq\Pqbar}}
	       {\rd\sigma_\text{(0,4)}^{\Pq\Pqbar}
		+\rd\sigma_\text{(0,4)}^{\aa}}\;,\qquad
\epsaa=1-\epsqq=
\frac{\rd\sigma_\text{(0,4)}^{\aa}}
	       {\rd\sigma_\text{(0,4)}^{\Pq\Pqbar}
		+\rd\sigma_\text{(0,4)}^{\aa}}\;.
\eeq
At NLO, the QCD correction factor in~\refeq{eq:delta_QCD_EW}
corresponds to 
\beq
  \label{eq:def_deltaQCD}
    \delta_\text{\QCD}
    \,=\, \frac{\rd\sigma_{(1,4)}^{\Pq\Pqbar}
	        +\rd\sigma_{(1,4)}^{g\Pq/g\Pqbar}}
	       {\rd\sigma_\text{(0,4)}^{\Pq\Pqbar}
		+\rd\sigma_\text{(0,4)}^{\aa}}
    \,=\, \epsqq\,\delta_\text{\QCD}^{\Pq\Pqbar}\,,
\eeq
where the relative correction 
\beq
\delta_\text{\QCD}^{\Pq\Pqbar}
    \,=\, \frac{\rd\sigma_{(1,4)}^{\Pq\Pqbar}
	        +\rd\sigma_{(1,4)}^{g\Pq/g\Pqbar}}
	       {\rd\sigma_\text{(0,4)}^{\Pq\Pqbar}}
\eeq
is restricted to the $q\bar q $ channel. Finally, for the EW correction
in~\refeq{eq:delta_QCD_EW} we have
\beq
  \label{eq:def_deltaEW}
  \begin{split}
    \delta_\text{\EW}
    \,=\, \frac{\rd\sigma_\text{(0,5)}^{\Pq\Pqbar}
                +\rd\sigma_\text{(0,5)}^{\Pa\Pq/\Pa\Pqbar}
                +\rd\sigma_\text{(0,5)}^{\aa}}
	       {\rd\sigma_\text{(0,4)}^{\Pq\Pqbar}
		+\rd\sigma_\text{(0,4)}^{\aa}}
    \,=\, \delta^{\Pq\Pqbar/\aa}_\text{\EW}\;.
  \end{split}
\eeq
Here, since the newly emerging $\Pa\Pq$ and $\Pa\Pqbar$ channels act as real 
emission corrections to both the \LO $\Pq\Pqbar$ and $\aa$ channels, 
it is not possible to unambiguously split the full EW correction into two parts
associated with the $q\bar q$ and $\aa$ channels.\footnote{
  The situation is analogous to the case of $t\bar{t}$ production
  at \NLO \QCD. At leading order a distinction can be made 
  between the $\Pq\Pqbar$- and $gg$-induced channels. At \NLO \QCD, 
  the emerging $\Pq g$- and $\Pqbar g$-induced channels act as real 
  corrections to both and therefore link both \LO 
  processes. An unambiguous assignement of the 
  $\Pq g$- and $\Pqbar g$-induced \NLO corrections to the $q\bar q$ and $gg$
  \LO channels is thus not possible.
}
Therefore, our definition of $\delta_\text{\EW}$ 
amounts to choosing not to assign arbitrary fractions of the $\Pa\Pq$- 
and $\Pa\Pqbar$-corrections to act as corrections to the \LO $\Pq\Pqbar$ 
and $\aa$ channels, but to define an overall \NLO \EW 
correction factor.

With the above definitions the multiplicative combination~\refeq{eq:EWxQCDfactorisation}
can be cast in the form
\beq
  \label{eq:multsplit}
  \begin{split}
  \rd\sigma_\text{\NLO \QCDtEW}
  =\;&
  \rd\sigma_\text{\LO}
  \left(1+\delta_\text{\QCD}\right)\left(1+\delta_\text{\EW}\right)\\
  =\;&
  \left[
    \rd\sigma_\text{(0,4)}^{\Pq\Pqbar}\left(1+\delta_\text{\QCD}^{\Pq\Pqbar}\right)
    +\rd\sigma_\text{(0,4)}^{\aa}
  \right]
  \left(
    1+\delta^{\Pq\Pqbar/\aa}_\text{\EW}
  \right)\;,
  \end{split}
\eeq
where
the relative weight of \QCD corrections in the different partonic channels
is manifestly respected.
In particular, the $\aa$ channel remains free of 
\QCD correction effects, consistent with its behaviour at \NLO \QCD. 

Alternatively, as is often done in the literature,
one may choose to regard the combination of the $\Pa\Pq$- and $\Pa\Pqbar$-induced 
\NLO \EW effects as a correction to the $\Pq\Pqbar$ channel, 
and to attribute the remnant NLO EW corrections to the 
$\aa$ chanel. With this {\it ad hoc} splitting,
\beq
  \label{eq:deltaEWsplitchannel}
  \begin{split}
    \delta_\text{\EW}^{\Pq\Pqbar}
    \;=\,\frac{\rd\sigma_\text{(0,5)}^{\Pq\Pqbar}
               +\rd\sigma_\text{(0,5)}^{\Pa\Pq/\Pa\Pqbar}}
              {\rd\sigma_\text{(0,4)}^{\Pq\Pqbar}}\,,\qquad
    \delta_\text{\EW}^{\aa}
    \;=\,\frac{\rd\sigma_\text{(0,5)}^{\aa}}
              {\rd\sigma_\text{(0,4)}^{\aa}}\;,
  \end{split}
\eeq
it is natural to adopt a channel--by-channel 
factorisation of EW and QCD corrections,
\beq
  \label{eq:multcombschemedepA}
  \begin{split}
  \rd\sigma_\text{\NLO \QCDotEW}
  =\;&
  \rd\sigma_\text{(0,4)}^{\Pq\Pqbar}
  \left(1+\delta_\text{\QCD}^{\Pq\Pqbar}\right)
  \left(1+\delta_\text{\EW}^{\Pq\Pqbar\vphantom{/}}\right)
  +\rd\sigma_\text{(0,4)}^{\aa}
   \left(1+\delta_\text{\EW}^{\aa\vphantom{/}}\right)\,.
  \end{split}
\eeq
While one may debate if~\refeq{eq:EWxQCDfactorisation} is more or less 
motivated than~\refeq{eq:multcombschemedepA}, we observe that, 
using
\beq
\delta_\text{\EW}\,=\, 
\epsqq\, \delta_\text{\EW}^{\Pq\Pqbar}+
\epsaa\, \delta_\text{\EW}^{\aa}\;,
\eeq
the difference between the two prescriptions can be cast in the form
\beq
\label{eq:multcombschemedepB}
\rd\sigma_\text{\NLO \QCDtEW}-\rd\sigma_\text{\NLO \QCDotEW}
 \; =\;
\rd\sigma_\text{\LO}\,  \epsqq\,\epsaa\,
   \delta_\text{\QCD}^{\Pq\Pqbar}
   \left(\delta_\text{\EW}^{\aa}-\delta_\text{\EW}^{\Pq\Pqbar\vphantom{/}}\right)\,.
\eeq
This indicates that the two prescriptions tend to coincide 
if either one \LO channel dominates, the \QCD correction 
is small, or both channels' \EW corrections are of the same size. 
In large regions of the phase space these conditions are simultaneously satisfied, 
and for all observables studied in the following the scheme 
dependence~\refeq{eq:multcombschemedepB} is found to be smaller that 5\,\textperthousand, 
in most cases even below 0.5\,\textperthousand, of the \LO cross section.
Both the size of the \EW corrections contributed by the three 
individual channels and the above scheme dependence are detailed 
in \refapp{app:splitcorr}.

In the multiplicative approach, which we deem 
our best prediction, the uncertainties are estimated by 
scaling the \NLO \QCD predictions with the 
relative \NLO \EW correction, 
\beq
\label{eq:deltaEWscale}
1+\delta_\EW(\mu_{\rR},\mu_{\rF}) 
=
\frac{\rd\sigma_\text{\NLO \EW}(\mu_{\rR},\mu_{\rF})}{\rd\sigma_\text{\LO}(\mu_{\rR},\mu_{\rF})},
\eeq
evaluated at the central scale.
This is justified by the fact that $\delta_\EW(\mu_{\rR},\mu_{\rF})$
is independent of $\mur$ and involves only a very weak $\muf$ 
dependence of  $\ord(\alpha)$, while the \LO \QCD $\muf$-dependence 
cancels out in the ratio.

As discussed in \refse{se:categorisation}, we include 
photon-induced contributions throughout, including 
$\aa\to 2\ell 2\nu$, $\aa\to 2\ell 2\nu\Pa$ and
$\Pa q\to 2\ell2\nu q$ channels at \NLO \EW. 
To assess the uncertainty arising from the 
choice of photon PDF we vary their parametrisation from 
their default (\CTQED) to that of \LUXQED and \NNPDFQED, 
while keeping the 
quark and gluon PDFs fixed, cf.\ \refse{se:setup:PDFs}.
The overall impact of photon-induced processes is illustrated by
switching off the photon PDF, both at \LO and \NLO \EW.

Additionally, as discussed in \refse{se:setup:yfs}, we investigate to which
degree exact \NLO \QCDtEW results can be reproduced by approximations
based on the combination of IR-subtracted virtual \EW corrections (\EWvirt) with
\YFS \QED resummation or, alternatively, with the \CSShower.  Such
approximation, denoted as \NLOQCDtEWYFS and \NLOQCDtEWCSS,
can be realised in realistic particle-level simulations using currently public
tools, and can be regarded as a first step towards \NLO \QCDpEW matching and merging.

In the following, we study various fiducial cross sections and differential distributions.
Physical observables involving charged leptons are known to 
be highly sensitive to \QED radiative corrections.
This should be avoided by using dressed leptons. To this end 
we recombine all leptons with nearly collinear photons that 
lie within a cone
\beq 
\Delta R_{\ell\Pa}=\sqrt{\Delta \phi^2_{\ell\Pa}
+\Delta \eta^2_{\ell\Pa}}<\Rrec = 0.1\;.
\eeq 
This dressing procedure captures the bulk of the 
collinear final-state radiation, while keeping contamination 
from large-angle photon radiation at a negligible level.

\begin{table}[t!]
  \begin{center}
    \begin{tabular}{ll}
      \multicolumn{2}{l}{Inclusive cuts} \\\hline
      $\missingET$		&> 20\,\GeV \\
      $p_{\rT,\ell^\pm}$ 	&> 20\,\GeV \\
      $|\eta_{\ell^\pm}|$	&< \phantom{0}2.5 \\
      $\Delta R_{\ell^+\ell^-}$	&> \phantom{0}0.2 \\
      $\HTjet$			&< \phantom{0}0.2\,$\HTl$ \\\hline
    \end{tabular}
  \end{center}
  \caption{
    Inclusive selection cuts for off-shell vector-boson pair production 
    in the $2\ell2\nu$ channel. The missing transverse momentum $\missingET$ 
    is calculated from the vector sum of neutrino momenta.
    \label{tab:cuts}
  }
\end{table}

In our analysis we apply a set of acceptance cuts, as 
listed in \refta{tab:cuts}, on the transverse momentum, 
pseudo-rapidity and angular separation of the dressed charged 
leptons and on the missing transverse momentum calculated 
based on the neutrino momenta, $\missingET=p_{\rT,\nu\bar \nu}$.

Inclusive vector-boson pair production receives large 
\NLO \QCD corrections in kinematic regions where one of the 
vector bosons might become soft. This effect is a variant 
of the well known `giant $K$-factors'~\cite{Rubin:2010xp}.
In order to suppress these large \QCD corrections that spoil the perturbative 
convergence we veto events with
\beq
\label{eq:jetveto}
 \HTjet  > 0.2~\HTl \,,
\eeq
where $\HTjet=\sum_{i\in {\rm jets}} p_{\rT,i}$ based on anti-$k_{\rm T}$ 
jets with $R=0.4$ and $p_{\rT}>30\,\GeV$. In practice, $\HTjet=p_{\rT,\jet}$
at \NLO \QCD.
A reliable inclusive prediction without such a jet veto requires the merging 
of $pp\to\llnn+0,1$\,jets at \NLO \QCDpEW, but goes beyond the 
scope of the present paper.
The complete analysis has been implemented in \Rivet \cite{Buckley:2010ar}.
For reference, we present the cross sections of the inclusive as well 
as three more exclusive event selections for both channels in 
\refapp{app:xsecs}.

\subsection[The different-flavour channel \texorpdfstring{$pp\to\llnndf$}{pp->emununu}]
           {The different-flavour channel $\boldsymbol{pp\to\llnndf}$}
\label{se:dfresults}

\begin{figure*}[t]
  \centering
  \includegraphics[width=\relplotwidth\textwidth]{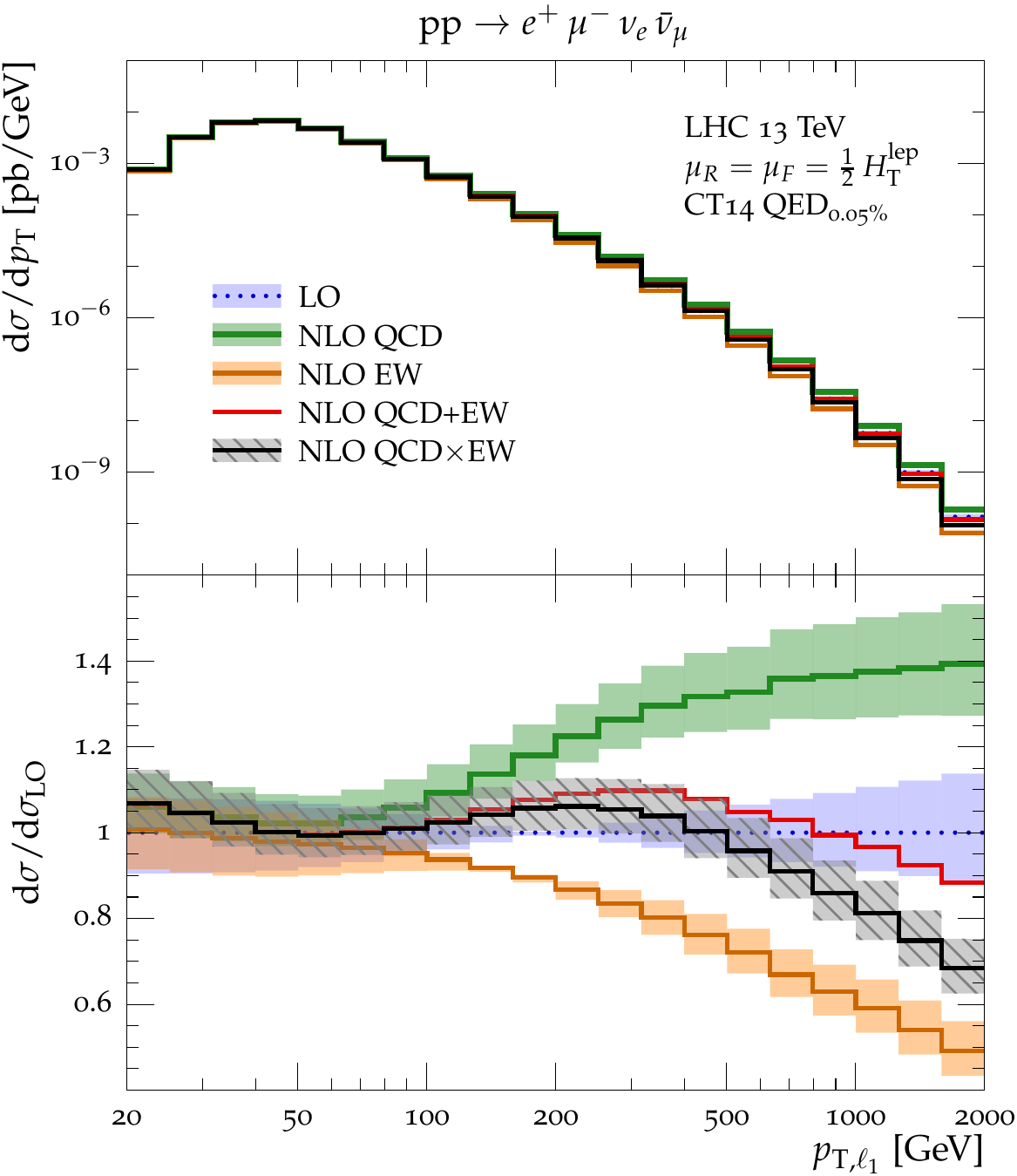}
  \qquad 
  \includegraphics[width=\relplotwidth\textwidth]{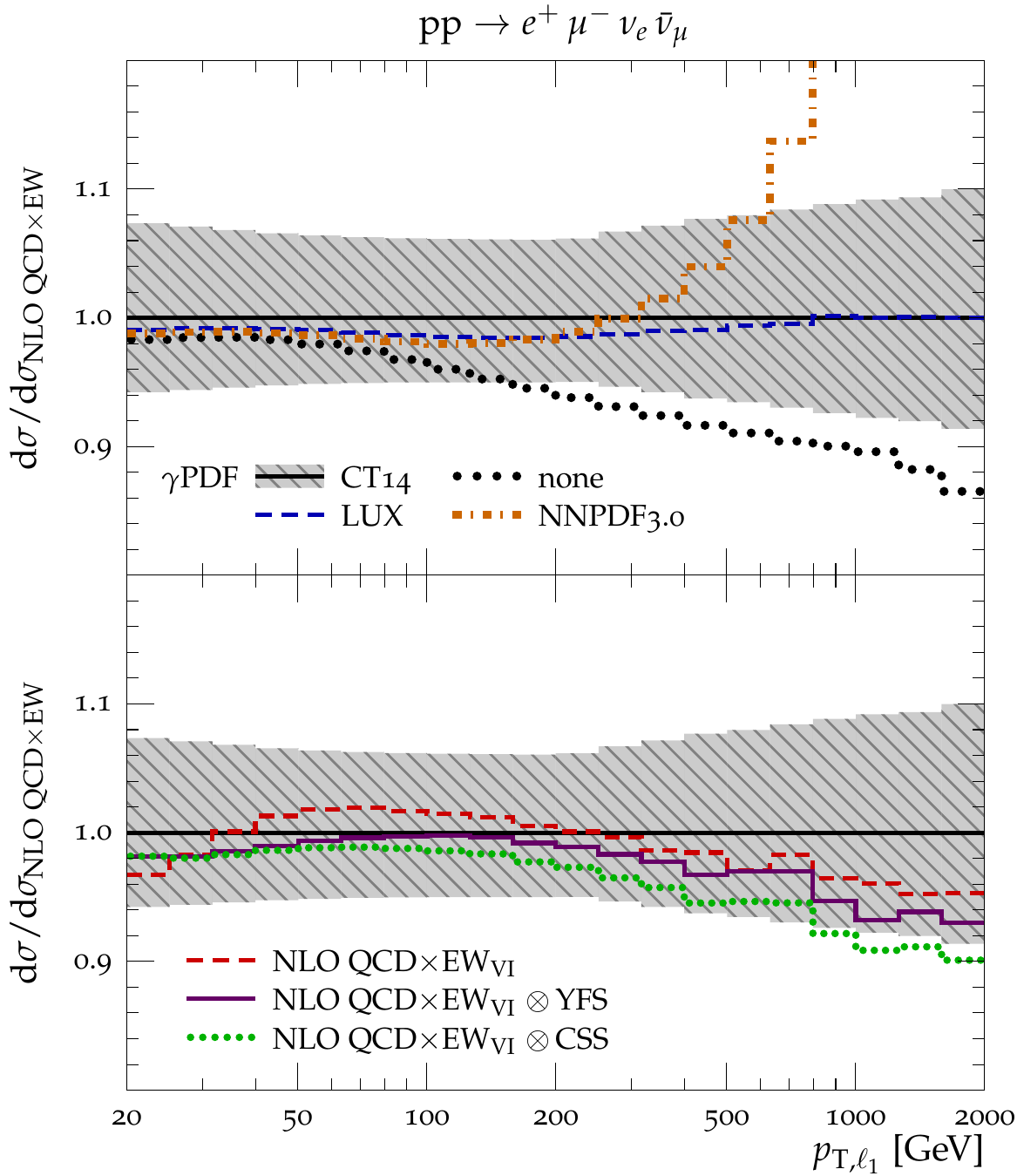}
  \caption{
    Distribution in the transverse momentum of the leading lepton, 
    $p_{\rT,\ell_1}$, for $pp\to\llnndf$ at 13\,\TeV. The 
    left panel shows the absolute predictions and relative corrections 
    with respect to \LO (including $\aa\to 2\ell2\nu$) for the nominal 
    \CTQED PDF. The bands correspond to factor-two scale variations. 
    The upper-right panel shows the effect, at \NLO \QCDtEW 
    level, of switching off \Pa-induced contributions or applying 
    photon densities from different current PDFs, while using quark 
    and gluon densities from the nominal \CTQED set throughout.
    The lower-right ratio shows the level of agreement of the 
    \NLO \QCDtEWvirt, \NLOQCDtEWYFS and \NLOQCDtEWCSS approximations
    with the exact \NLO \QCDtEW calculation.
    \label{fig:emvv_pTl1}
    \vspace*{5mm}
  }
\end{figure*}

Differential distributions for  $pp\to\llnndf$
are presented in \reffis{fig:emvv_pTl1}{fig:emvv_dPhi}.
In every figure, the left plot shows absolute predictions as 
well as relative \NLO corrections with scale-variation bands.
The upper-right ratio plot quantifies the importance of 
photon-induced contributions as well as the effect of using 
different \aPDF{}s, while the lower-right ratio plot
compares exact \NLO results against the
\NLOQCDtEWYFS and \NLOQCDtEWCSS approximations.
For reference, we also show the pure fixed-order \NLO \QCDtEWvirt 
approximation, which includes only the IR-subtracted part of 
virtual \EW corrections and lacks any differential description of 
\QED real corrections.


\begin{figure*}[t]
  \centering
  \includegraphics[width=\relplotwidth\textwidth]{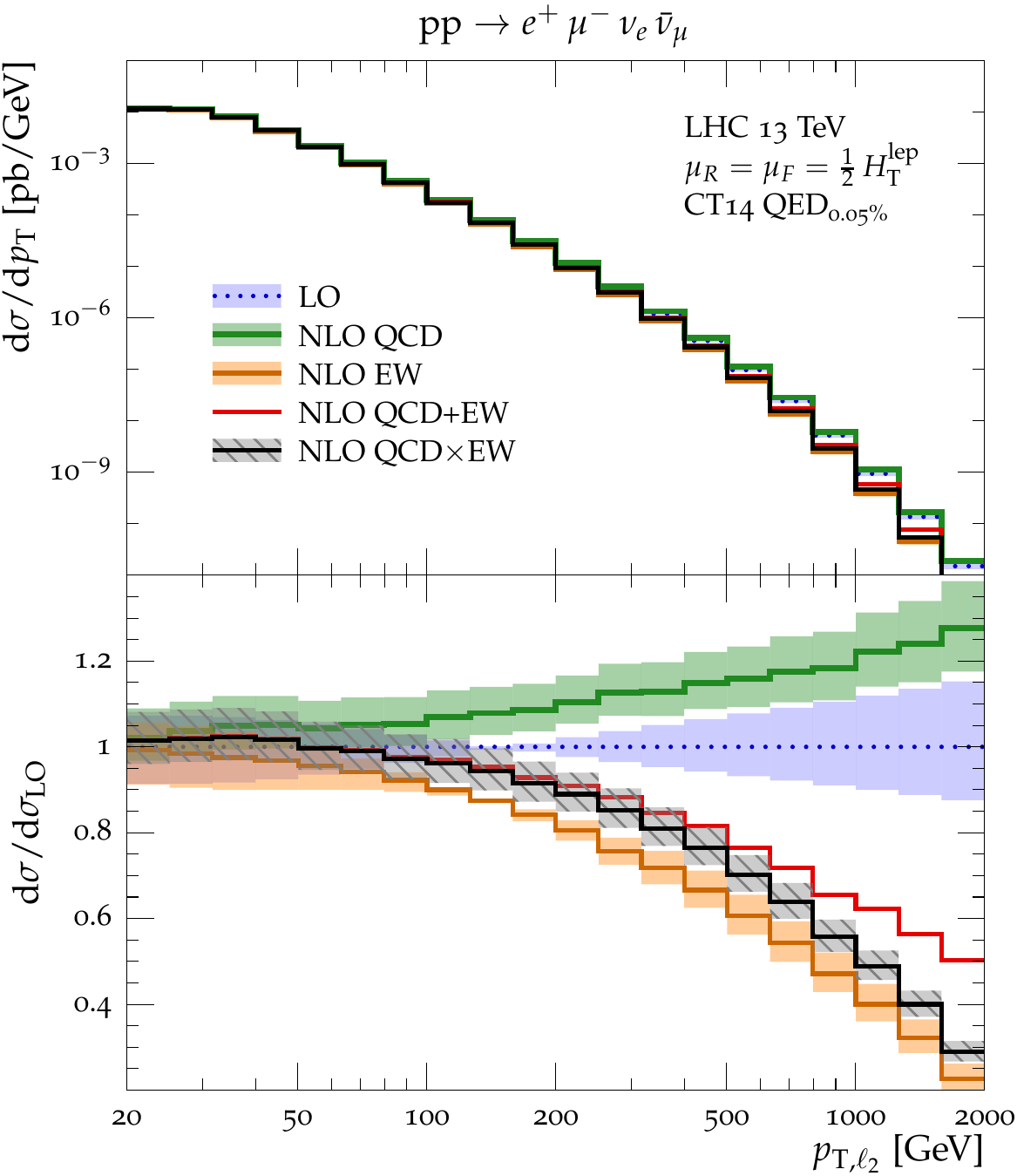}
  \qquad 
  \includegraphics[width=\relplotwidth\textwidth]{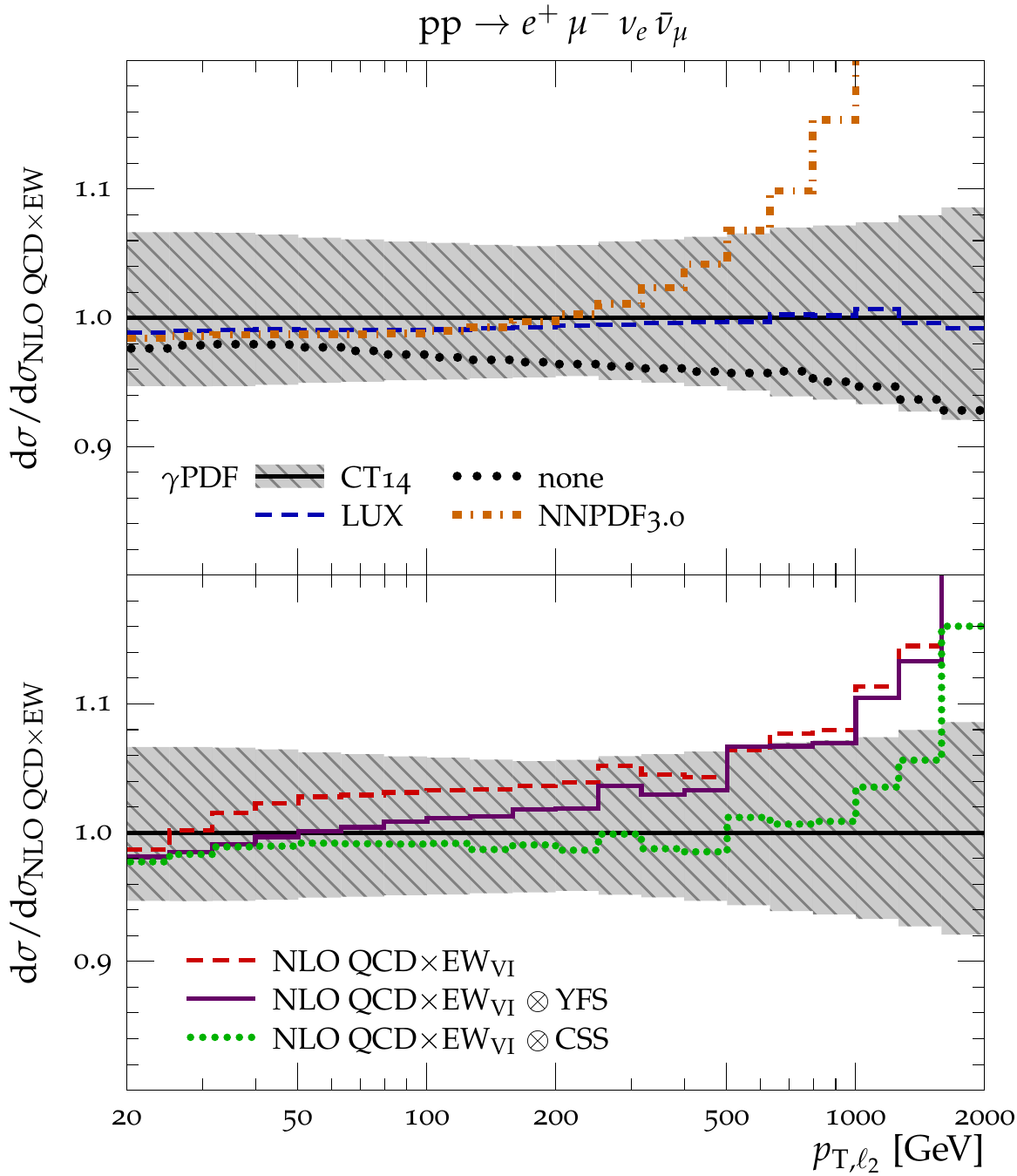}
  \caption{
    Distribution in the transverse momentum of the subleading lepton, 
    $p_{\rT,\ell_2}$, for $pp\to\llnndf$ at 13\,\TeV. Details as in 
    \reffi{fig:emvv_pTl1}.
    \label{fig:emvv_pTl2}
    \vspace*{5mm}
  }
\end{figure*}

In \reffis{fig:emvv_pTl1}{fig:emvv_mll} we present distributions 
in the transverse momenta of the leading 
and subleading leptons, $p_{\rT,\ell_1}$ and $p_{\rT,\ell_2}$,
the total missing transverse momentum, $\missingET$, and the  
invariant mass of the $\Pe^+\Pmu^-$ pair, $m_{\ell\ell}$.
The leading and subleading lepton are defined by their ordering 
in transverse momentum, irrespective of their charge.
\EW corrections to these observables feature the typical Sudakov
behaviour, with small effects below 100\,\GeV and large negative corrections
at the \TeV scale.  In the tails of the lepton-$p_\rT$ and $m_{\ell\ell}$
distributions \NLO \EW corrections can reach and even largely exceed $-50\%$.
The dominant effects originate from $q\bar q\to \wpwm$ topologies with
resonant \PW bosons, and the strong enhancement of \EW Sudakov corrections is
induced by the high $p_\rT$ and the large SU(2) charges of the \PW bosons.
In the presence of \EW corrections of several tens of percent, fixed-order
\NLO predictions should be supplemented by a resummation of Sudakov
logarithms.  As a rough indication of the possible magnitude of higher-order
\EW effects, we observe that na\"ive exponentiation can turn \NLO \EW corrections
of $-50$--80\% into an overall all-order \EW correction of $-40$--55\%.  We also
note that \EW corrections of this magnitude appear in a kinematic range that
cannot be probed with decent statistics at the LHC.  Nevertheless, such
phase-space regions would play an important role at a 100\,\TeV $pp$
collider~\cite{Mangano:2016jyj}.

Due to the presence of the jet veto~\refeq{eq:jetveto}, 
the impact of \QCD corrections in \reffis{fig:emvv_pTl1}{fig:emvv_mll} 
is rather mild at energies below $\MW$, and grows only up to
$+10$--40\% in the tails.  While the actual size of \QCD $K$-factors 
depends on the scale choice, we recall that, in general, \QCD 
corrections to $pp\to \llnndf$  receive
sizeable real-emission contributions in the absence of jet
vetoes~\cite{Grazzini:2016ctr}. 
Scale uncertainties at \NLO \QCD
are rather constant and somewhat below 10\%.

Due to their opposite sign, \QCD and \EW corrections cancel against each other
to a certain extent.  At the same time, in regions where both \QCD and \EW
corrections are well beyond 10\%, contributions of relative 
$\ord(\alphaS\alpha)$ become relevant.  
Such \NNLO \QCDtEW effects are estimated in our predictions by
means of the multiplicative combination of \NLO corrections 
\refeq{eq:EWxQCDfactorisation}, which is well justified if \EW corrections
are dominated by Sudakov logarithms and \QCD radiation is
softer than the characteristic scale of the $q\bar q\to 2\ell2\nu$
\EW subprocess.
Comparing the additive and multiplicative combination of
\QCD and \EW corrections in \reffis{fig:emvv_pTl1}{fig:emvv_mll}, 
we find that contributions of relative $\ord(\alphaS\alpha)$ can exceed 10\% in the tails.  
Among the virtues of a multiplicative combination of \QCD and \EW corrections,
it is worth pointing out that \NLO \EW corrections are implicitly supplemented by \QCD
radiation, resulting, for instance, in a reasonable behaviour with respect
to possible jet vetoes.
At the same time, it should be stressed that,
for a more reliable
assessment of $\ord(\alpha\alphaS)$ corrections,
an approach like \NLO \QCDpEW merging~\cite{Kallweit:2015dum} is certainly preferable. 


\begin{figure*}[t]
\centering
   \includegraphics[width=\relplotwidth\textwidth]{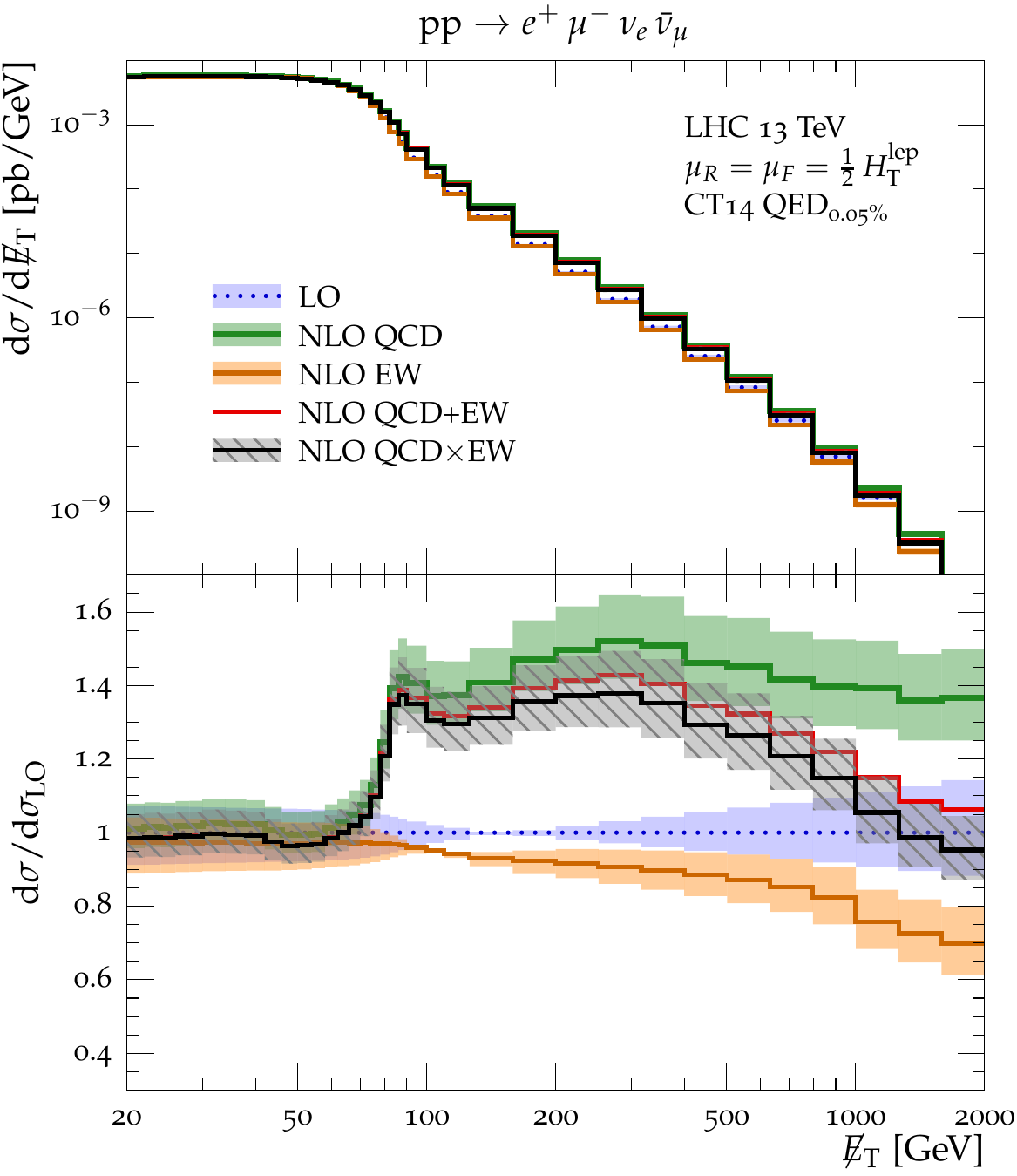}
   \qquad 
   \includegraphics[width=\relplotwidth\textwidth]{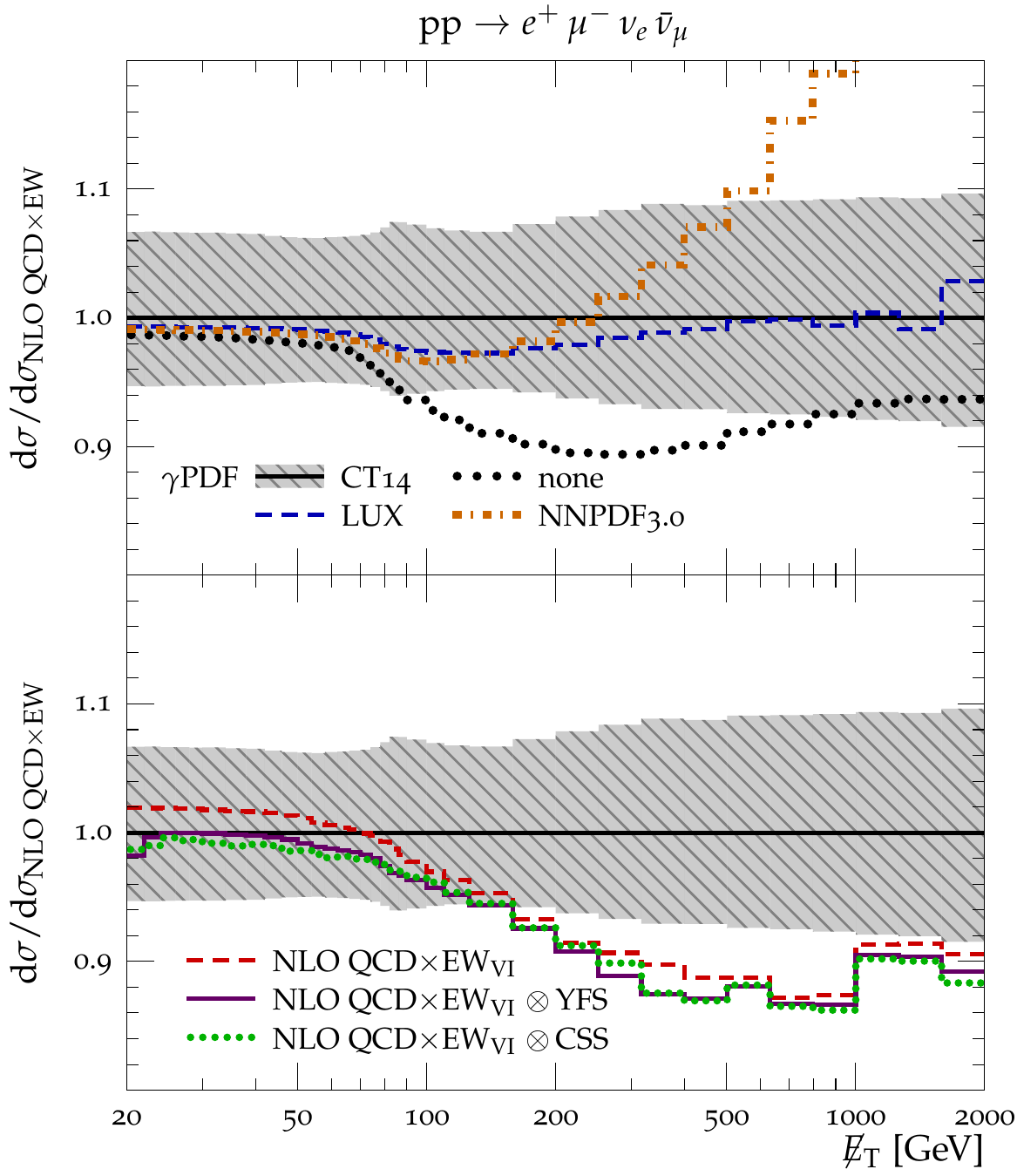}
  \caption{
    Distribution of the missing transverse momentum, \missingET, for 
    $pp\to\llnndf$ at 13\,\TeV. Details as in \reffi{fig:emvv_pTl1}.
    \label{fig:emvv_pTmiss}
    \vspace*{5mm}
  }
\end{figure*}

The behaviour of the  \missingET distribution 
(\reffi{fig:emvv_pTmiss}) deserves a few additional comments.  First,
in the tail of this distribution we observe that Sudakov \EW effects are
less pronounced than in other observables.  This is due to the fact that
requiring a high-$\pT$ $\nu_{\Pe}\bar\nu_{\Pmu}$ pair forces the \PW bosons
into the off-shell regime.  As a result, Sudakov logarithms arise only from
\EW interactions between the on-shell final-state leptons and, like in Drell-Yan
processes, they turn out to be less enhanced than in $pp\to\wpwm$. 
Second, the \QCD $K$-factor features a sizeable enhancement characterised by 
a rather sharp threshold at $\missingET\sim \MW$.  This is related to the
fact that, in  $pp\to\wpwm$ at LO, $p_{\rT,\wpwm}=0$ strongly disfavours 
the production of a
$\nu_{\Pe}\bar\nu_{\Pmu}$ pair with $p_{\rT}>\MW$.
Therefore, the \wpwm transverse momentum induced by \NLO \QCD radiation results in
a sizeable enhancement in the $\missingET>\MW$ region (see~\cite{Grazzini:2016ctr}).

Photon-induced contributions in \reffis{fig:emvv_pTl1}{fig:emvv_mll} can
reach up to 5--20\%, depending on the observable.  The largest  
effects are typically observed in the \TeV tails. 
The \Pa-induced contributions to the \missingET distribution, however, 
approach 10\% already at 200\,\GeV, an effect that can already be observed 
at \LO. This is due to the presence of non-resonant diagrams 
that are absent in the $q\bar q$ channels. 
They can populate this 
phase-space region, which is disfavoured as soon as the neutrinos need 
to be produced through an $s$-channel \PW propagator. 
This effect is further increased by real-emission channels of type 
$\Pa q\to 2\ell2\nu q$, which are strongly
enhanced at $\missingET\ge \MW$, similarly as for \QCD radiative effects.

Comparing different photon PDFs, for all observables we find a fairly good agreement 
between \CTQED and \LUXQED \aPDF{}s, with differences that never exceed the level of 
five percent.
Conversely, the usage of the \NNPDFQED \aPDF yields 
similar inclusive cross sections as \CTQED and \LUXQED, but 
much bigger \Pa-induced contributions in the tails.
Nevertheless the differences are consistent with the large uncertainty 
of the photon density in \NNPDFQED, 
while using the other PDF sets leads to a \aPDF uncertainty
well below the overall \QCD scale uncertainty.
The largest \Pa-induced effects are observed in the tail of the
$m_{\ell\ell}$ distribution, where the dominant contribution originates from
$\aa\to\wpwm$ topologies with $t$-channel poles in the
forward/backward regions.  We note that relaxing 
rapidity cuts on charged leptons, which act as a cut-off on $t$-channel poles,
would further enhance $\aa\to\wpwm$ contributions.


\begin{figure*}[t]
\centering
  \includegraphics[width=\relplotwidth\textwidth]{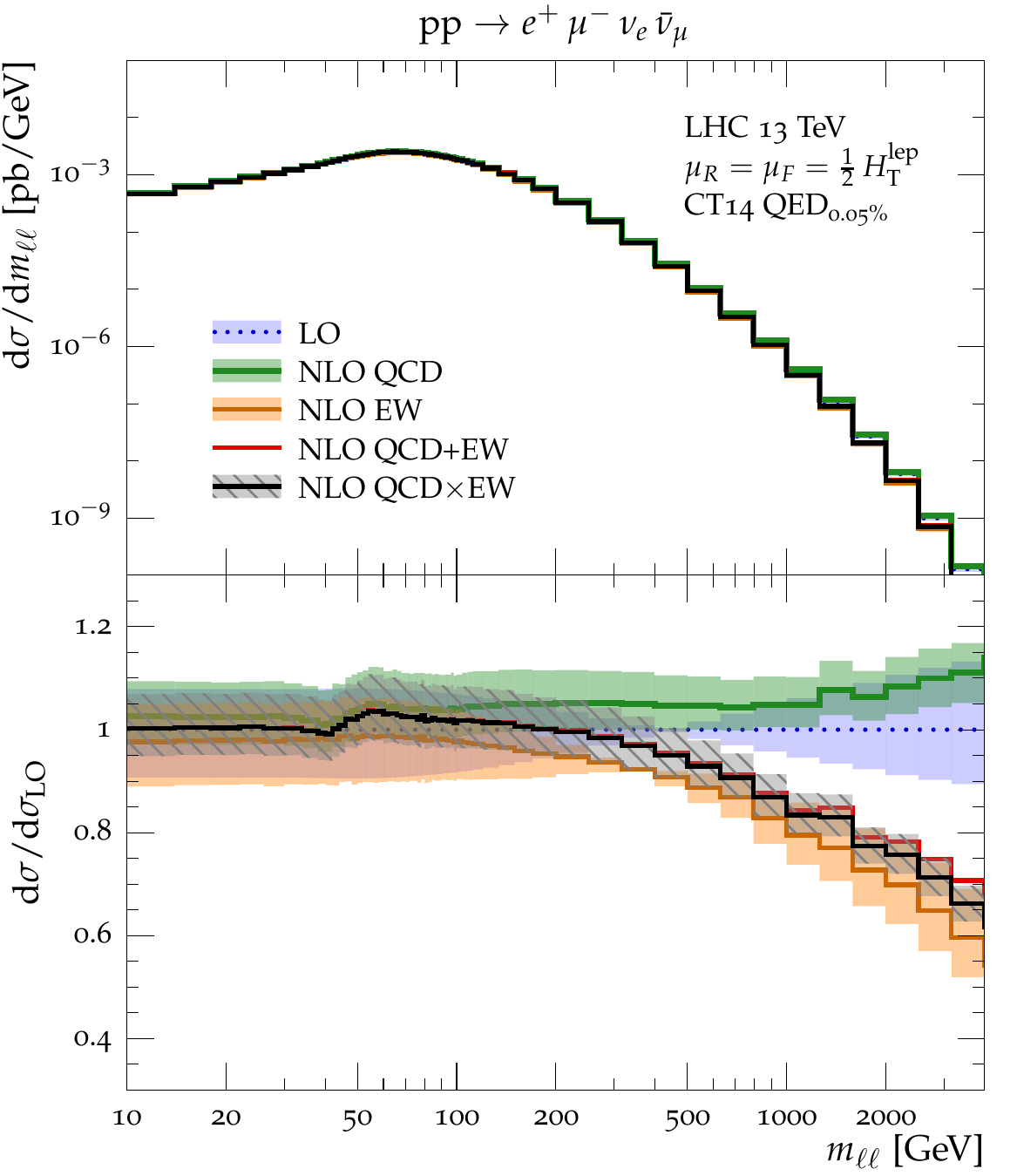}
  \qquad 
  \includegraphics[width=\relplotwidth\textwidth]{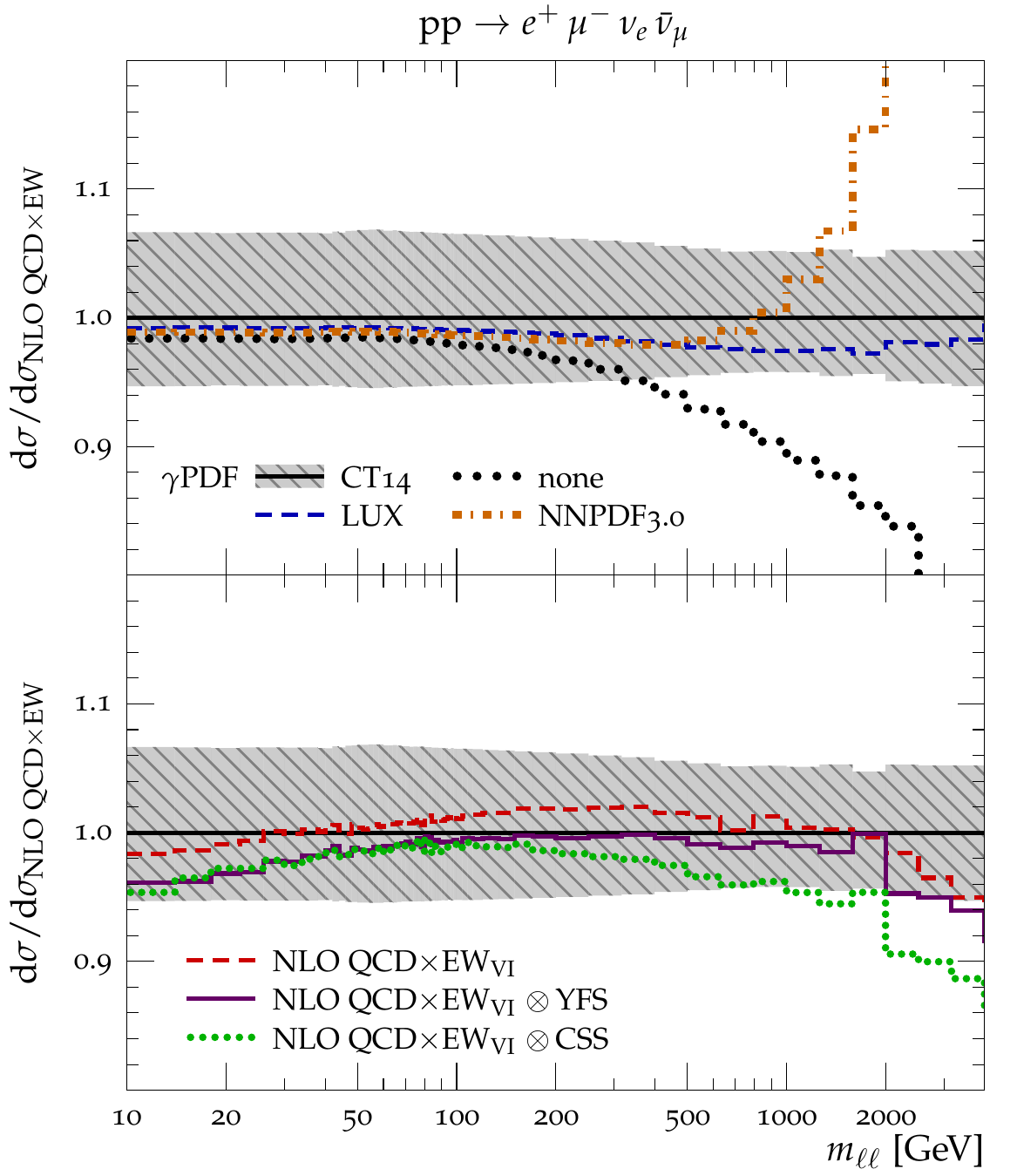}
  \caption{
    Distribution in the invariant mass of the $\Pe^+\Pmu^-$  pair, $m_{\ell\ell}$, 
    for $pp\to\llnndf$ at 13\,\TeV. Details as in \reffi{fig:emvv_pTl1}.
    \label{fig:emvv_mll}
    \vspace*{5mm}
  }
\end{figure*}

Comparing the \NLOQCDtEWYFS and \NLOQCDtEWCSS approximations against 
exact \NLO \QCDtEW results, in \reffis{fig:emvv_pTl1}{fig:emvv_mll} 
we observe agreement at the few-percent
level for $p_{\rT,\ell_1}$ and $m_{\ell\ell}$, while in the tails of the
$p_{\rT,\ell_2}$ and \missingET distributions the error of the \NLOQCDtEWYFS
approximation can exceed 10\%.  This can be attributed to the fact that the 
\YFS resummation as implemented in \Sherpa does not 
account for initial-state QED radiation in the $q\bar q$ channels and 
neglects the $\Pa q$ channels. The \CSShower, on the other hand, 
describes these configurations, but lacks accuracy due to its dipole 
structure. In any case, both approximations improve the pure fixed-order approximation 
of \NLO \QCDtEWvirt.

\begin{figure*}[t!]
\centering
  \includegraphics[width=\relplotwidth\textwidth]{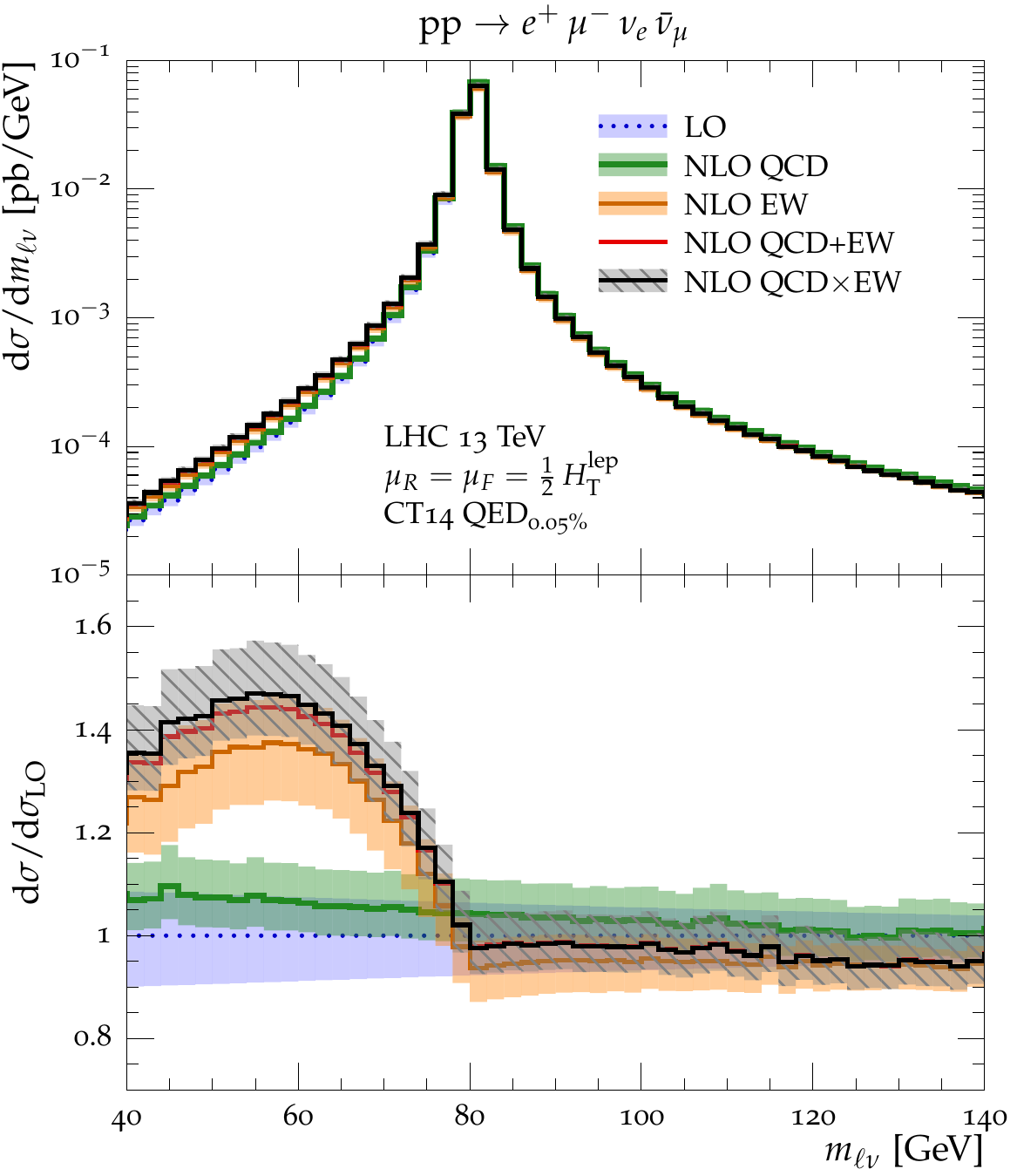}
  \qquad 
  \includegraphics[width=\relplotwidth\textwidth]{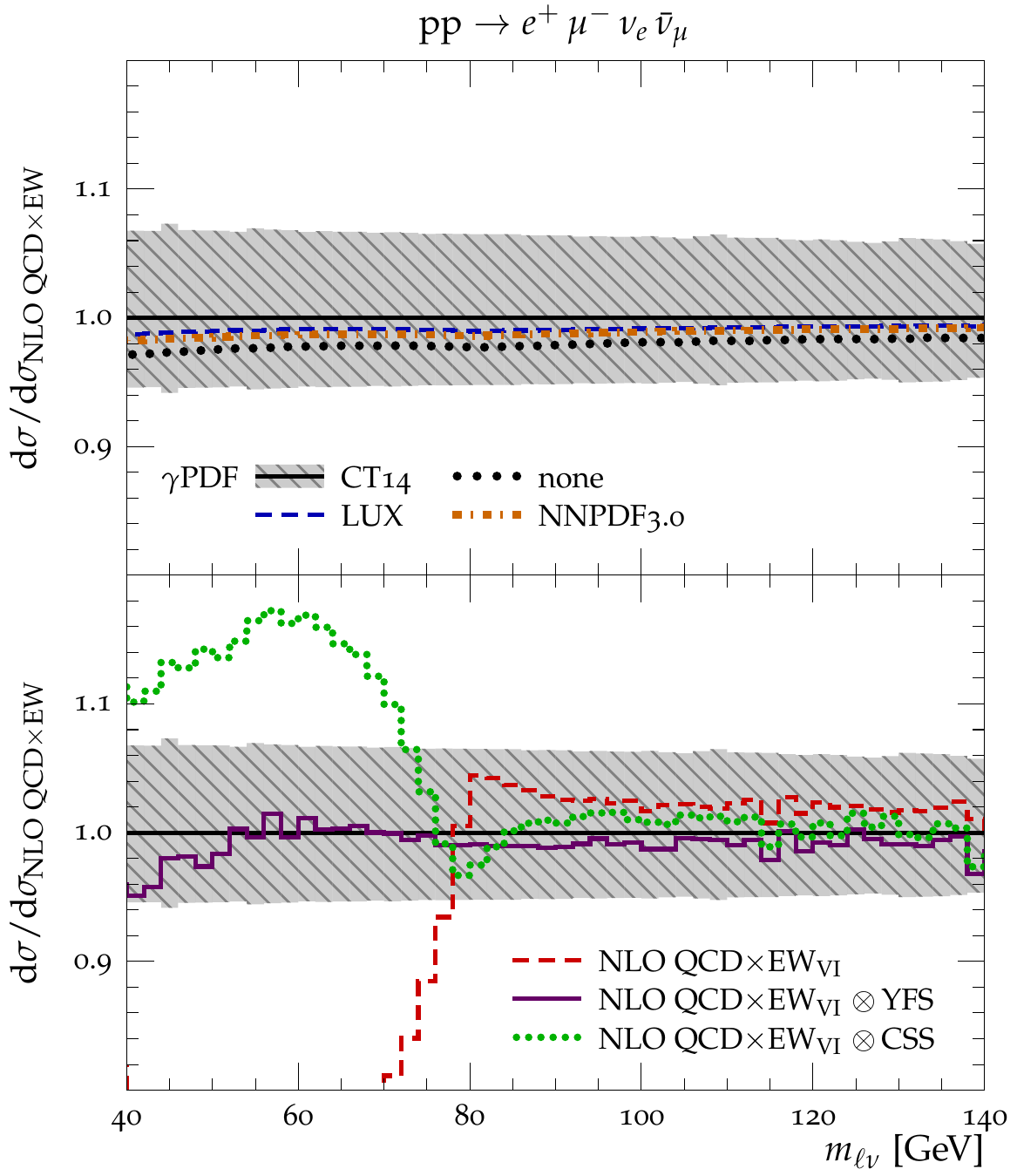}
  \caption{
    Distribution in the invariant mass of the matching 
    lepton-neutrino pair, $m_{\ell\nu}$, for $pp\to\llnndf$ at 13\,\TeV.
    Details as in \reffi{fig:emvv_pTl1}.
    \label{fig:emvv_mln}
    \vspace*{5mm}
  }
\end{figure*}

\reffis{fig:emvv_mln}{fig:emvv_mllnn} illustrate distributions in the
\PW-boson mass, $m_{\ell\nu}$, and in the \wpwm invariant mass,
$m_{\ell\ell\nu\nu}$.  Such observables are not experimentally accessible,
but they provide valuable insights into the resonance structure of
$pp\to\llnndf$ and into the behaviour of \EW corrections.
Focussing on the $m_{\ell\nu}$ and $m_{\ell\ell\nu\nu}$ regions 
of the $W\to \ell\nu$ and $\PZ\to 2\ell2\nu$ peaks 
and the $\wpwm\to \ell\ell \nu\nu$ threshold, we observe that 
\QCD corrections are almost insensitive to the
presence of \EW resonances and thresholds.
Photon-induced contributions are on the level of 1--3\%, while
\EW corrections feature sizeable shape 
distortions due to \Pa bremsstrahlung off the charged leptons. Such 
shape corrections can be understood
as a net migration of events from the peak and threshold regions towards the
low-mass tails and, in the case of the $m_{\ell\ell\nu\nu}$ distribution,
towards the local minimum above the $\PZ\to \ell\ell\nu\nu$ peak.  
In these observables, apart from the region of very high $m_{\ell\ell\nu\nu}$,
the \NLOQCDtEWYFS approximation is found to reproduce exact results with fairly 
good accuracy. In particular, in the off-shell regime, \ie for $m_{\ell\nu}<\MW$
or $m_{\ell\ell\nu\nu}<2\MW$, the offset between 
\NLO \QCDtEWvirt approximation and exact results indicates the presence of
\QED radiation effects beyond 10\%, which turn out to be well described by the \YFS approach.
The remaining differences are below 5\% or so. They can be attributed 
to higher-order corrections, missing in the fixed-order calculations, 
and to ambiguities related to the YFS resummation for highly off-shell decays.
In contrast, \QED radiative corrections to $m_{\ell\nu}$ and  $m_{\ell\ell\nu\nu}$
are strongly overestimated in the \NLOQCDtEWCSS approach. This is most likely due to the fact that
the \CSShower is unaware of resonance structures.

\begin{figure*}[t!]
\centering
  \includegraphics[width=\relplotwidth\textwidth]{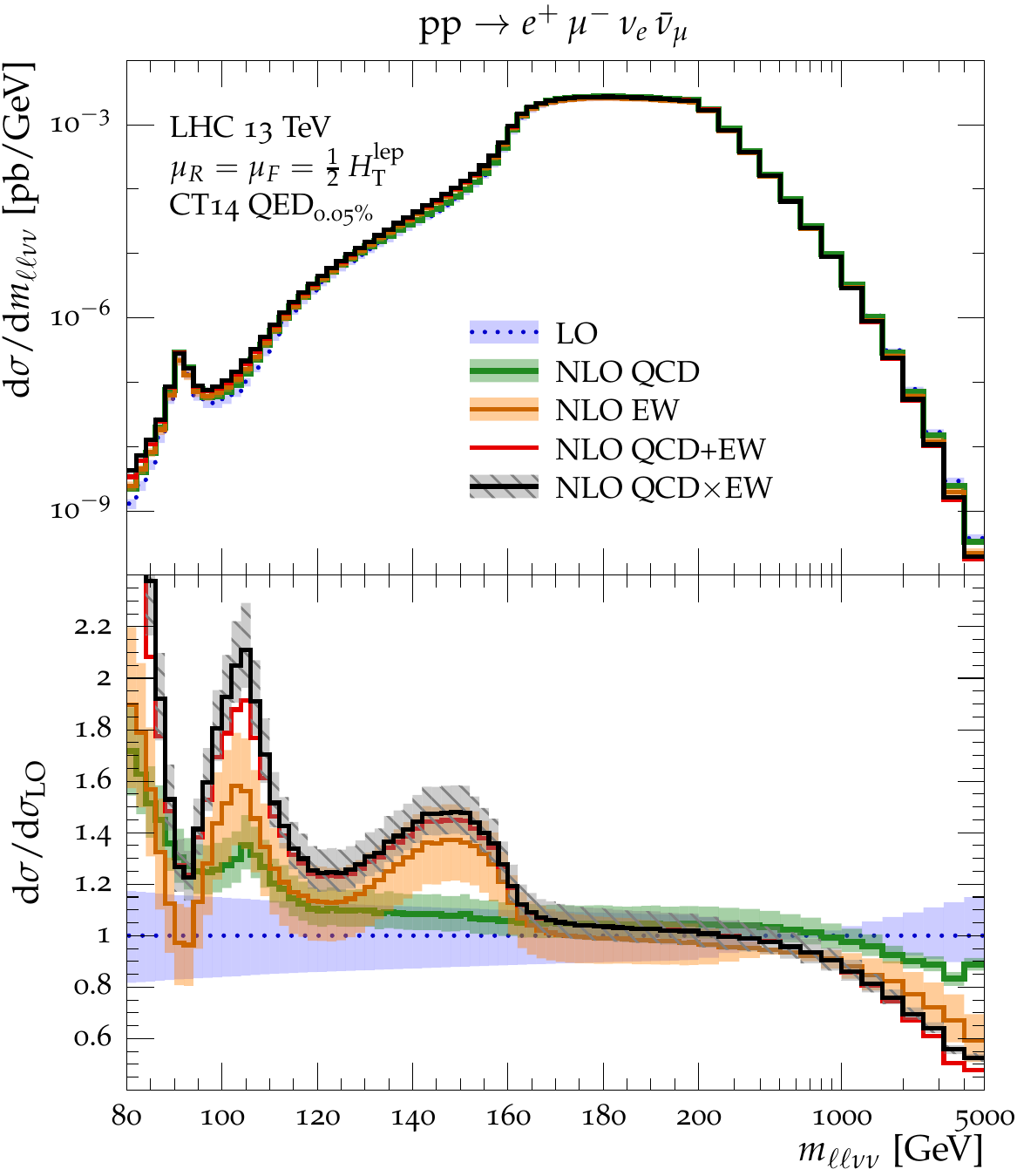}
  \qquad 
  \includegraphics[width=\relplotwidth\textwidth]{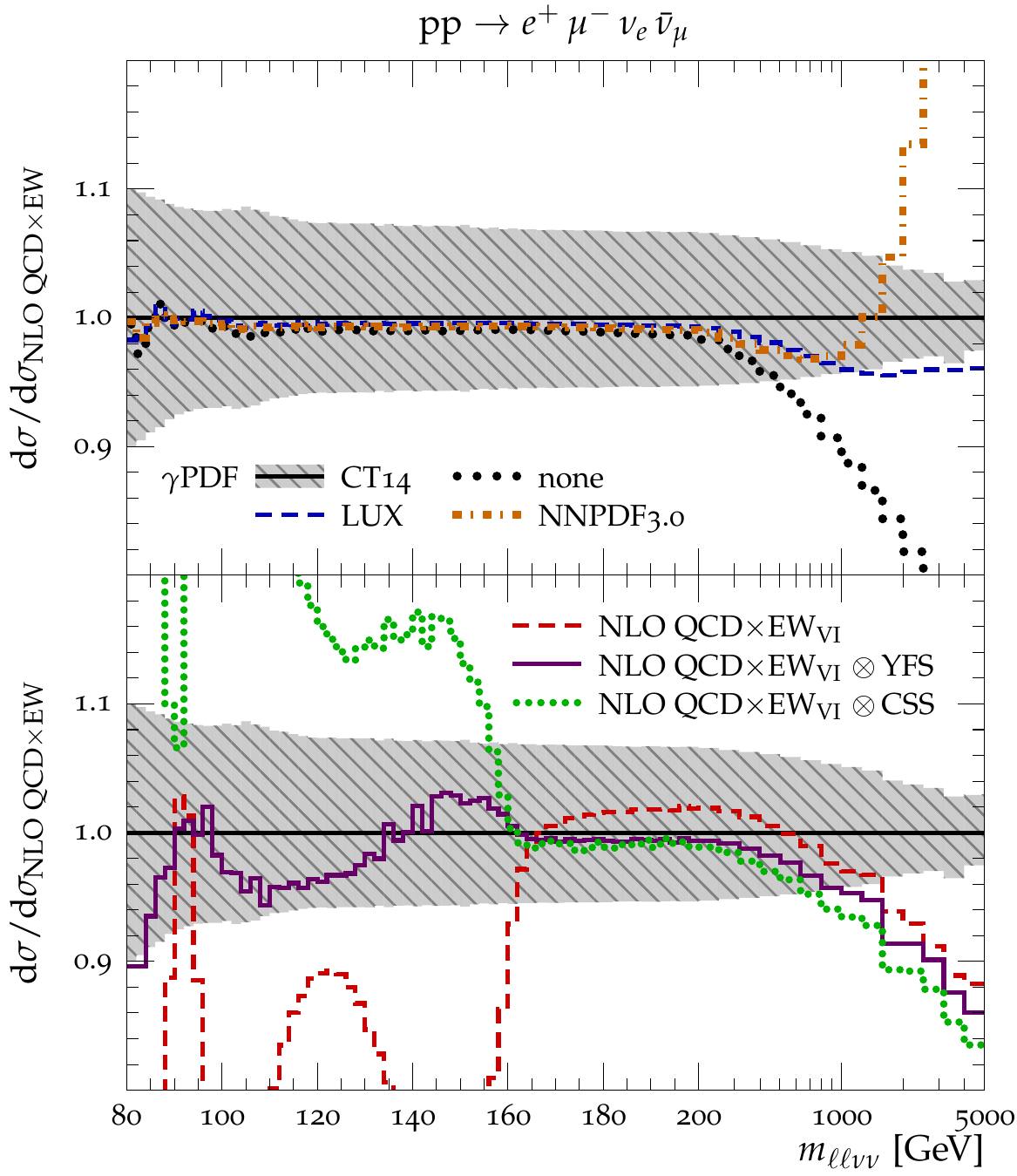}
  \caption{
    Distribution in the invariant mass of all four final 
    state leptons and neutrinos, $m_{\ell\ell\nu\nu}$, for $pp\to\llnndf$ at 
    13\,\TeV. Details as in \reffi{fig:emvv_pTl1}.
    \label{fig:emvv_mllnn}
    \vspace*{5mm}
  }
\end{figure*}

\reffi{fig:emvv_mllnn} also displays the multi-\TeV region of the 
$m_{\ell\ell\nu\nu}$ distribution, where  large negative \EW Sudakov corrections are observed, as well
as $\Pa$-induced contributions  beyond 10\%, with large deviations between the 
different \aPDF sets. 
At the same time the \NLOQCDtEWYFS and 
\NLOQCDtEWCSS predictions grow gradually worse when compared with 
the exact \NLO \QCDtEW calculation due, respectively, to the missing or limited 
accuracy in the description of $\Pa\to q\bar q$-splittings in the initial state.

\begin{figure*}[t]
\centering
   \includegraphics[width=\relplotwidth\textwidth]{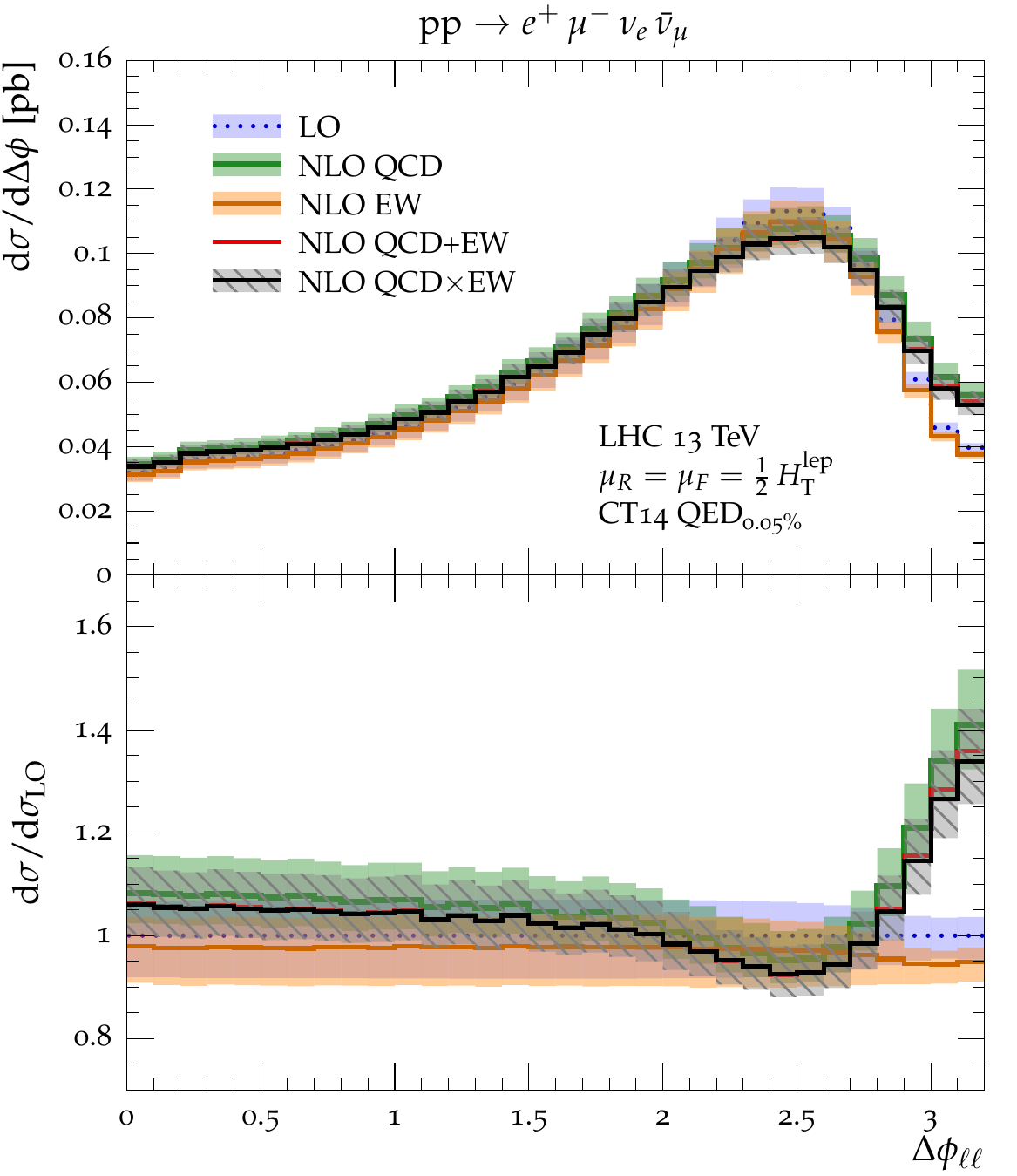}
   \qquad 
   \includegraphics[width=\relplotwidth\textwidth]{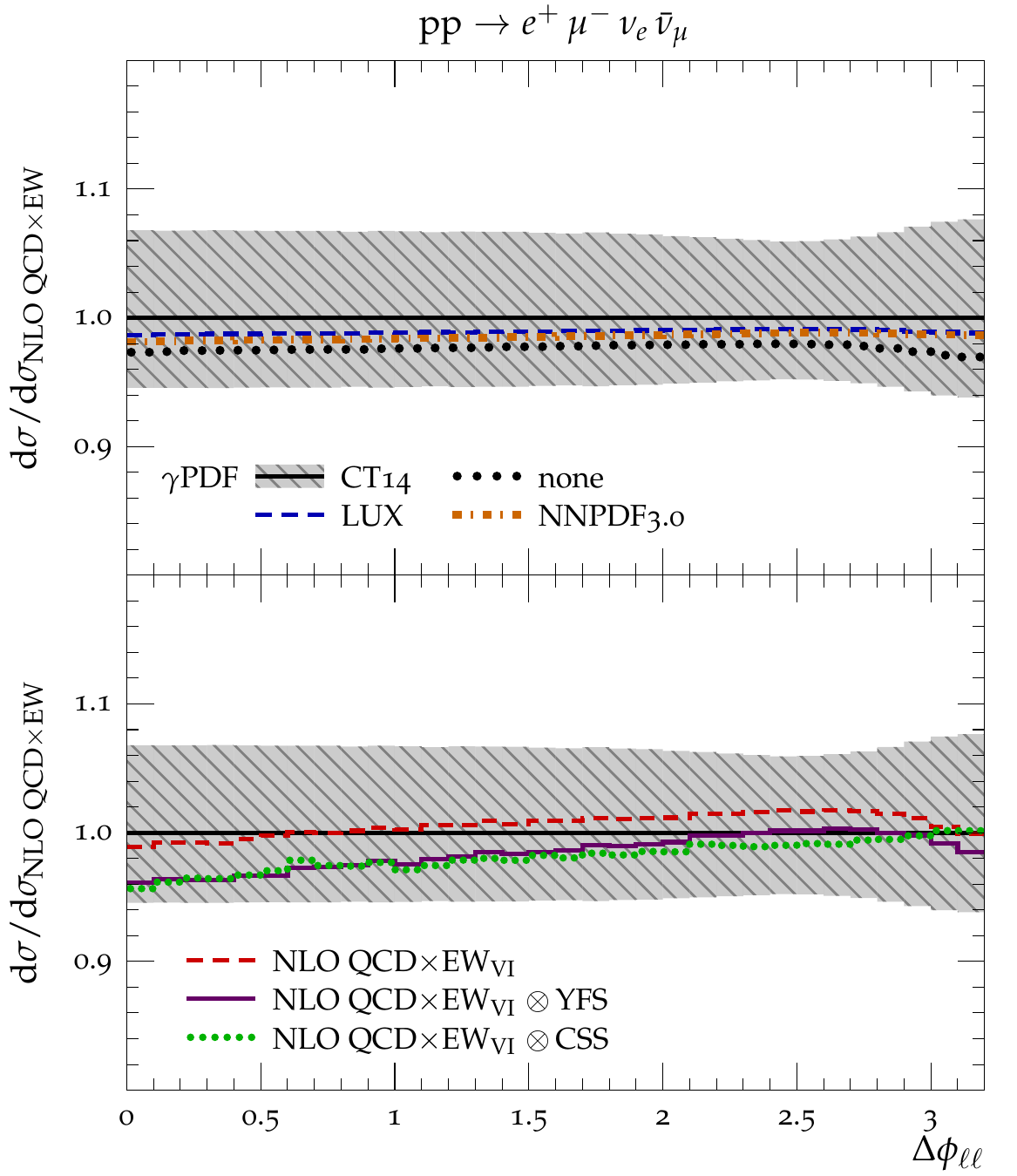}
  \caption{
    Distribution in the azimuthal separation of the $e^+\mu^-$ pair, 
    $\Delta\phi_{\ell\ell}$, for $pp\to\llnndf$ at 13\,\TeV. 
    Details as in \reffi{fig:emvv_pTl1}.
    \label{fig:emvv_dPhi}
    \vspace*{5mm}
  }
\end{figure*}

Finally, in \reffi{fig:emvv_dPhi} we show the distribution in the azimuthal
separation of the $\Pe^+\Pmu^-$ pair, $\Delta\phi_{\ell\ell}$.  For this
observable, \EW corrections and \Pa-induced effects are almost flat and
similarly small as for the integrated cross section.  As for \QCD
corrections, we observe a pronounced kinematic dependence for
$\Delta\phi_{\ell\ell}\to \pi$.  This can be understood as a statistical
effect related to the migration of events form highly populated to poorly 
populated bins.


\subsection[The same-flavour channel \texorpdfstring{$pp\to\llnnsf$}{pp->eenunu}]
           {The same-flavour channel $\boldsymbol{pp\to\llnnsf}$}
\label{se:sfresults}

In this section we discuss results for $pp\to\llnnsf$ at 13\,\TeV, including all 
neutrino flavours, i.e.
\beqar
\label{eq:sfsum}
\rd\sigma(pp\to\llnnsf) &=&
\sum_{\ell=\Pe,\Pmu,\Ptau} \rd\sigma(pp\to \Pe^+\Pe^-\nu_\ell\nu_\ell).
\eeqar
As discussed in \refse{se:anatomy},
the $\llnnsfwwzz$ channel receives contributions from 
\zz and \ww diboson resonances, while the channels with $\Pmu$- and $\Ptau$-neutrinos
involve only \zz resonances.
In order to disentangle the individual contributions of \ww and \zz resonances 
to the full cross section \refeq{eq:sfsum}, we define
\beq
\label{eq:WWSFcont}
\begin{split}
\rd\sigma(pp\to\ww\to\llnnsf)
=&\; 
\rd\sigma(pp\to\ww\to\llnnsfwwzz)\\
=&\;
\rd\sigma(pp\to \llnndf)\;,
\end{split}
\eeq
and
\beqar
\label{eq:ZZSFcont}
\rd\sigma(pp\to\zz\to\llnnsf) &=& 3\times\rd\sigma(pp\to\llnnsfzzone),
\eeqar
where \zz or \ww
resonances are excluded by selecting $2\ell2\nu$ flavour configurations that admit only  
interactions between $\ell^+\nu$ and $\ell^-\bar\nu$ final states or 
$\ell^+\ell^-$ and $\nu\bar\nu$
final states, respectively.
The \ww cross section \refeq{eq:WWSFcont} is dominated by \ww diboson resonances
and is free from \zz resonances. 
By definition, it includes 
all resonant and non-resonant topologies that 
contribute to $pp\to\llnndf$, and it receives contributions only 
from the $pp\to\llnnsfwwzz$ channel.
Similarly, the \zz cross section \refeq{eq:ZZSFcont}
is dominated by \zz diboson resonances
and is free from \ww resonances.
It involves only resonant and non-resonant topologies that 
contribute to $pp\to\llnnsfzzone$, and it
receives contributions from all neutrino flavours.
The various neutrino-flavour contributions to \refeq{eq:sfsum} are related to 
\refeqs{eq:WWSFcont}{eq:ZZSFcont} through
\beq
\label{eq:sfv_mu_tau}
\rd\sigma(pp\to\llnnsfzzone) 
=\rd\sigma(pp\to\Pe^+\Pe^-\nu_\Ptau\bar\nu_\Ptau)
=\frac{1}{3}\,\rd\sigma(pp\to\zz\to\llnnsf)\,,
\eeq
and the following separation holds
\beq
\label{eq:sfv_e}
\begin{split}
\rd\sigma(pp\to\llnnsfwwzz)
=&\;
\rd\sigma(pp\to\ww\to\llnnsf)
+\frac{1}{3}\,\rd\sigma(pp\to ZZ\to\llnnsf)\\
&{}+\rd\sigma_{\mathrm{int}},
\end{split}
\eeq
where $\rd\sigma_{\mathrm{int}}$ stands for the interference between 
topologies of \ww and \zz type.
As we will see, the splitting of the DF cross section into a \ww
and a \zz channel (and the interference of the two) is very instructive
in order to understand the shapes and higher-order corrections of 
certain kinematic distributions, which are affected to a
different extent by these two dominant contributions in different regions of
phase space.


\begin{figure*}[t]
  \centering
  \includegraphics[width=\relplotwidth\textwidth]{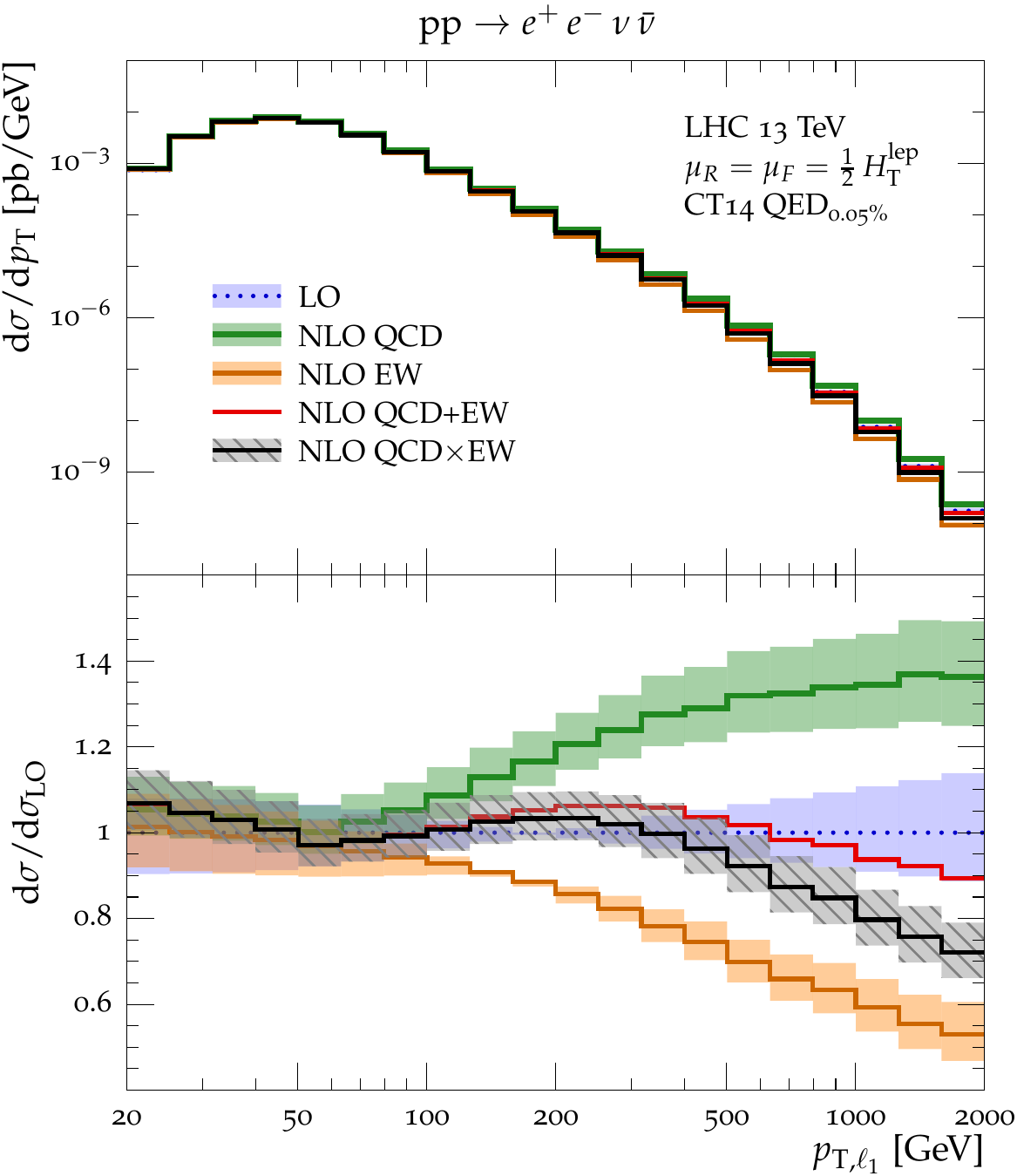}
  \qquad 
  \includegraphics[width=\relplotwidth\textwidth]{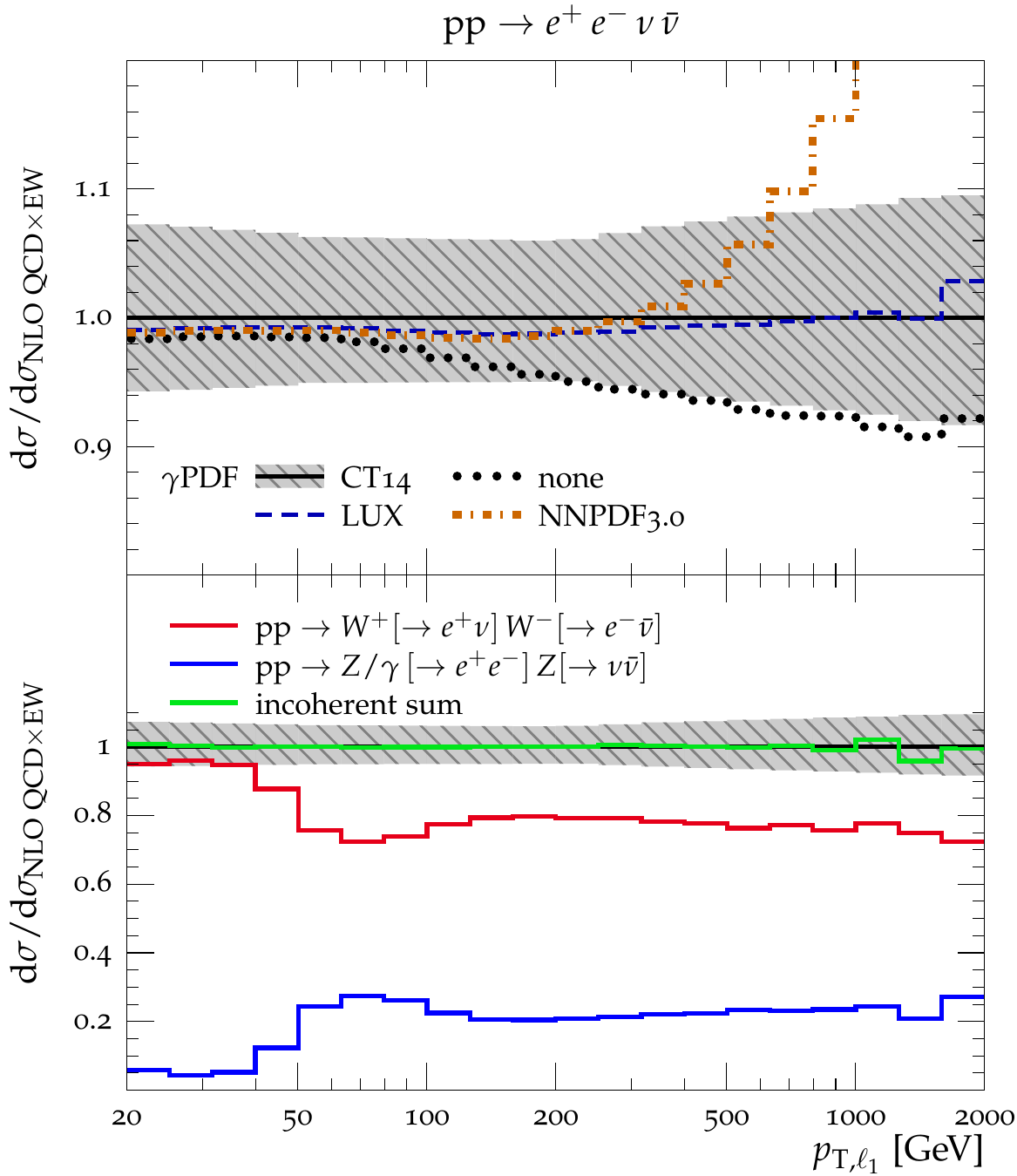}
  \caption{
    Distribution in the transverse momentum of the leading lepton, 
    $p_{\rT,\ell_1}$, for $pp\to\llnnsf$ at 13\,\TeV. All neutrino flavours, 
    $\nu=\nu_e,\nu_\nu,\nu_\tau$, are included.
    Left and upper-right plots as in \reffi{fig:emvv_pTl1}.
    The lower-right ratio plot shows the
    relative weight of the 
    $W^+W^-\to\llnnsf$ and $ZZ\to \llnnsf$ contributions, as defined in 
    \refeqs{eq:WWSFcont}{eq:ZZSFcont}, as well as their incoherent sum
    \refeq{eq:WWZZSFsum}. 
    \label{fig:eevv_pTl1}
    \vspace*{5mm}
  }
\end{figure*}

A selection of differential distributions is presented 
in \reffis{fig:eevv_pTl1}{fig:eevv_dPhi}.
Similarly as in \refse{se:dfresults}, 
in every figure we illustrate \NLO \QCD and \EW predictions with 
corresponding $K$-factors (left plot)
as well as $\Pa$-induced effects (upper-right plot). Since the \NLOQCDtEWYFS 
and \NLOQCDtEWCSS approximations behave similarly as for the different-flavour process, we
do not show corresponding plots.\footnote{
  It should be noted, however, that the \NLOQCDtEWYFS and \NLOQCDtEWCSS 
  approximations reproduce the generally subdominant \zz processes to 
  much higher precision in the \TeV regime than the dominant \ww processes.
}
Instead, in the lower-right panels we quantify the
relative importance of the \ww and \zz contributions defined 
in \refeqs{eq:WWSFcont}{eq:ZZSFcont}, as well as their incoherent sum,
\beq
\label{eq:WWZZSFsum}
\begin{split}
\lefteqn{\hspace{-2ex}\rd\sigma(pp\to\ww\to\llnnsf)+\rd\sigma(pp\to\zz\to\llnnsf)}\\
=&\;
\sum_{\ell=\Pe,\Pmu,\Ptau}\!\!\!\rd\sigma(pp\to\Pe^+\Pe^-\nu_\ell\nu_\ell)
-\rd\sigma_{\mathrm{int}}\;.
\end{split}
\eeq

\begin{figure*}[t]
\centering
  \includegraphics[width=\relplotwidth\textwidth]{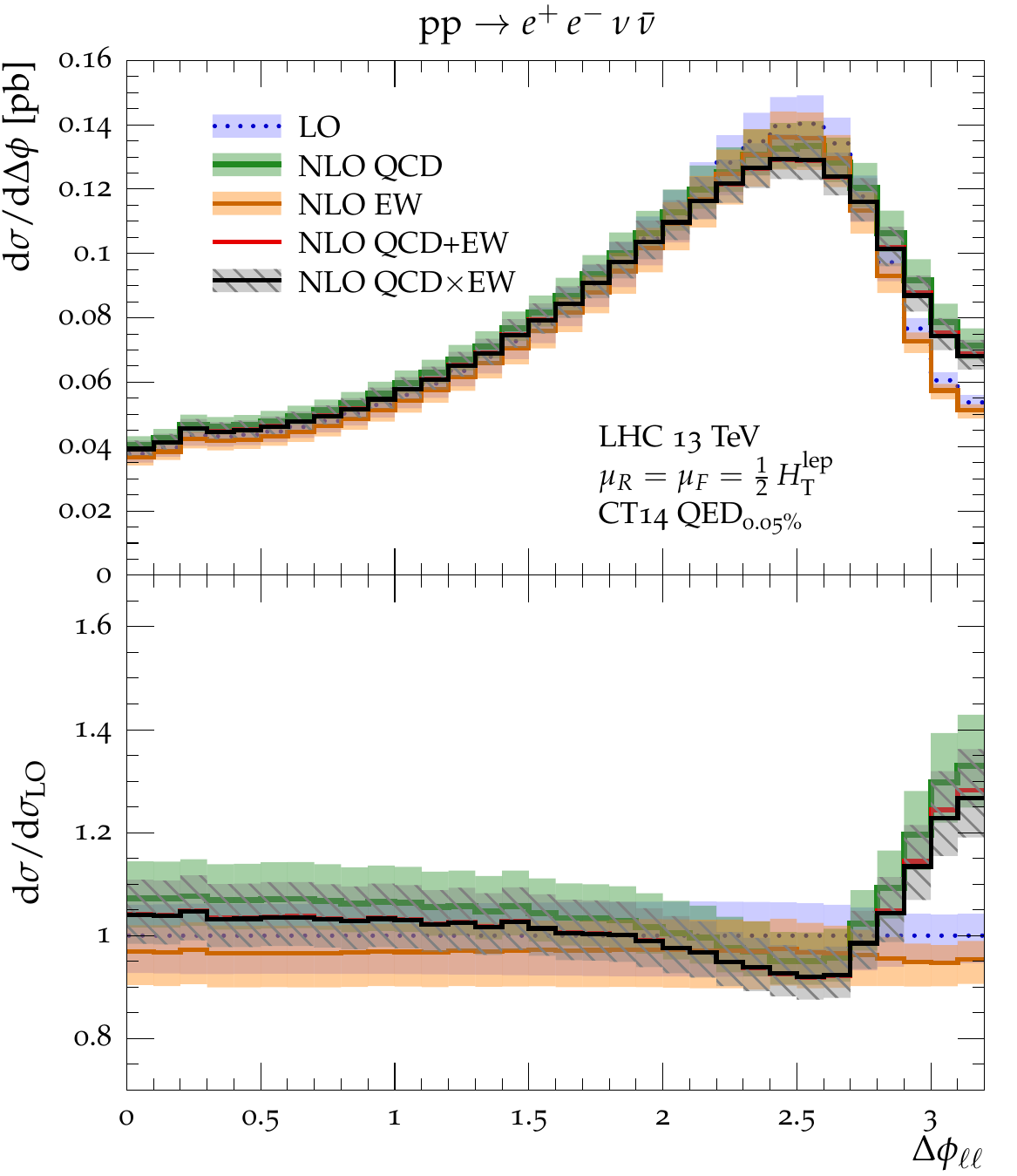}
  \qquad 
  \includegraphics[width=\relplotwidth\textwidth]{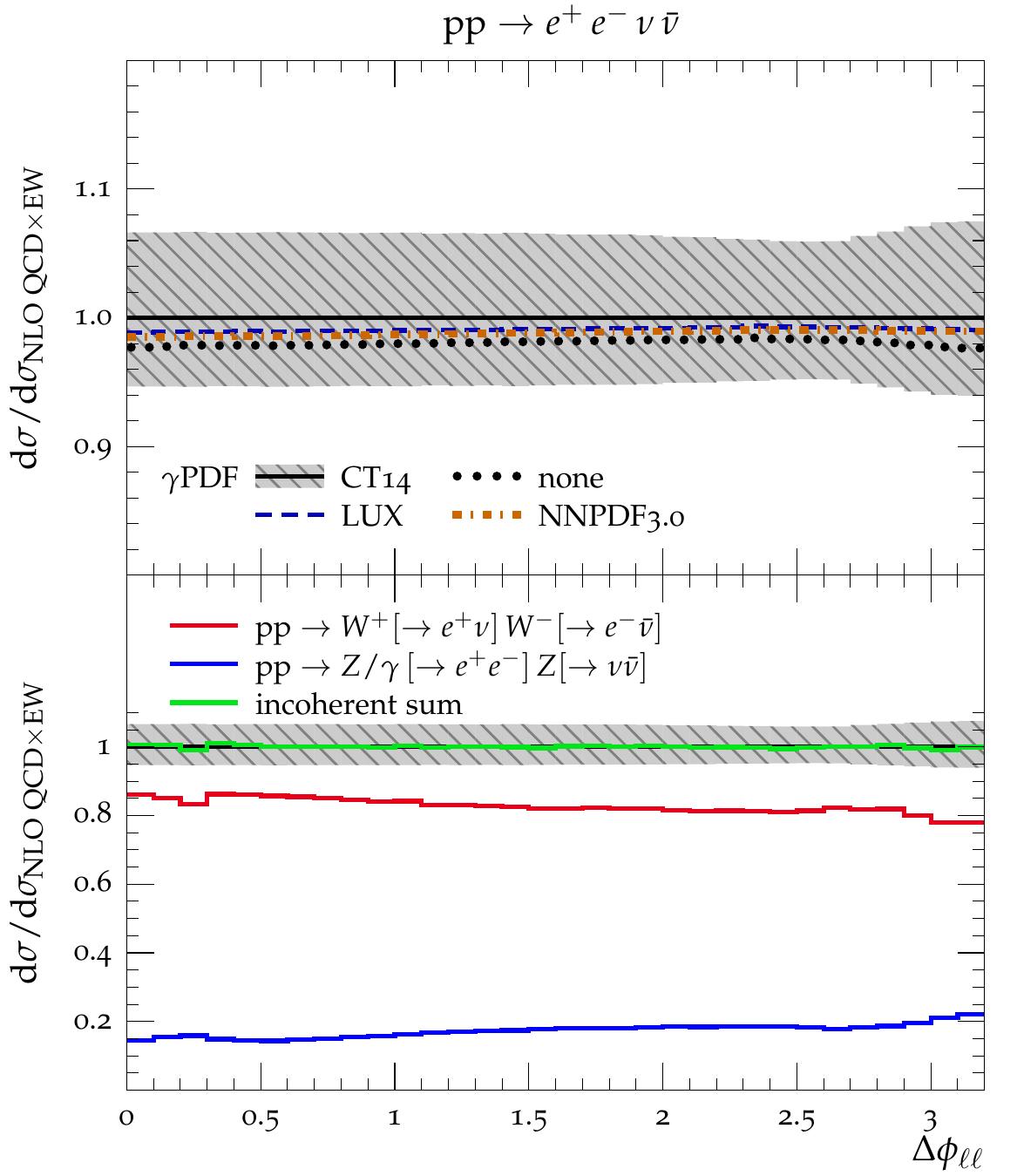}
  \caption{
    Distribution in the azimuthal separation of the $e^+e^-$ pair, 
    $\Delta\phi_{\ell\ell}$, for $pp\to\llnnsf$ at 13\,\TeV. 
    Details as in \reffi{fig:eevv_pTl1}.
    \label{fig:eevv_dPhi}
    \vspace*{5mm}
  }
\end{figure*}

The most striking evidence emerging from 
\reffis{fig:eevv_pTl1}{fig:eevv_mll} is that
the incoherent sum 
\refeq{eq:WWZZSFsum} provides an excellent approximation of the full $\llnnsf$
cross section at \NLO \QCDtEW level. 
In fact, in all considered observables, 
apart from the far off-shell tail 
of the four-lepton invariant mass distribution $m_{\ell\ell\nu\nu}$ 
shown in \reffi{fig:eevv_dPhi}, 
interference effects are so suppressed that they
cannot be resolved at all with the available Monte Carlo statistics.


\begin{figure*}[t]
  \centering
  \includegraphics[width=\relplotwidth\textwidth]{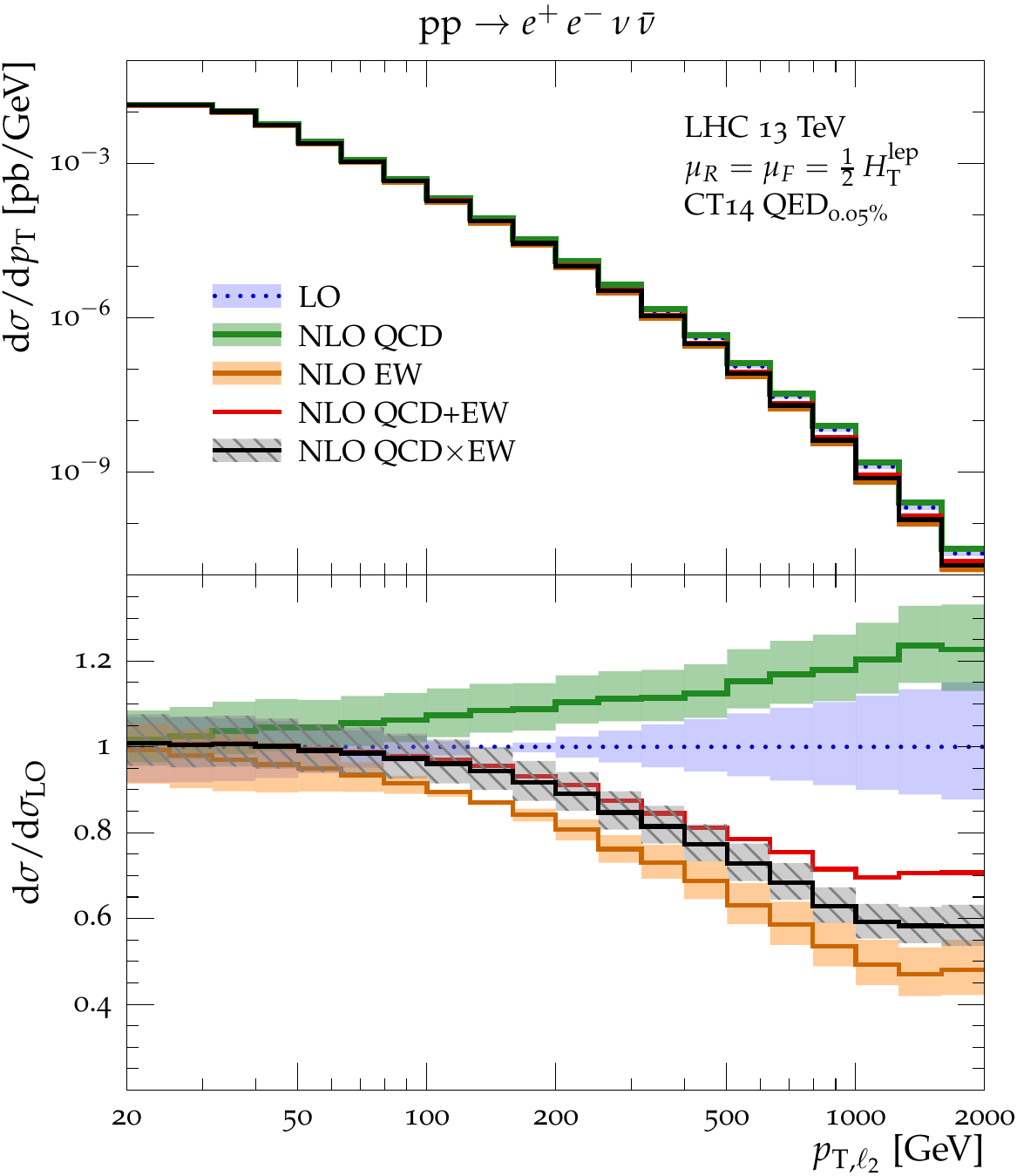}
  \qquad 
  \includegraphics[width=\relplotwidth\textwidth]{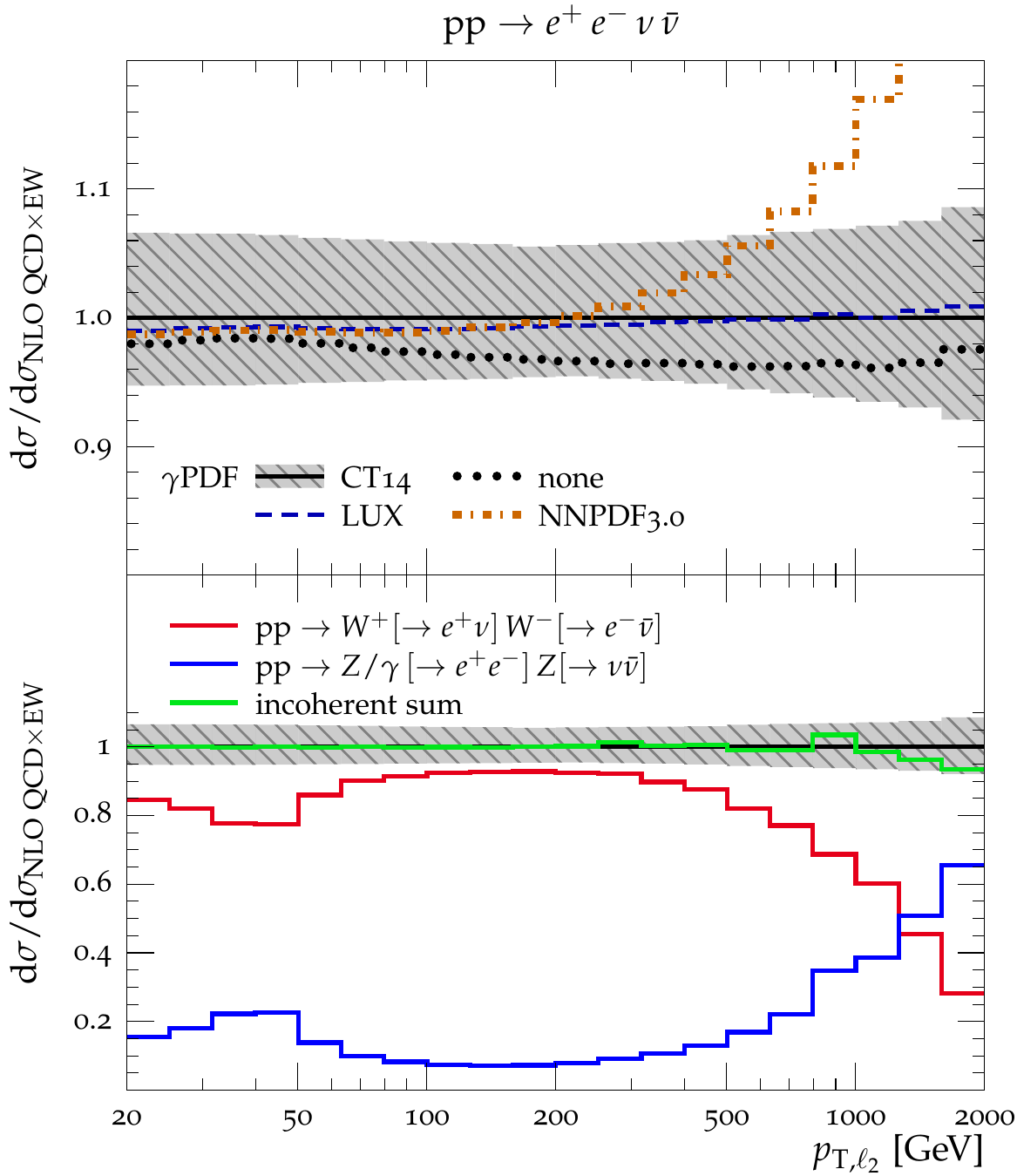}
  \caption{
    Distribution in the transverse momentum of the subleading lepton, $p_{\rT,\ell_2}$, for 
    $pp\to\llnnsf$ at 13\,\TeV. Details as in \reffi{fig:eevv_pTl1}.
    \label{fig:eevv_pTl2}
    \vspace*{5mm}
  }
\end{figure*}

\begin{figure*}[t!]
\centering
  \includegraphics[width=\relplotwidth\textwidth]{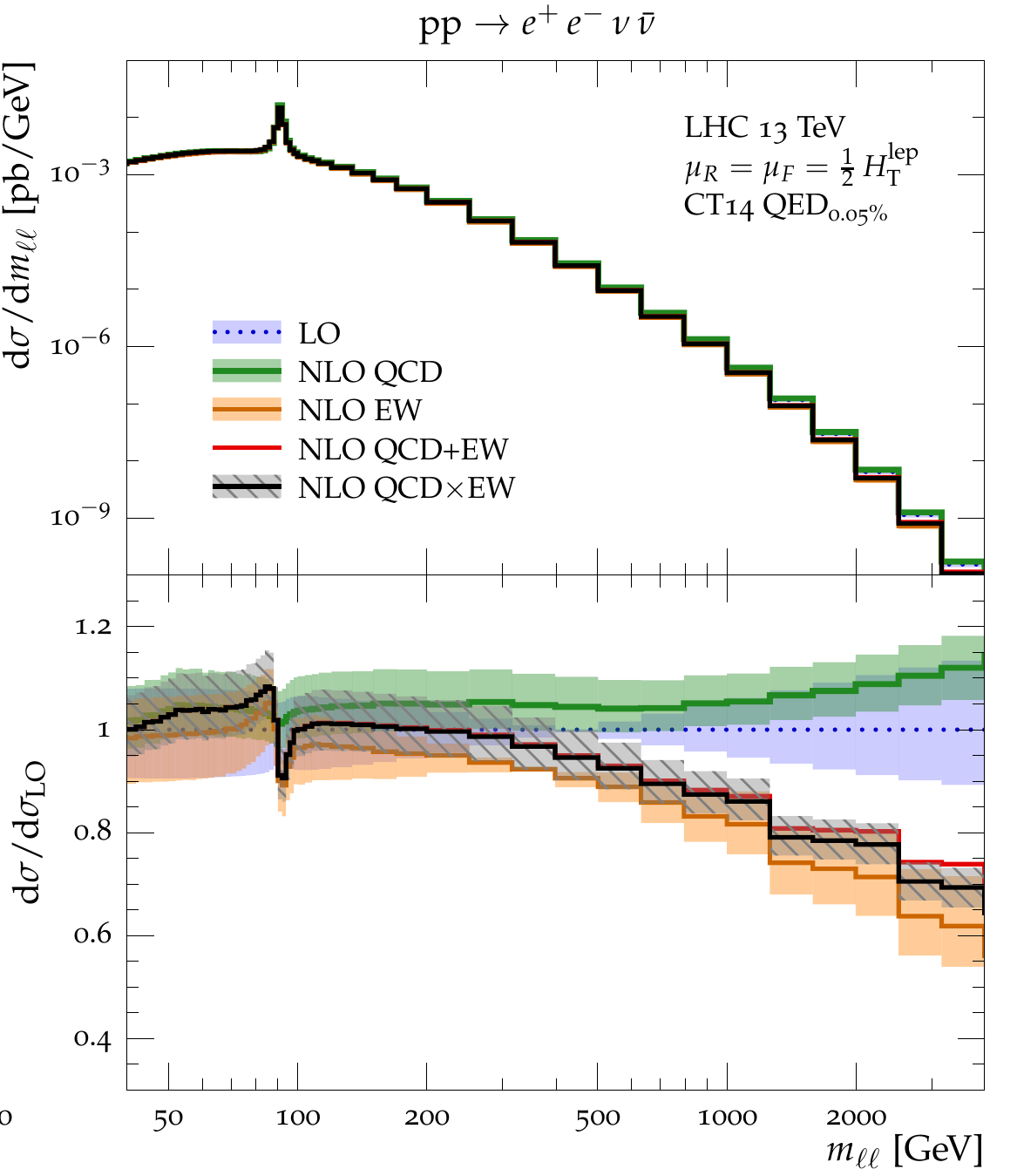}
  \qquad 
  \includegraphics[width=\relplotwidth\textwidth]{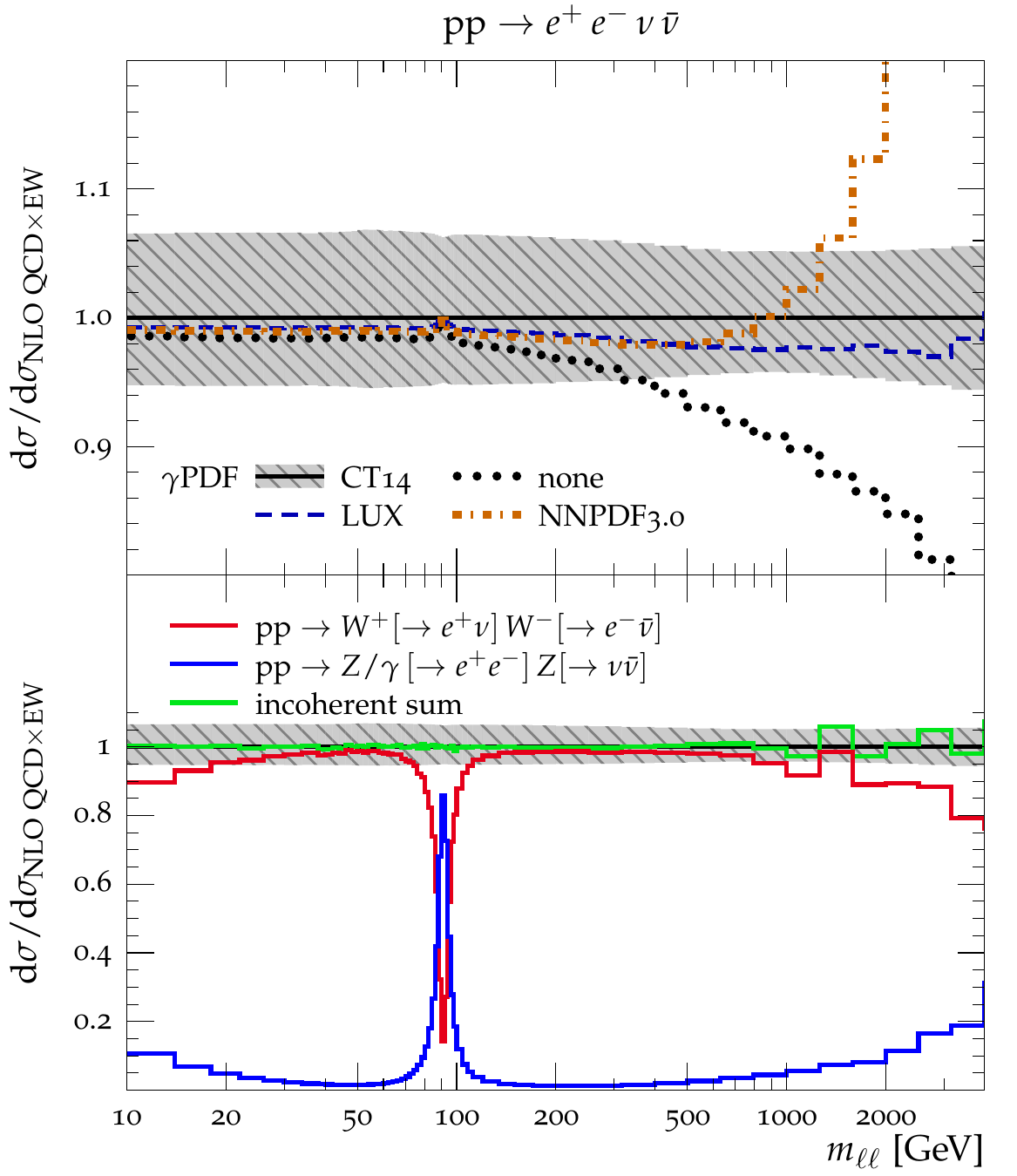}
  \caption{
    Distribution in the $e^+e^-$ pair, $m_{\ell\ell}$, for $pp\to\llnnsf$ at 13\,\TeV.
    Details as in \reffi{fig:eevv_pTl1}.
    \label{fig:eevv_mll}
    \vspace*{5mm}
  }
\end{figure*}

The integrated $\llnnsf$ cross section, the distributions
in $p_{\rT,\ell_1}$ (\reffi{fig:eevv_pTl1}), $p_{\rT,\ell_2}$
(\reffi{fig:eevv_pTl2}), $m_{\ell\ell}$
(\reffi{fig:eevv_mll}), and $\Delta\phi_{\ell\ell}$
(\reffi{fig:eevv_dPhi}) are dominated by \ww resonances in the 
majority of the plotted range. In those regions it is not surprising
to observe that \QCD and \EW corrections behave very similarly as 
in the different-flavour case discussed in \refse{se:dfresults}.
Vice versa, in the presence of sizeable \zz contributions, radiative
corrections can behave in a different manner as compared to the different-flavour
case.  For example, this is observed in the tail of the $p_{\rT,\ell_2}$ distribution 
beyond 1\,\TeV. 
There, \zz resonance contributions become as important as \ww ones, resulting in a
reduction of the magnitude and a change in shape of the
\EW corrections.
Similarly, the size of the contributions from $\Pa$-induced 
processes is reduced as compared to $pp\to\llnndf$ (\reffi{fig:emvv_pTl2}).

\begin{figure*}[t!]
\centering
  \includegraphics[width=\relplotwidth\textwidth]{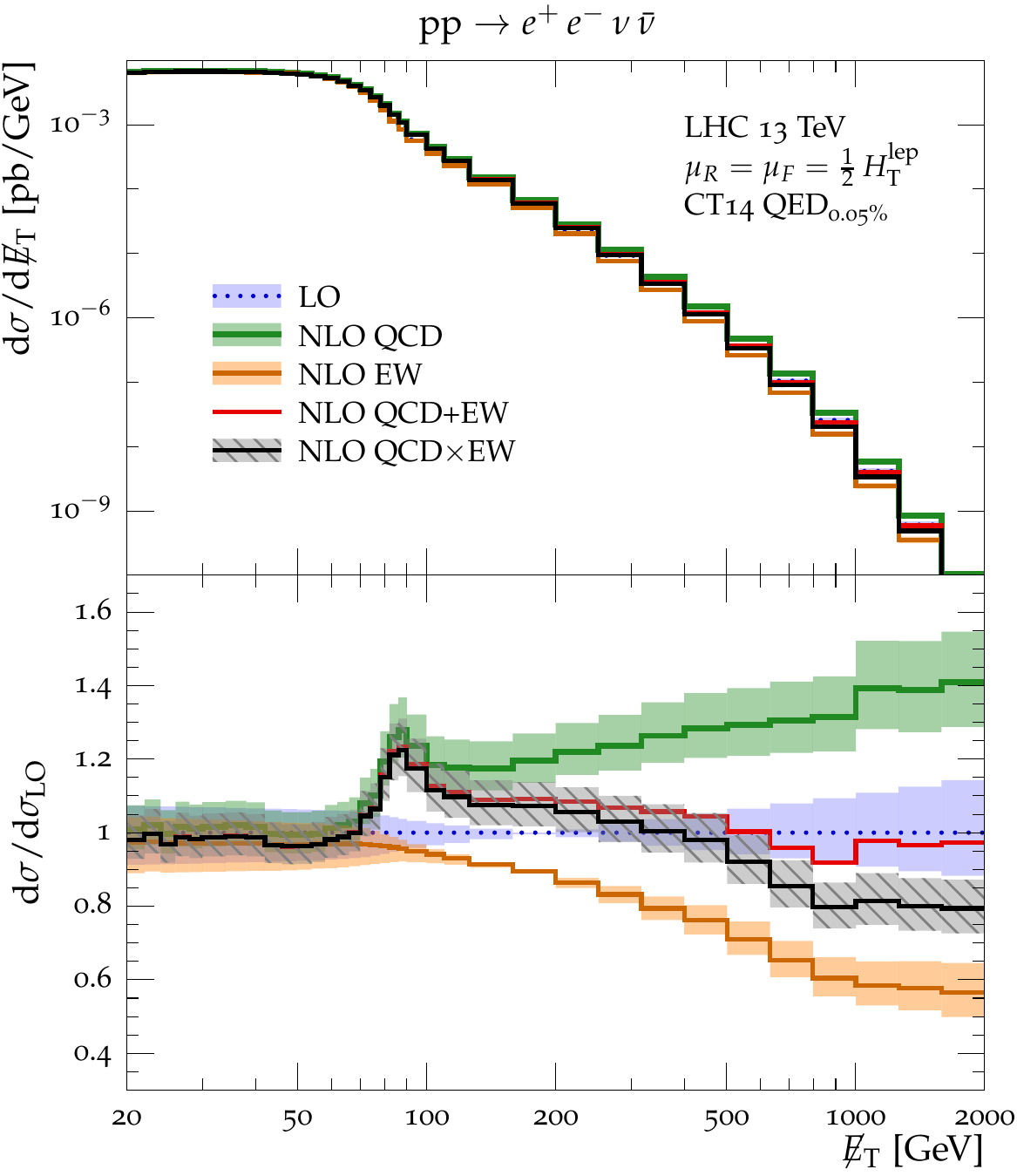}
  \qquad 
  \includegraphics[width=\relplotwidth\textwidth]{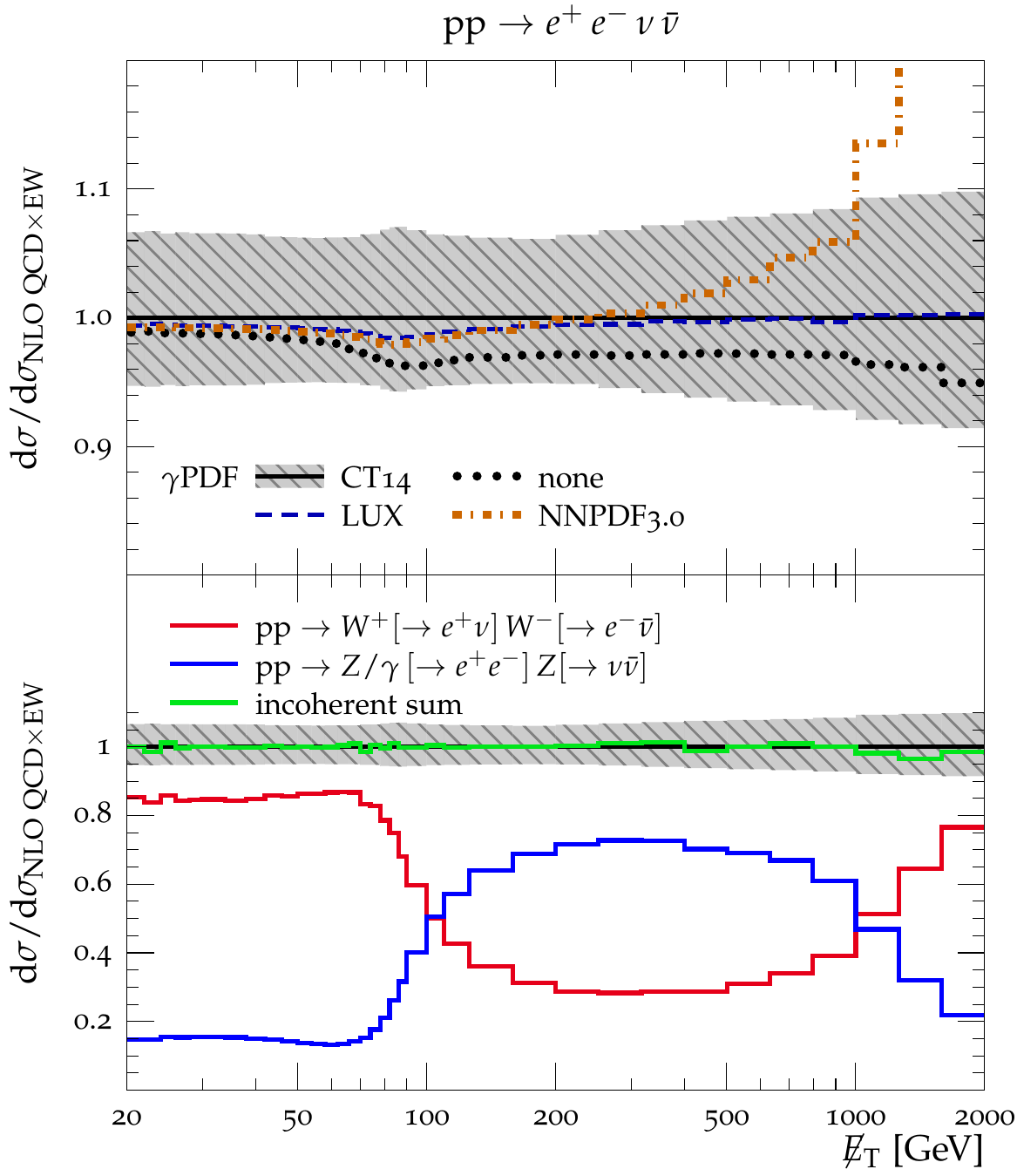}
  \caption{
    Distribution in the missing transverse momentum, \missingET, for 
    $pp\to\llnnsf$ at 13\,\TeV. Details as in \reffi{fig:eevv_pTl1}.
    \label{fig:eevv_pTmiss}
    \vspace*{5mm}
  }
\end{figure*}

In the \missingET distribution (\reffi{fig:eevv_pTmiss}) we observe a more
intriguing interplay between \ww and \zz resonances.  While \ww
topologies represent the main contribution at low and very high \missingET,
the region between 100\,\GeV and 1\,\TeV is dominated by \zz resonances.
This is related to the fact---already observed in the different-flavour 
case---that the production of a $\nu\bar\nu$ system via \ww resonances
is strongly suppressed for
$\missingET>\MW$.  In the $pp\to \llnnsf$ channel, this suppression 
manifests itself through the enhancement of \zz contributions, 
where large \missingET can directly arise through a boosted 
\PZ boson decaying to $\nu\bar\nu$.
In contrast, due to the absence of \zz resonances, 
in the $\llnndf$ channel the suppression of \ww resonances
leads to the enhancement of radiative effects at \NLO \QCD and
\NLO \EW. Vice versa, due to the opening of \zz resonances,
in the $\llnnsf$ channel we observe smaller \NLO \QCD and photon-induced contributions 
and larger negative \NLO \EW corrections.


The invariant mass of the $\Pe^+\Pe^-$ pair (\reffi{fig:eevv_mll}) represents a
powerful discriminant between \ww and \zz channels. On the one hand, 
most of the spectrum is driven by \ww contributions and 
behaves very similarly as for the corresponding different-flavour 
observable shown in \reffi{fig:emvv_mll}. On the other hand,
in the vicinity of $m_{\ell\ell}\approx \MZ$, the \zz channel 
gives rise to a sharp $\PZ\to\Pe^+\Pe^-$ peak 
well above the \ww continuum. In this region
photon radiation off the charged leptons induces significant 
distortions of the \PZ line shape that
are obviously not present in the DF case.
Such shape corrections are qualitatively similar 
to those in \reffi{fig:emvv_mln}
for the $m_{\ell\nu}$ spectrum. 
However, since $\PZ\to\Pe^+\Pe^-$ decays involve two charged leptons,
we find an even more significant reduction of the peak cross section. Moreover,
due to the presence of a large \ww background,
the positive \NLO \EW $K$-factor 
below the peak turns out to be much less pronounced than in
$m_{\ell\nu}$.

\begin{figure*}[t]
\centering
  \includegraphics[width=\relplotwidth\textwidth]{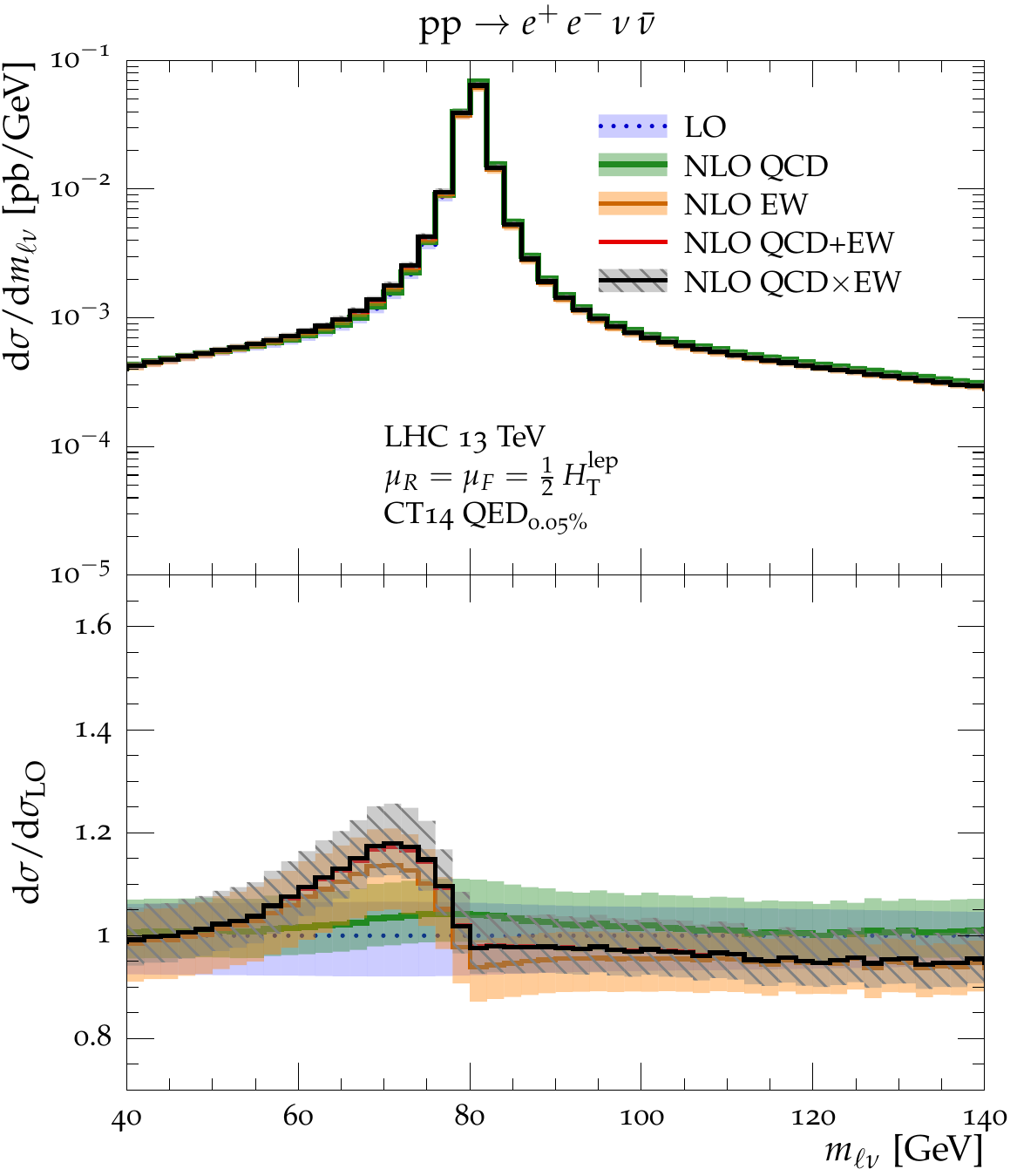}
  \qquad 
  \includegraphics[width=\relplotwidth\textwidth]{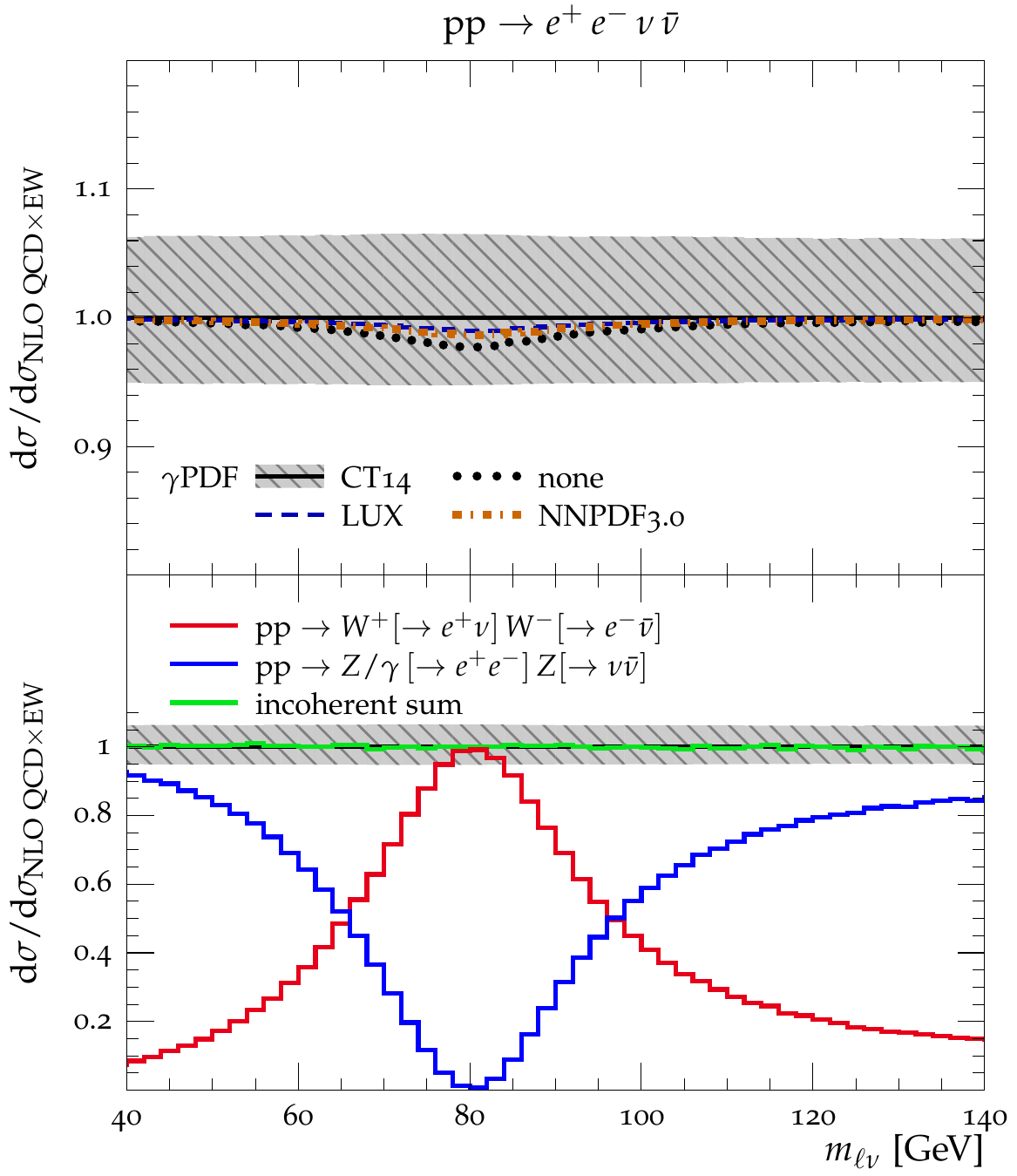}
  \caption{
    Distribution in the invariant mass of one matching 
    lepton-neutrino pair, $m_{\ell\nu}$, for $pp\to\llnnsf$ at 13\,\TeV.
    Details as in \reffi{fig:eevv_pTl1}.
    \label{fig:eevv_mln}
    \vspace*{5mm}
  }
\end{figure*}

Similarly, the experimentally unobservable $m_{\ell\nu}$ distribution shown in 
\reffi{fig:eevv_mln} for the SF case, while dominated by \ww 
resonant channels near the \PW resonance, receives large contributions
from \zz channels on either side. Consequently, the large \NLO \EW 
corrections below the \PW peak in the \ww channel, dominated by real 
photon radiation, cf.\ \reffi{fig:emvv_mln}, are much smaller as they are diluted by the very 
small corrections for the \zz channel.

\begin{figure*}[t]
\centering
 \includegraphics[width=\relplotwidth\textwidth]{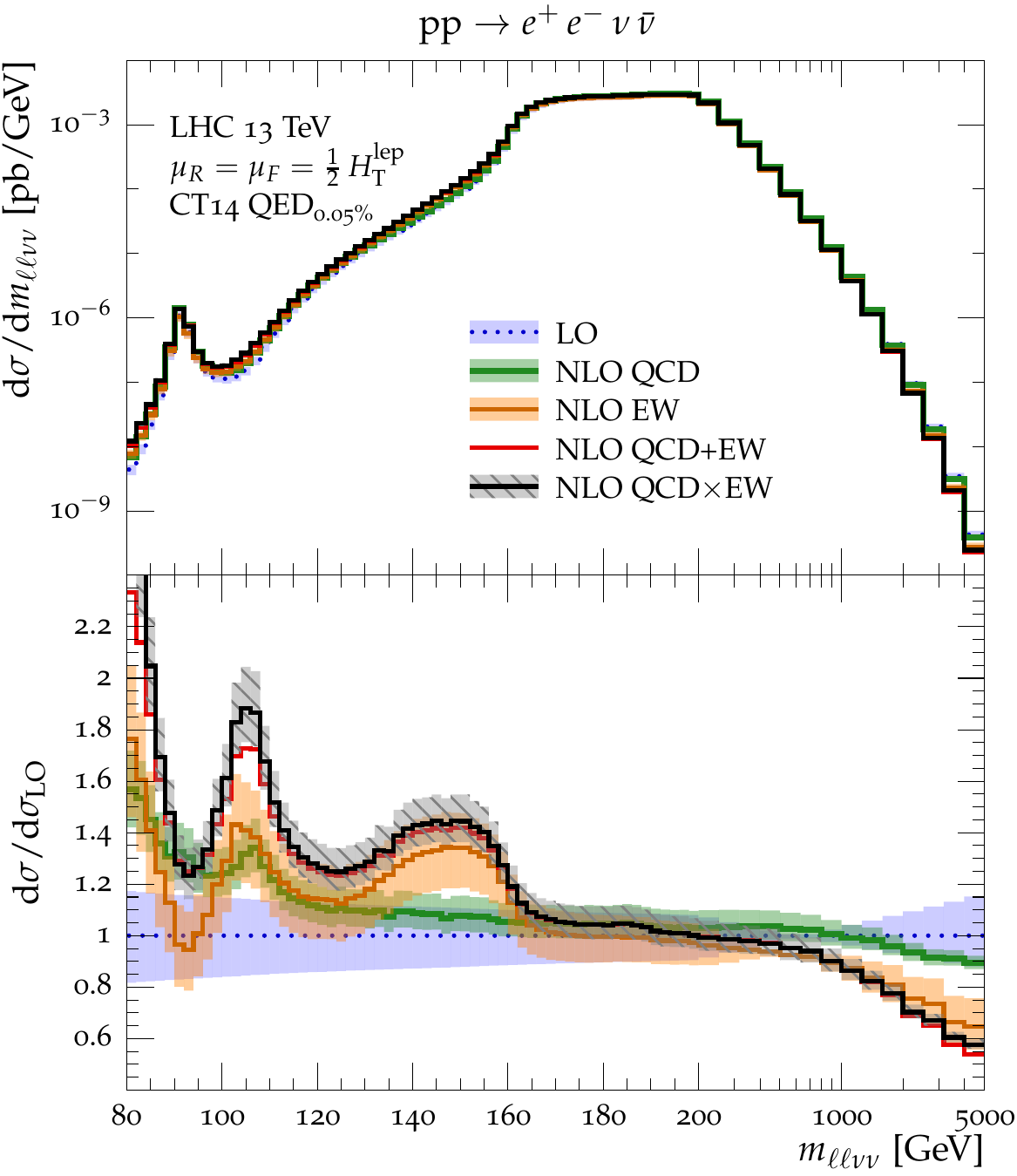}
 \qquad 
 \includegraphics[width=\relplotwidth\textwidth]{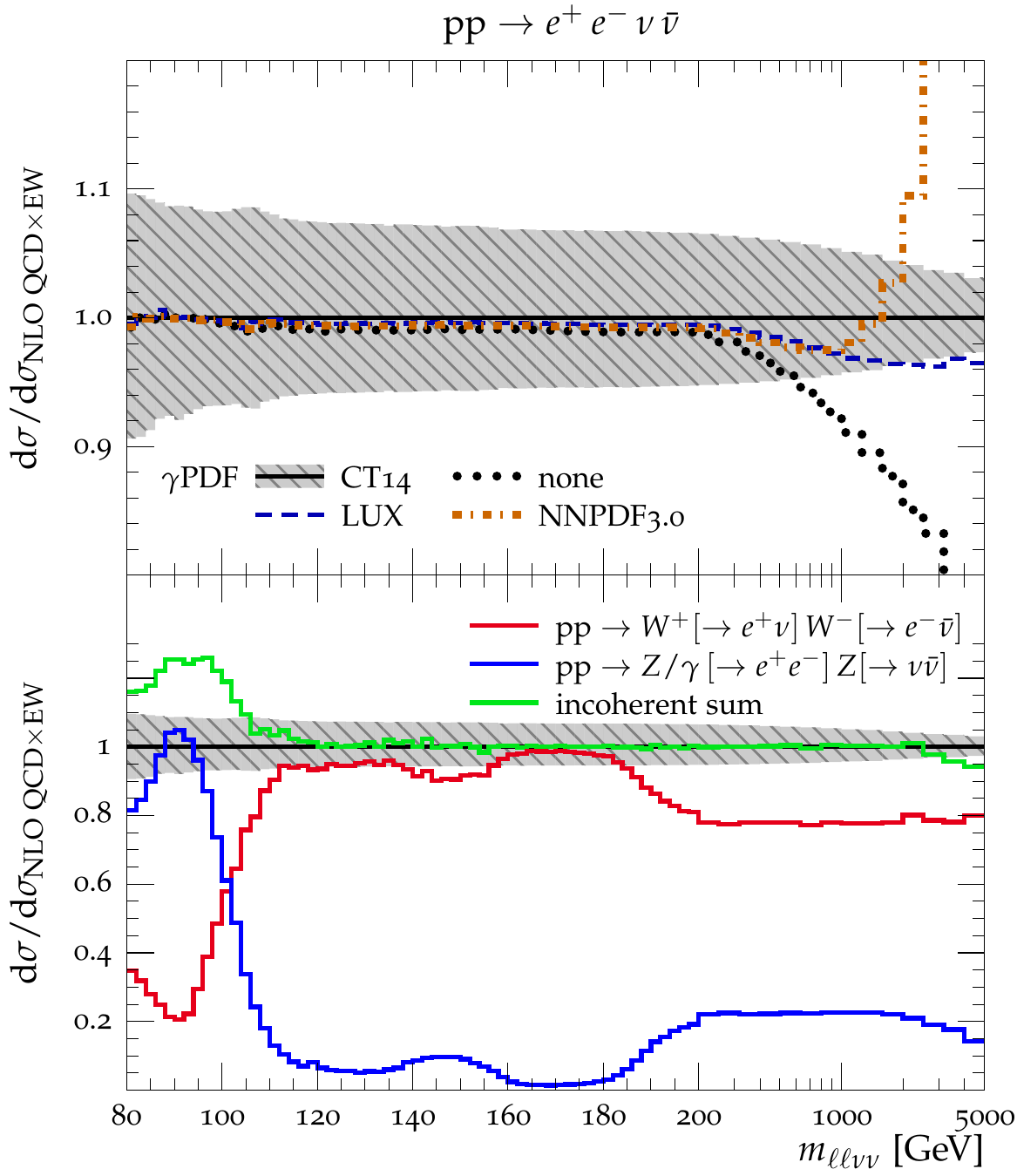}
 \caption{
   Distribution in the  invariant mass of all four final 
   state leptons and neutrinos, $m_{\ell\ell\nu\nu}$, for $pp\to\llnnsf$ at 
   13\,\TeV. Details as in \reffi{fig:eevv_pTl1}.
   \label{fig:eevv_mllnn}
    \vspace*{5mm}
 }
\end{figure*}

The equally unobservable four-lepton invariant mass, $m_{\ell\ell\nu\nu}$, 
displayed in \reffi{fig:eevv_mllnn}, shows similar features as its 
DF counterpart. Again, the reason is the dominance of the \ww channels 
over much of its range. Only at very low invariant masses, near the 
the \PZ-pole, the importance of the \zz channels increases up to becoming dominant. 
This is the only regime out of all 
observables considered in this paper, where a visible 
interference effect between the \ww and \zz channels can be observed, reaching up to 
-25\% on the \PZ-pole itself. This observation can be explained by 
the fact that this is the only region where at least one of the gauge 
bosons is forced off shell and both the \ww and \zz channels populate the 
same surviving resonance, cf.\ the diagrams of \reffi{fig:BorndiagramsWW}c 
and \reffi{fig:BorndiagramsZZ}b.

Finally, \reffi{fig:eevv_dPhi} shows the azimuthal separation 
of both charged leptons. Here, due to the dominance of the \ww 
channel throughout we observe very similar effects as already documented in 
\reffi{fig:emvv_dPhi} for the DF case.

Similarly as for the DF case, we have checked that the \NLO 
\QCDtEW predictions of \reffis{fig:eevv_mll}{fig:eevv_dPhi} are 
reproduced with sufficient accuracy by the \NLOQCDtEWYFS approximation.


\clearpage


\clearpage
\section{Summary and conclusions}
\label{se:conclusions}

We have presented \NLO \QCD and \EW  predictions for $\llnndf$ and $\llnnsf$
production at the LHC. These reactions are 
representative of all possible 
diboson processes 
$pp \to\ww/\zz\to\ell^+_i\ell^-_j\nu\bar\nu$, 
which lead to signatures with two leptons of opposite charge
plus missing transverse energy.
Due to the large SU(2) charges of the intermediate \PW and \PZ bosons, the
underlying $q\bar q\to \PV\PV $ subprocesses 
induce huge \EW Sudakov effects
at high energy. As a result, in various observables we find negative \EW corrections beyond
$-50\%$ at the \TeV scale.
Also \QCD corrections can be sizeable, and in order to account
for unknown \NNLO contributions of $\ord(\alphaS\alpha)$ in an approximate
way, we have advocated a factorised combination of \NLO \QCD and \EW
corrections.

Photon-induced processes have been computed at \NLO \EW, taking into account 
the channels of type $\aa\to 2\ell2\nu$,
$\aa\to 2\ell2\nu\Pa$, and $\Pa q \to2\ell 2\nu q$. 
In the tails of the $m_{\ell\ell}$ and leading-lepton $p_\rT$  distributions,
such contributions can become important. In particular, the $\missingET$ 
distribution receives 
large \Pa-induced corrections starting already from about
100\,\GeV due to a suppression of the \LO $q\bar q\to\ww$ process in this region.
%
%
With the poorly constrained
photon density of the \NNPDFQED fit, \Pa-induced
processes can be strong enough to compensate the negative corrections of
Sudakov type.  However, based on the more precise photon densities in the 
\CTQED and \LUXQED PDFs, \Pa-induced contributions can reach at most
10--20\% and remain clearly subleading with respect to \EW Sudakov logarithms.

For observables that are inclusive with respective to \QED radiation, \NLO \EW
corrections can be described with good accuracy by a so-called \EWvirt
approximation, which includes only IR-subtracted virtual \EW corrections and
is particularly suitable in the context of multi-jet merging.  However, for
observables depending on charged leptons, also radiative \QED effects can
play an important role.  Thus we have studied the possibility of 
augmenting the \EWvirt approximation through \QED radiation effects
generated via \YFS soft-photon resummation or, alternatively, 
by the Catani--Seymour dipole-based DGLAP-type resummation of the \CSShower.
In general, both approaches provide reasonably accurate results.
More precisely, both approaches describe the high-energy regions 
on a similar level with deviations being typically smaller than 10\%, 
while the \YFS resummation implementation in \Sherpa 
also preserves the existing resonance structure.

Radiative corrections in $2\ell 2\nu$ channels with same and opposite lepton
flavour behave in a fairly similar way.  This can be understood in the light
of the respective resonance structures.  On the one hand, $pp\to\llnndf$ 
solely contains \wpwm resonances, while $pp\to \llnnsf$ involves both \wpwm
and \zz resonances.  On the other hand, possible interferences between \wpwm
and \zz topologies turn out to be completely negligible for all relevant observables.
Moreover, \wpwm contributions to $pp\to\llnnsf$ are widely dominant with respect to 
\zz ones. 
This is the reason why \QCD and \EW corrections 
behave very similarly in $\llnnsf$  and $\llnndf$ production.
Nevertheless, we have pointed out that 
\NLO effects can still be quite sensitive to the flavour structure in certain observables.
This can for example be observed in the $m_{\ell\ell}$ and \missingET distributions in
correspondence to the occurrence of \zz dominated regions that originate, respectively, from the
$\PZ\to \ell^+\ell^-$ resonance and due to the suppression of \wpwm
topologies for $\missingET>\MW$.

Concerning the treatment of hard scattering processes with external photons at \NLO \EW,
in \refapp{app:subren} we have presented 
a  general analysis of the interplay between the definition of the coupling
$\alpha$ for  external photons, the renormalisation of 
the photon wave function, and the renormalisation of the \aPDF. 
In particular, we have pointed out that,
in order to avoid large logarithms associated with $\Delta \alpha(M_Z^2)$,
the coupling $\alpha$ for final- and initial-state photons should be defined, respectively,
at high energy
and in the Thomson limit, $Q^2\to0$.
In practice, 
at energies of the order of the
\EW scale or above,
initial-state photon couplings can be parametrised 
in the $G_\mu$ scheme or in the $\alpha(M_Z)$
scheme, while $\alpha(0)$
should not be used.

The tools that have been used in this project---\Sherpa, \Munich, and
\OpenLoops---implement automated \NLO \QCDpEW algorithms that are applicable
to any Standard Model process and will be made publicly available in the
near future.

\acknowledgments
MS is grateful for illuminating discussions with G.\ Salam, C.\ Schmidt and 
C.--P.\ Yuan.
This research was supported in part by the Swiss
National Science Foundation (SNF) under contracts PP00P2-128552 
and BSCGI0-157722
as well as by the Research Executive Agency (REA) of the European Union
under the Grant Agreements PITN--GA--2010--264564 ({\it
LHCPhenoNet}), PITN--GA--2012--315877 ({\it MCnet}), 
PITN--GA--2012--316704 ({\it HiggsTools}) and the ERC Advanced Grant
MC@NNLO (340983).

\begin{appendix}

\section{Infrared subtraction, \texorpdfstring{\boldaPDF}{photon-PDF} 
         renormalisation and definition of \texorpdfstring{$\boldsymbol{\alpha}$}{alpha}}
\label{app:subren}

This Appendix starts, in~\refse{app:CS}, with a general documentation of 
the implementation of Catani--Seymour subtraction at NLO \EW in \Sherpa and \Munich.
This serves as a basis for the discussion, 
in~\refses{app:ypdfrenormalisation}{app:alpharenormalisation},
of the cancellation of light-fermion mass singularities
in processes with external photons. Such cancellations involve a subtle interplay between 
the definition of the coupling $\alpha$,
the renormalisation of
the \aPDF, and the photon wave-function renormalisation.
In particular we point out that, in order to avoid a logarithmic sensitivity
to light-quark and lepton masses, the coupling of 
on-shell final-state photons should be defined in the limit of vanishing 
$Q^2$, while for initial-state photons a definition of 
$\alpha$ at the \EW scale or at $\muf^2\sim \hat s$ should be used.
This was first noticed in~\cite{Harland-Lang:2016lhw}, 
based on arguments related to the evolution of the \aPDF 
at \LO, and is confirmed by our explicit 
analysis at \NLO \EW.

\subsection{Catani--Seymour subtraction at \texorpdfstring{$\ord(\alpha)$}{O(alpha)}}
\label{app:CS}

In this section we present the implementation of Catani--Seymour subtraction
at \NLO \EW in \Sherpa and \Munich.  While the construction of Catani--Seymour
dipoles for \QED radiation has been discussed in detail
in~\cite{Dittmaier:1999mb,Dittmaier:2008md,Gehrmann:2010ry}, our
implementation relies on the direct transposition of the
original $\ord(\alphaS)$ subtraction formalism~\cite{Catani:1996vz,Catani:2002hc} to
$\ord(\alpha)$.  In the following, we provide the 
complete set of formulae that permits to obtain $\ord(\alpha)$ dipoles 
from the results of~\cite{Catani:1996vz} for massless partons,  thereby extending the schematic
description given in~\cite{Kallweit:2014xda}. 
Moreover,  we point out some subtle aspects related to leptonic contributions and external
photons, which are relevant for the cancellation of fermion-mass logarithms
discussed in~\refses{app:ypdfrenormalisation}{app:alpharenormalisation}.

{
\renewcommand{\arraystretch}{1.5}
\begin{table}[tb]
\begin{center}
\begin{tabular}{cccccc}
dipole type  & $I$            & $J$             & $K$           &  splitting     & $\hat{\bom V}_{IJ,K}$  \\\hline
final--final     & $i\in \Sout$   &  $j\in\Sout$  & $k\in\Sout$  &  $ij\to i+j$   & ${\bom V}_{ij,k}$ \\
final--initial      & $i\in \Sout$   &  $j\in\Sout$  &  $b\in\Sin$   &  $ij\to i+j$   & $x_{ij,b}^{-1}\;{\bom V}^b_{ij}$ \\
initial--final      & $a\in \Sin$    &  $j\in\Sout$  &  $k\in\Sout$  &  $a\to (aj)+j$   & $x_{jk,a}^{-1}\;{\bom V}^{aj}_k$ \\
initial--initial       & $a\in \Sin$    &  $j\in\Sout$  &  $b\in\Sin$   &  $a\to (aj)+j$   & $x_{j,ab}^{-1}\;{\bom V}^{aj,b}$
\end{tabular}
\end{center}
\caption{Correspondence between the generic splitting kernels $\hat{\bom V}_{IJ,K}$ of
\refeq{eq:alldipoles} and the kernels ${\bom V}$ of~\cite{Catani:1996vz}.
For initial-state and final-state partons we use 
specific indices  $a,b \in \Sin$ and $i,j,k\in \Sout$.
Moreover, since the emittee $J\in\Sout$ we identify $J=j$.
The terms  with initial-state emitters, $I=a\in \Sin$, describe 
splittings $a\to (aj)+j$, while final-state emitters, $I=i\in \Sout$,
corresponds to  splittings $(ij)\to i+j$. 
The spectators $K$ can be either 
initial-state ($K=b$) or final-state ($K=k$) partons.
\label{tab:CSsplittingkernels}}
\end{table}
}

Let us consider the $\ord(\alpha)$ corrections to a $2\to m$ hard-scattering process.
The subtraction term for the singularities stemming from photon- or
fermion-bremsstrahlung in the $(m+1)$-parton phase space 
has the general form
\beqar
\label{eq:alldipoles}
\rd \sigma^{A} &=& -\!\!\sum_{I\in \Sinout}
\sum_{J\in \Sout}
\frac{1}{2p_I p_J}
\sum_{K\neq I,J}
\qtilde{\IJtilde}{K}\;
\hat{\bom V}_{IJ,K}
\otimes
\bornIJ\;,
\eeqar
where $\Sinout=\Sin\cup\,\Sout$ is the full set of 
initial-state ($\Sin$) and final-state ($\Sout$) partons. 
Each term in the triple sum over $I$, $J$ and $K$ describes 
$1/(p_I p_J)$ singularities arising from the exchange of a soft parton $J$ 
between an emitter $I$ and a spectator $K$, as well as collinear 
singularities involving the partons $I,J$.
The relevant splitting kernels $\hat{\bom V}_{IJ,K}$ in \refeq{eq:alldipoles}
are convoluted with the 
reduced Born cross section $\bornIJ$,  
where the partons $I$ and $J$ are clustered into a single parton $\IJtilde$ according to the 
respective splitting process.\footnote{For details of the ${\bom V}\otimes \rd\sigma^{B}$ convolution,
such as kinematic  mappings, 
we refer to~\cite{Catani:1996vz}.}
The various types of splitting kernels 
are listed in \refta{tab:CSsplittingkernels}.
In general we consider $\ord(\alpha)$ emissions off 
quarks and leptons, generically denoted as $f=q,\bar q,\ell^-,\ell^+$, 
as well as photon splittings.
Explicit expressions for $\hat{\bom V}_{IJ,K}$ 
corresponding to 
$f \to f\Pa$ 
and $\Pa\to f\bar f$ splittings   can be obtained from the corresponding 
QCD kernels~\cite{Catani:1996vz} for $q \to q g$, $\bar q \to \bar q g$, and
$g\to q\bar q$ splittings
by replacing 
\beq\label{eq:subtraction_replacements}
  \alpha_s\longrightarrow\alpha\,,\quad\;
  C_F \longrightarrow Q_f^2\,,\quad\;
  T_R \longrightarrow \ncqsf\,,\quad\;
  n_f\,T_R \longrightarrow \sum_f \ncqsf\,,\quad\;
  C_A\longrightarrow 0\,,
\eeq
where $N_{{\rm c},f}=1$ for leptons and 3 for quarks.
As discussed in \refapp{app:ypdfrenormalisation}, all terms  
$\sum_f \ncqsf$, which arise from massless fermion-loop insertions in the photon propagator
or related real-emission contributions, should include both
quarks and charged leptons. 
The matrix $\qtilde{\IJtilde}{K}$ in \refeq{eq:alldipoles} collects the charge factors 
of the partons $\IJtilde$ and $K$.
It is related to the 
colour-insertion operators of~\cite{Catani:1996vz} via
\beq
\label{eq:QEDcorrelA}
  \frac{\mathbf{T}_{\IJtilde}\cdot\mathbf{T}_K}{\mathbf{T}_{\IJtilde}^2}
  \longrightarrow 
  \qtilde{\IJtilde}{K}.
\eeq
If emitter $\IJtilde$ is a charged fermion, we simply have
\beq
\label{eq:QEDcorrelB}
\qtilde{\IJtilde}{K}
=
\frac{Q_{\IJtilde}\, Q_K}{Q^2_{\IJtilde}}  \qquad\text{for $\IJtilde=f\in\Sinout$},
\eeq
where $Q$ is the incoming charge, \eg $Q=\mp 1\,(\pm 1)$ for an incoming\,(outgoing) 
$\ell^\mp$.
For a photon emitter, $\IJtilde=\Pa$, \refeq{eq:QEDcorrelB} is not applicable
due to $Q_\Pa=0$. This situation occurs for 
final-state $\Pa\to f\bar f$ splittings and initial-state 
$f\to \Pa f$ splittings, which involve only collinear 
singularities that are insensitive to the electromagnetic 
charge of the spectator $K$. In fact, the role of the spectator is merely to 
absorb the recoil resulting form the splitting process, and the  matrix 
$\qtilde{\IJtilde}{K}$ in \refeq{eq:alldipoles} 
distributes the recoil to the various spectators based on the identity
\beq
\label{eq:QEDcorrelC}
\sum_{K\neq \IJtilde} \qtilde{\IJtilde}{K}=-1,
\eeq
which is a manifestation of the charge-conservation relation $\sum_K Q_K=0$.
Since $\qtilde{\Pa}{K}$ does 
not need to be related to the actual charges of the spectators $K$,
any matrix $\qtilde{\Pa}{K}$ that obeys \refeq{eq:QEDcorrelC} 
guarantees a consistent IR subtraction.
The choice implemented in \Sherpa and \Munich for initial-state photon emitters
reads
\beq
\label{eq:QEDcorrelD}
\qtilde{a}{K} =  -\delta_{b,K} \qquad \mbox{for $a=\Pa\in\Sin$}.
\eeq
Here, $a$ and $b$ denote the two initial-state partons,
\ie the recoil
of initial-state $\Pa\to f\bar f$ splittings is absorbed by the
initial-state partner $b$ of the emitter photon $a$.

Final-state $\Pa\to f\bar f$ splittings 
should not be considered for processes with identified on-shell photons.
However, they should be taken into account when
photons are not distinguished from $f\bar f$ pairs.
In order to account for both cases in a flexible way,
we introduce a discriminator $\epsFSgammai$ for every 
final-state photon, defined as
\beq
\label{eq:FSgffswitch}
\epsFSgammai = 
\left\{\begin{array}{l}
1 \\
0 \\
\end{array}\right.
\qquad
\text{when final-state $\gammai\to f\bar f$ splittings are}\quad 
\left\{\begin{array}{l}
\text{allowed}\\
\text{disallowed}\\
\end{array}\right..
\eeq
The $\epsFSgammai$ can be set individually for each photon, taking care 
that the prescription is infrared safe. Of course, if multiple photons fulfil 
the identification criteria simultaneously the assignment has to be 
properly symmetrised.~\footnote{
  One example may be the production of an isolated photon accompanied by 
  a jet, which at $\ord(\alpha^2)$ can be described by 
  $\Pq\Pqbar\to\Pa\Pa$. Now, once an isolated photon is found, for 
  which we set $\epsFSgammai=0$, the remaining photon forming the jet 
  at \LO is allowed to split, thus its $\epsFSgammai=1$. 
}
The charge correlation matrix in \Sherpa and \Munich 
is then chosen as\footnote{
  Other recoil strategies for initial and final state photon splittings 
  are possible and a number of generic choices is implemented in \Sherpa. 
}
\beq
\label{eq:QEDcorrelE}
\qtilde{\gammai}{K} 
= -\frac{1}{2}\,\epsFSgammai\big(\delta_{a,K}+\delta_{b,K}\big)  
   \qquad\text{for every $\gammai\in\Sout$}\,.
\eeq
In this way, when final-state $\gammai\to f\bar f$ splittings are allowed the 
resulting recoil is shared by the two initial-state 
partons $a$ and $b$.

\begin{table}[tb]
\begin{center}
\scalebox{0.93}{
{\renewcommand{\arraystretch}{2.5}
 \hspace*{-2mm}
\begin{tabular}{c|c|c|c}
$AB$ & $P^{AB}(x)$ & ${\overline K}^{AB}(x)$ & ${\widetilde K}^{AB}(x)$ \\
\hline
$f\Pa$       
& $\displaystyle{Q_f^2\, \frac{1 + (1-x)^2}{x}}$    
& $\displaystyle{P^{f\Pa}(x) \ln\frac{1-x}{x} + Q^2_f\, x}$    
& $\displaystyle{P^{f\Pa}(x) \ln(1-x)}$    
\\[1mm]
\hline
$\Pa f$      
& $\displaystyle{\ncqsf \left[ x^2 + (1-x)^2 \right]}$    
& $\displaystyle{P^{\Pa f}(x) \ln\frac{1-x}{x} + 2\ncqsf\, x(1-x)}$    
& $\displaystyle{P^{\Pa f}(x) \ln(1-x)}$    
\\[1mm]
\hline
$f f$  
& $\displaystyle{Q_f^2 \left( \frac{1 + x^2}{1-x} \right)_+}$    
& $\displaystyle{Q_f^2 \left[\bar G^{ff}(x)\vp
- \delta(1-x) \left( 5 - \pi^2 \right) \right]}$   
& $\displaystyle{Q_f^2 \left[ \tilde G^{ff}(x)
- \frac{\pi^2}{3} \delta(1-x) \right]}$    
\\[1mm]
\hline
$\Pa\Pa$ 
& $\displaystyle{\gamma_\Pa\; \delta(1-x)}$    
& $\displaystyle{-\frac{8}{3}\; \gamma_\Pa\; \delta(1-x)}$    
& $0$    
\\
\hline
\end{tabular}}}
\end{center}
\caption{Explicit expressions for $P^{AB}$,
${\overline K}^{AB}$, and ${\widetilde K}^{AB}$
in~\refeq{eq:KP} for all relevant combinations of 
photons and fermions, $f=q,\bar q, \ell^+, \ell^-$, with the auxiliary functions
$\bar G^{ff}(x)= \left(\frac{2}{1-x} \ln\frac{1-x}{x}\right)_+ - (1+x) \ln\frac{1-x}{x} + (1-x)$
and 
$\tilde G^{ff}(x)=\left(\frac{2}{1-x} \ln (1-x)\right)_+ -(1+x)\ln (1-x)$.
\label{tab:CSingredients}}
\end{table}

The cancellation of soft and collinear singularities against virtual corrections 
requires the analytic integrals of the dipole terms  \refeq{eq:alldipoles} 
supplemented by PDF-factorisation counterterms. 
This leads to~\cite{Catani:1996vz}
\beq
\label{eq:IKP}
\begin{split}
  \int_1\rd\sigma^{A}_{ab} + \sigma^{\mathrm{CT}}_{ab}(\muf) 
  \,=&\;\;
  {\bom I}(\{p\};\epsilon)\; \rd\sigma^{\mathrm{B}}_{ab}(p_a,p_b)
  \\ 
  &\;{}
  +\int_0^1\rd x \sum_{a'}\left[{\bom P}(\{p\};x,\muf^2)+{\bom K}(x)\right]^{a,a'}\rd\sigma^{\mathrm{B}}_{a'b}(xp_a,p_b)
  \\ 
  &\;{}
  +\int_0^1\rd x \sum_{b'}\left[{\bom P}(\{p\};x,\muf^2)+{\bom K}(x)\right]^{b,b'}\rd\sigma^{\mathrm{B}}_{ab'}(p_a,xp_b)\;,
\end{split}
\eeq
where the various $\rd\sigma^{\mathrm{B}}_{ab}$ terms denote 
reduced Born cross sections that result from the clustering of an unresolved parton.
All IR divergences are captured by the ${\bom I}$ operator. For 
massless fermions at $\ord(\alpha)$ it reads
\beq
\label{Iapp}
{\bom I}(\{p\};\ep) = -
\frac{\alpha}{2\pi}\;
\ceps\sum_{I\in \Shatinout} {\cal V}_I(\ep)
\sum_{\genfrac{}{}{0pt}{}{K\in \Shatinout}{K \neq I}} \qtilde{I}{K}\; 
\left( \frac{\mud^2}{2 p_I\cdot p_K} \right)^{\ep}\;,
\eeq
with
\beq
\label{eq:UVpole}
\ceps= \frac{{(4\pi)^\ep}}{\Gamma(1-\ep)} = 
1+\epsilon\Big[\ln(4\pi)-\gamma_{\mathrm{E}}\Big]+\ord(\epsilon^2)\;,
\eeq
\beq
\label{VIapp}
{\cal V}_{I}(\ep) = Q_I^2 \left( \frac{1}{\ep^2} -
\frac{\pi^2}{3} \right) + \gamma_I \;\frac{1}{\ep}
+ \gamma_I + K_I\;,
\eeq
and
\beq
\label{gaapp}
\gamma_f = \frac{3}{2}\; Q_f^2\;,\qquad
\gamma_\Pa = - \frac{2}{3} \sum_{f} \ncqsf\;,\qquad
K_f = Q^2_f \left( \frac{7}{2} - \frac{\pi^2}{6} \right)\;,\qquad
K_\Pa = \frac{5}{3}\;\gamma_\Pa\;,
\eeq
for $f=q,\bar q, \ell^-, \ell^+$.
The fermion sum in~\refeq{gaapp} runs over massless fermions and includes a single term per 
fermion--antifermion pair. 
As discussed in \refapp{app:ypdfrenormalisation},
all massless leptons and quarks should be taken into account,
\ie
\beq
\label{eq:gammagamma}
\gamma_\Pa
= - \frac{2}{3}\, \sum_{f} \ncqsf
= -\frac{6\nmassless{\ell}+8\nmassless{u}+2\nmassless{d}}{9},
\eeq
where $\nmassless{\ell}$, $\nmassless{u}$ and  $\nmassless{d}$
are the number of massless leptons and quarks of type up and down, respectively.

The ${\bom P}$ and ${\bom K}$ 
operators in \refeq{eq:IKP} read
\beqar
\label{eq:KP}
&&{\bom P}^{a,a'}(\{p\};x;\mu_F^2) =
\frac{\alpha}{2\pi} \;P^{aa'}(x) 
\sum_{\genfrac{}{}{0pt}{}{K\in\Sinout}{K \neq a'}} \qtilde{a'}{K}
\;\ln \frac{\mu_F^2}{2 x p_a p_K}\;,\\
&&{\bom K}^{a,a'}(x)
= \frac{\alpha}{2\pi}
\left\{ \frac{}{} {\overline K}^{aa'}(x) 
+ \;\delta^{aa'} \; \sum_{i\in\Shatout} \qtilde{i}{a'}\;
\gammai \left[
\left( \frac{1}{1-x} \right)_+ + \delta(1-x) \right]
\right\} -
\frac{\alpha}{2\pi} \qtilde{a'}{b}\;
{\widetilde K}^{aa'}(x)\;,\nonumber
\eeqar
where $b$ stands for the initial-state partner of $a$.
All relevant ingredients are specified in \refta{tab:CSingredients}.
Note that $\Shatinout=\Shatin\cup\Shatout$  in~\refeq{Iapp} and~\refeq{eq:KP}
should be understood as the incoming and outgoing partons
of the relevant Born sub-process.
The ${\bom P}$ and ${\bom K}$ operators are free from soft and collinear singularities. 
The former depends on the factorisation-scale $\muf$ introduced via the PDF 
counterterm, while the latter depends on the factorisation scheme. The result 
\refeq{eq:KP} corresponds to the case of two initial-state hadrons
in the $\overline{\mathrm{MS}}$ scheme and can be easily translated to 
the  DIS scheme \cite{Catani:1996vz}.

\subsubsection*{Processes with resolved photons}
For hard processes with resolved photons in the final state,
real-emission processes corresponding to 
final-state $\Pa\to f\bar  f$ splittings
and related subtraction terms should 
be omitted at $\ord(\alpha)$. This is achieved by setting
$\epsFSgammai=0$ in \refeq{eq:QEDcorrelE}.
Consequently, in the subtraction term \refeq{eq:alldipoles} we have
\beq
\label{eq:qtildeFSgammaA}
\qtilde{\wtffbar}{K}=0\qquad\mbox{if}\quad \wtffbar\equiv\Pa\in\Sout,
\qquad\mbox{and}\quad
\qtilde{\IJtilde}{\Pa}=0\qquad\mbox{if}\quad\Pa\in\Sout.
\eeq
Thus, external photons contribute to~\refeq{eq:alldipoles}
only through $\IJtilde\to I+J$ final-state splittings of type $f\to f\Pa$, while they can contribute to 
all types of 
$I\to \IJtilde+J$ initial-state splittings,
\ie $\Pa \to f\bar f$, $f\to \Pa f$, and $f\to f \Pa$.
In analogy to~\refeq{eq:qtildeFSgammaA}, for the matrix (identical)  $\qtilde{I}{J}$
that enters the ${\bom I}$, ${\bom K}$ and ${\bom P}$ operators we have 
\beq
\label{eq:qtildeFSgammaB}
\qtilde{\Pa}{K}=\qtilde{I}{\Pa}=0\qquad\mbox{if}\quad\Pa\in\Sout.
\eeq
Thus, resolved final-state photons can be completely excluded from the sums over 
$I\in\Sout$ and $K\in\Sout$ in \refeq{Iapp} and \refeq{eq:KP}, and external photon contributions
to ${\bom I}$, ${\bom K}$ and ${\bom P}$ arise only through
\beq
\label{eq:qtildeISgamma}
\qtilde{\Pa}{b} =  - 1 \qquad \mbox{for $\Pa,b\in\Sin$},
\eeq
\ie from dipoles with initial-state emitters $a=\Pa$ and initial-state 
spectators $b$.

\subsubsection*{Processes with unresolved photons}

For hard processes with unresolved photons in the final state,
the  {\bom I}, {\bom P} and  {\bom K} operators 
in \refeq{Iapp} and \refeq{eq:KP},
contain, as compared to the case of resolved photons,
the following additional contributions from final-state $\Pa\to f\bar f $ splittings,
\beqar
\label{IKPsubt}
\DeltaFSgamma {\bom I}(\{p\};\ep) &=&
{\bom I}(\{p\};\ep)
-\Big[{\bom I}(\{p\};\ep)\Big]_{\epsFSgammai=0}
\nonumber\\
&=& 
\frac{\alpha}{4\pi}\;
\gamma_\Pa\;\ceps
\sum_{\gammai\in\Sout}\epsFSgammai
\left[
2\left(\frac{1}{\ep}+\frac{8}{3}\right)
+\sum_{K\in \Sin } 
\ln\left(\frac{\mud^2}{2 p_i\cdot p_K} \right)\right],
\nonumber\\
\DeltaFSgamma {\bom K}^{a,a'}(x)
&=&
{\bom K}^{a,a'}(x)
-
\Big[{\bom K}^{a,a'}(x)\Big]_{\epsFSgammai=0}
\nonumber\\
&=&
-\frac{\alpha}{4\pi}\;
\gamma_\Pa\;\delta^{aa'}  
\sum_{\gammai\in\Sout}\epsFSgammai
\left[
\left( \frac{1}{1-x} \right)_+ + \delta(1-x) \right],
\nonumber\\
\DeltaFSgamma {\bom P}^{a,a'}(\{p\};x;\mu_F^2)
&=& 
{\bom P}^{a,a'}(\{p\};x;\mu_F^2)
-
\Big[{\bom P}^{a,a'}(\{p\};x;\mu_F^2)\Big]_{\epsFSgammai=0}
\nonumber\\
&=&
0\;.
\eeqar

\subsection[\texorpdfstring{$\aPDF$}{Photon-PDF} renormalisation]
           {\texorpdfstring{$\boldaPDF$}{Photon-PDF} renormalisation}
\label{app:ypdfrenormalisation}

External-photon contributions to the 
${\bom I}$ operator \refeq{Iapp} yield the collinear poles
\beq
\label{eq:Igammasing}
{\bom I}(\{p\};\ep)\Big|_{\Pa,\mathrm{sing}} =
\frac{\alpha}{2\pi}\ceps\left[
\ngammain + \ngammaouteps\right]
\frac{\gamma_\Pa}{\ep}\;,
\eeq
where $\ngammain$ and $\ngammaouteps=\sum_i\epsFSgammai$ are 
the number of incoming photons and outgoing unresolved photons.
When final-state $\Pa\to f\bar f$ splittings
are disabled ($\epsFSgammai=0$), real bremsstrahlung at $\ord(\alpha)$
is free from collinear 
$\gamma_\Pa/\ep$ poles originating 
from final-state photon emitters, and the only pole contributions 
are due to initial-state photons in \refeq{eq:Igammasing}.
Such collinear singularities arise through the $\ord(\alpha)$ 
renormalisation of the photon PDF,
\beq
\label{eq:gammaPDF_ren}
\hat\Pa(x,\muf) = \Pa(x) 
- \frac{\alpha}{2\pi}\left[\frac{\ceps}{\ep}+\lograt{\mud^2}{\muf^2}\right]
\int_x^1\frac{\rd y}{y}\left\{
\sum_{f} P_{\Pa f}\Big(\frac{x}{y}\Big)\Big[f(y) + \bar f(y)\Big]
+
P_{\Pa\Pa}\Big(\frac{x}{y}\Big) \Pa(y)
\right\}.
\eeq
Here, $\mud$ is the scale of the dimensional regularisation. The term
proportional to $P_{\Pa f}$ absorbs collinear singularities arising from
real-emission processes where the (off-shell) initial-state photon
originates from $f\to \Pa f$ splittings. 
The remaining term is due to the \aa splitting function 
(see~\refta{tab:CSingredients}),
\beq
\label{eq:Pgammagamma}
P_{\Pa\Pa}\left(\frac{x}{y}\right) = \gamma_\Pa\;
\delta\left(1-\frac{x}{y}\right)\,,
\eeq
which consists only of virtual fermion-loop contributions
associated with the photon wave function renormalisation. 
It can be understood as a negative correction to the \aPDF that 
compensates real $\Pa\to f\bar f$ splittings.
The corresponding splitting functions
are related via the momentum sum rule
\beq
\label{eq:sumrule}
\int_0^1 \rd z\, z  \left\{P_{\Pa\Pa}(z)+\sum_{f}\Big[P_{f\Pa}(z)+P_{\bar f\Pa}(z)
\Big]\right\}=1.
\eeq 
Including also the logarithmic dependence on $\muf$, 
which appears in the ${\bom P}$ operator in \refeq{eq:KP},
the effect of the \aPDF renormalisation 
can be summarised through an overall renormalisation factor,
\beq
\label{eq:gammaPDF_renc}
\delta Z_{\Pa,\PDF}=
\frac{\alpha}{2\pi}\,\gamma_\Pa \left[\frac{\ceps}{\ep}+\lograt{\mud^2}{\muf^2}\right]\,,
\eeq
for each initial-state photon.

\subsection*{Contributions form $\Pa\to \ell^+\ell^-$ splittings}

Photon splittings into $q\bar q$ and
$\ell^+\ell^-$ should be included on the same footing
at $\ord(\alpha)$.
Thus, as
pointed out above, the photon anomalous dimension of \refeq{eq:gammagamma}
should include both quark and lepton contributions. 
This should be clear, since $\gamma_\Pa$ represents 
contributions of virtual type, and different kinds of 
fermion loops are indistinguishable.  Moreover, omitting leptonic
contributions to $\gamma_\Pa$ would jeopardise the cancellations
of fermion-mass singularities between \refeq{eq:gammaPDF_renc} 
and the virtual corrections to the hard cross section
(see~\refapp{app:alpharenormalisation}).

Since $\gamma_\Pa$ in \refeq{eq:gammaPDF_renc} arises from the 
renormalisation of the \aPDF of \refeq{eq:gammaPDF_ren},
virtual $\Pa\to \ell^+\ell^-$ splittings should be taken into account
also in the evolution of $\Pa(x,\muf)$. In addition, for consistency with
the sum rule \refeq{eq:sumrule}, also real $\Pa\to \ell^+\ell^-$ splittings and thus
lepton distributions should be included in the PDF evolution.
While this is desirable from the theoretical viewpoint,
the effect of $\Pa\to \ell^+\ell^-$ splittings 
hardly exceeds 1\% in the photon PDF~\cite{Schmidt:2015zda}
and is completely negligible in the quark PDFs. 
Moreover, lepton-induced processes are extremely suppressed 
at the LHC~\cite{Bertone:2015lqa}.
Thus, excluding 
$\Pa\to \ell^+\ell^-$ splittings from the PDF evolution, as in the 
\CTQED set used in the nominal predictions in this paper, 
is well justified.

\subsection{Definition and renormalisation of \texorpdfstring{$\alpha$}{alpha} in processes with external photons}
\label{app:alpharenormalisation}

The collinear singularities in \refeq{eq:gammaPDF_renc}
have to be combined with corresponding singularities 
that arise from the 1-loop counterterms associated with the
renormalisation of the photon wave function ($\delta Z_{AA}$)
and of the electromagnetic coupling $\alpha$.
Such counterterms yield a universal correction factor
\beq
\label{eq:Zgammavirt}
\Zgammavirt= \frac{\delta \alpha}{\alpha} + \delta Z_{AA}
\eeq
for each external (incoming or outgoing) photon in the hard scattering process. In the following,
in order to articulate the interplay between the 
renormalisation of $\alpha$ and the cancellation of collinear singularities,
we will focus on the contributions from light fermions
with $0\le m_f < \MZ$, which can be either treated in 
dimensional regularisation or using finite fermion masses.
While all massless and massive fermions are assumed to contribute to the
virtual corrections and to the ultraviolet renormalisation,
only massless fermions are assumed to be included in the Catani-Seymour subtraction
and in the \aPDF renormalisation.

The photon wave function counterterm reads,
\beq
\label{eq:dZAA}
\delta Z_{AA} = -\Pi^{\Pa\Pa}(0)=-\Pi_\light^{\Pa\Pa}(0)-\Pi_\heavy^{\Pa\Pa}(0)\;, 
\eeq
where \textit{light} and \textit{heavy} refer, respectively, to light-fermion and top-quark plus 
bosonic contributions.
The UV and collinear singularities in \refeq{eq:dZAA}
can be separated from each other 
by rewriting 
\beq
\label{eq:PiAA}
\Pi_\light^{\Pa\Pa}(0) = \Pi_\light^{\Pa\Pa}(\MZ^2) +\Delta \alpha (\MZ^2)\;.
\eeq
Here\footnote{For simplicity, in the following we omit mass-suppressed terms of
$\ord(m_f^2/\MZ^2)$ 
from light 
fermions with  $0<m_f<\MZ$. However such terms are typically included
in realistic calculations, as it is the case for the calculation presented in this paper.}
\beq
\label{eq:PiMZ}
\Pi_\light^{\Pa\Pa}(\MZ^2)= -\frac{\alpha}{2\pi}\;\gamma_\Pa
\left[\frac{\ceps}{\ep}+\lograt{\mud^2}{\MZ^2}+\frac{5}{3}\right]
\eeq
represents the UV divergent piece, while all collinear singularities are
contained in
\beq
\label{eq:Delta_alpha}
\begin{split}
\Delta \alpha (\MZ^2)
\,=&\;
\Pi_\light^{\Pa\Pa}(0)-\Pi_\light^{\Pa\Pa}(\MZ^2)
\\
\,=&\;
\frac{\alpha}{2\pi}\;\gamma_\Pa 
\left[\frac{\ceps}{\ep}+\lograt{\mud^2}{\MZ^2}+\frac{5}{3}\right]
-\frac{\alpha}{3\pi}\sum_{f\in\mferm} \ncqsf 
\left[\ln\left(\frac{m_f^2}{\MZ^2}\right)+\frac{5}{3}\right]\;,
\end{split}
\eeq
where the anomalous dimension $\gamma_\Pa$, defined in \refeq{eq:gammagamma}, accounts for 
all massless fermion loops, while the sum over $f\in \mferm$ includes all light
fermions with $0<m_f<\MZ$. As is well known,  
$\Delta \alpha (\MZ^2)$ is associated with the running of $\alpha$ from $Q^2=0$ to $Q^2=\MZ^2$.
In order to arrive at a finite expression 
for $\Delta \alpha (\MZ^2)$, 
all fermions could be treated as massive, in which case $\gamma_\Pa=0$. Alternatively,
hadronic contributions to $\Delta \alpha (\MZ^2)$ can be obtained via dispersion relations.
However, we advocate the approach of choosing an appropriate definition 
of  $\alpha$, 
such as to cancel all singularities associated with $\Delta\alpha(\MZ^2)$ in the final result.
As detailed in the following, such a definition depends 
on the presence of resolved external photons in the processes at hand.

\subsection*{Resolved final-state photons}
In processes with resolved on-shell photons that do not split into 
$f\bar f$ pairs the collinear singularity 
from $\delta Z_{AA}$ remains uncancelled unless the electromagnetic coupling is 
renormalised in the on-shell scheme.
Thus, $\alpha$ should be defined as the photon coupling in the
on-shell limit $q^2\to 0$. The resulting counterterm 
is related to the photon wave-function renormalisation via~\cite{Denner:1991kt}
\beq \label{eq:dalphaOS}
\frac{\delta \alpha(0)}{\alpha(0)}= - \delta Z_{AA} -
\frac{\sinw}{\cosw}\,\delta Z_{ZA}
= \Pi^{\Pa\Pa}(0)-2\frac{\sinw}{\cosw}\frac{\Sigma_\rT^{AZ}(0)}{\MZ^2}\;,
\eeq
where $\thetaw$ is the weak mixing angle. Light-fermion contributions to 
\refeq{eq:dalphaOS} read
\beq \label{eq:dalphaOSlight}
\frac{\delta \alpha(0)}{\alpha(0)}\Big|_\light= 
 \Pi_\light^{\Pa\Pa}(0)\;,
\eeq
since the $\Sigma_\rT^{AZ}(0)$ term
receives only bosonic contributions.
This yields, for each on-shell photon in the final state,
\beq
\label{eq:ZgammavirtOS}
\Zgammavirt\Big|_{\OS,\light}
= \Bigg[\frac{\delta \alpha(0)}{\alpha(0)}+\delta Z_{AA}\Bigg]_{\light}=0\;,
\eeq
while using the $\alpha(\MZ)$ scheme, cf.\ \refeqs{eq:alphaMZ}{eq:dalphaMZ},
would lead to
\beq
\Zgammavirt\big|_{\MZ,\light}
= \Bigg[\frac{\delta \alpha(\MZ^2)}{\alpha(\MZ^2)}+\delta Z_{AA}\Bigg]_{\light}=-\Delta\alpha (\MZ^2)\;.
\eeq
Thus, as is well known, in order to 
avoid fermion-mass singularities from $\Delta\alpha (\MZ^2)$
in the hard cross section, 
the couplings of on-shell (resolved) final-state photons
should be parametrised in terms of $\alpha(0)$.

\subsection*{Initial-state photons and unresolved final-state photons}

In the case of initial-state photons, 
virtual contributions to the \aPDF renormalisation~\refeq{eq:gammaPDF_ren} 
are designed such as to absorb the collinear singularity 
of $\delta Z_{AA}$. Thus, by construction, the combination 
\beq
\begin{split}
\delta Z_{AA}\Big|_{\light}+\delta Z_{\Pa,\PDF} 
\,=&\; 
-\Pi_\light^{\Pa\Pa}(\MZ^2)
-\frac{\alpha}{2\pi}\;\gamma_\Pa
 \left[\lograt{\muf^2}{\MZ^2}+\frac{5}{3}\right] 
\\
\,=&\; 
\frac{\alpha}{3\pi}\sum_{f\in\mferm} \ncqsf \left[\ln\left(\frac{m_f^2}{\MZ^2}\right)+\frac{5}{3}\right]\;,
\end{split}
\eeq
is free from $1/\epsilon$ mass singularities, and there
is no need to adopt the $\alpha(0)$ scheme. 
In fact, expressing the coupling of initial-state photons 
in terms of 
\beq
\label{eq:alphaMZ}
\alpha(\MZ^2)= \frac{\alpha(0)}{1-\Delta \alpha(\MZ^2)}\;,
\eeq
with counterterm
\beq
\label{eq:dalphaMZ}
\frac{\delta \alpha(\MZ^2)}{\alpha(\MZ^2)}= 
\frac{\delta \alpha(0)}{\alpha(0)}
-\Delta \alpha(\MZ^2)
=  \Pi_\light^{\Pa\Pa}(\MZ^2)
+\Pi_\heavy^{\Pa\Pa}(0)-2\frac{\sinw}{\cosw}\frac{\Sigma_\rT^{AZ}(0)}{\MZ^2}\;,
\eeq
results in the overall initial-state photon factor
\beqar
\label{eq:ZgammavirtPDFa}
\Zgammavirt\big|_{\MZ,\light} +\delta Z_{\Pa,\PDF} 
&=&
\Bigg[\frac{\delta \alpha(\MZ^2)}{\alpha(\MZ^2)} + \delta Z_{AA}\Bigg]_{\light}+\delta Z_{\Pa,\PDF}
\\
&=&
-\frac{\alpha}{2\pi}\,\gamma_\Pa\left[\lograt{\muf^2}{\MZ^2}+\frac{5}{3}\right]
+\frac{\alpha}{3\pi}\sum_{f\in\mferm} \ncqsf \left[\ln\left(\frac{m_f^2}{\MZ^2}\right)+\frac{5}{3}\right],
\nonumber
\eeqar
which is manifestly free from $1/\epsilon$ fermion-mass singularities, while,
as usual, those degrees of freedom that do not contribute as active fermions
in the PDF evolution give rise to logarithms of $m_f$ in the 
hard-scattering cross section.
Vice versa, using the $\alpha(0)$ scheme for initial-state photons
would lead to the divergent result
\beq
\begin{split}
\Zgammavirt\big|_{\OS,\light} +\delta Z_{\Pa,\PDF}
\,=&\;
\Bigg[\frac{\delta \alpha(0)}{\alpha(0)} + \delta Z_{AA}\Bigg]_\light + \delta Z_{\Pa,\PDF}
\\
\,=&\;{}
\Zgammavirt\big|_{\MZ,\light} +\delta Z_{\Pa,\PDF} 
+\Delta\alpha(\MZ^2)
\\
\,=&\;
\frac{\alpha}{2\pi}\;\gamma_\Pa \left[\frac{\ceps}{\ep}+\lograt{\mud^2}{\muf^2}\right].
\end{split}
\eeq
A fully analogous cancellation mechanism applies also to 
unresolved final-state photons, where 
the term proportional to $\ngammaouteps$ in \refeq{eq:Igammasing}, 
which originates from final-state $\Pa\to f\bar f$ splittings, 
plays a similar role as the \aPDF counterterm for initial-state photons.
 
Thus, in order to  avoid fermion-mass singularities in the hard cross section, 
the couplings of initial-state photons and unresolved final-state photons
should be parametrised in terms of $\alpha(\MZ^2)$ or any other 
scheme where $\alpha$ is defined at a hard scale, such as the 
$G_\mu$-scheme or a running $\alpha(\mur^2)$ with $\mur^2\sim \hat s$.
For the case of initial-state photons, this was first pointed out in~\cite{Harland-Lang:2016lhw} 
based on arguments related to the PDF evolution.

\section{Flavour-number scheme conversion}
\label{se:conversion}

As discussed in \refse{se:setup:PDFs}, in order to avoid single-top contributions,
we compute parton-level cross sections using $m_b>0$ and omitting external $b$-quarks, 
both in the initial and in the final state.
This approach corresponds to the four-flavour scheme,
and can be consistently used in combination with five-flavour PDFs by applying a simple
scheme conversion~\cite{Cacciari:1998it}, which 
amounts to the following  substitution at the level of squared Born matrix elements,
\beq
\label{eq:qcdschemetransf}
{\mathcal B}_{ij}  \to  \left\{1+\frac{\alphaS}{3\pi}\,\TR\left[
n^{(\alphaS)}_{ij}\;\theta(\mur^2-\Mb^2)\log\left(\frac{\Mb^2}{\mur^2}\right)
- n^{(g)}_{ij}\;\theta(\muf^2-\Mb^2)\log\left(\frac{\Mb^2}{\muf^2}\right)
\right]\right\}\mathcal{B}_{ij}\;.
\eeq
Here $ij\in\{q\bar q, g q, g\bar q, gg\}$ are the initial-state \QCD partons, while
$n^{(\alphaS)}_{ij}$ and $n^{(g)}_{ij}$ are, respectively, the power of
$\alphaS$ and the number of initial-state gluons in the channel at hand.
For the process of interest in this paper, $pp\to 2\ell2\nu$, 
initial-state gluons do not contribute at Born level, and in 
the $q\bar q$ channel we have $n^{(\alphaS)}_{q\bar q}=n^{(g)}_{q\bar q}=0$.
Thus, as far as \QCD partons are concerned, the scheme conversion of 
\refeq{eq:qcdschemetransf} is trivial. However, the $\aa\to 2\ell 2\nu$ channel 
requires a non-zero scheme transformation,   
\beq
\label{eq:aaschemetransf}
{\mathcal B}_{\aa}  \to  \left[1-\frac{2\alpha}{3\pi}\,\NC Q^2_b\; 
\theta(\muf^2-\Mb^2)\log\left(\frac{\Mb^2}{\muf^2}\right)
\right]\mathcal{B}_{\aa}\;,
\eeq
which involves a single term, related to the scheme dependence of the photon PDF.
Note that there is no scheme-conversion term associated with the electromagnetic 
coupling required since, usually,  $\alpha$ is not defined in the $\MSbar$ scheme.

\section{Electroweak corrections by parton luminosity}
\label{app:splitcorr}

\begin{figure*}[t!]
  \centering
  \includegraphics[width=\relplotwidth\textwidth]{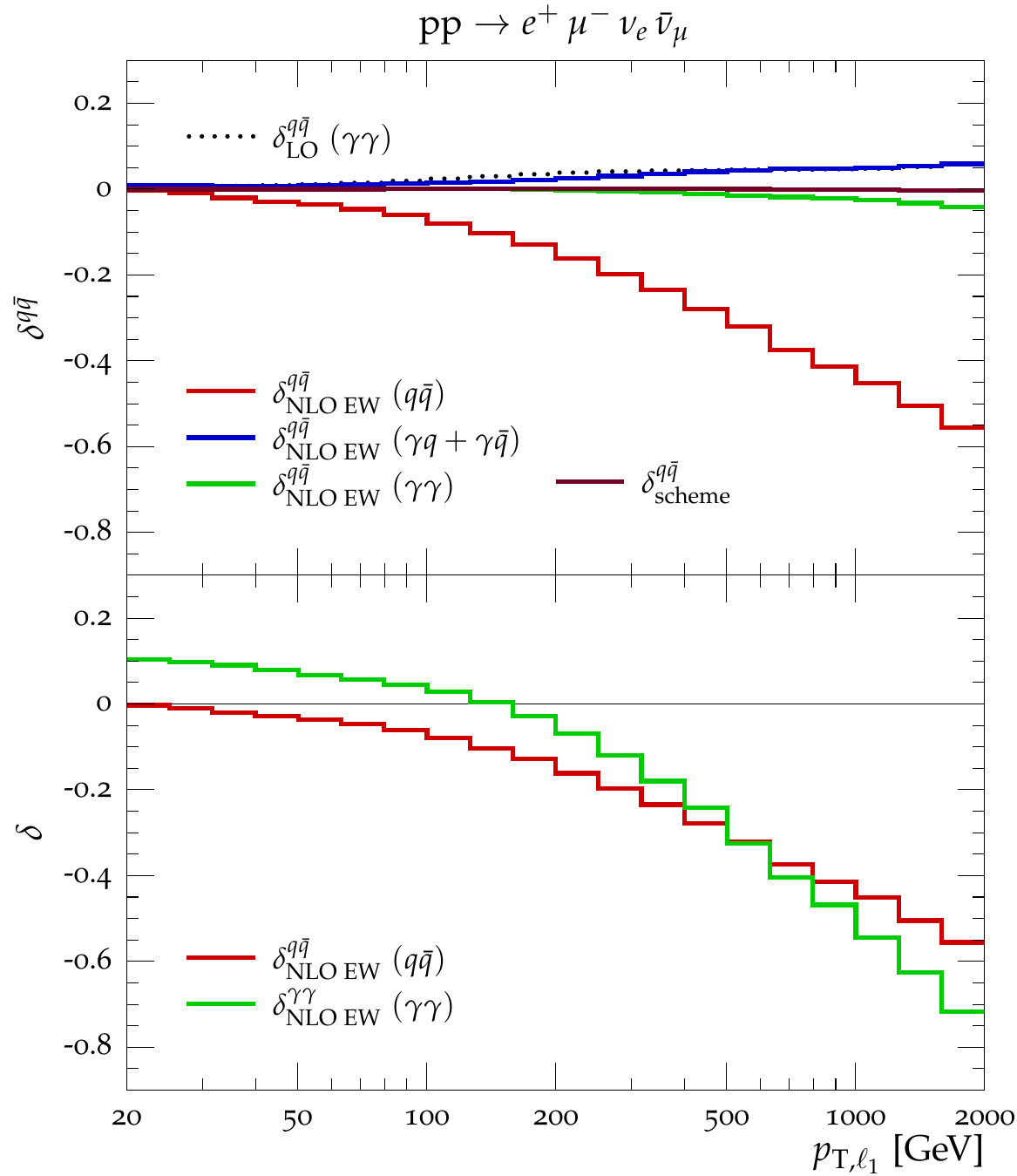}
  \qquad 
  \includegraphics[width=\relplotwidth\textwidth]{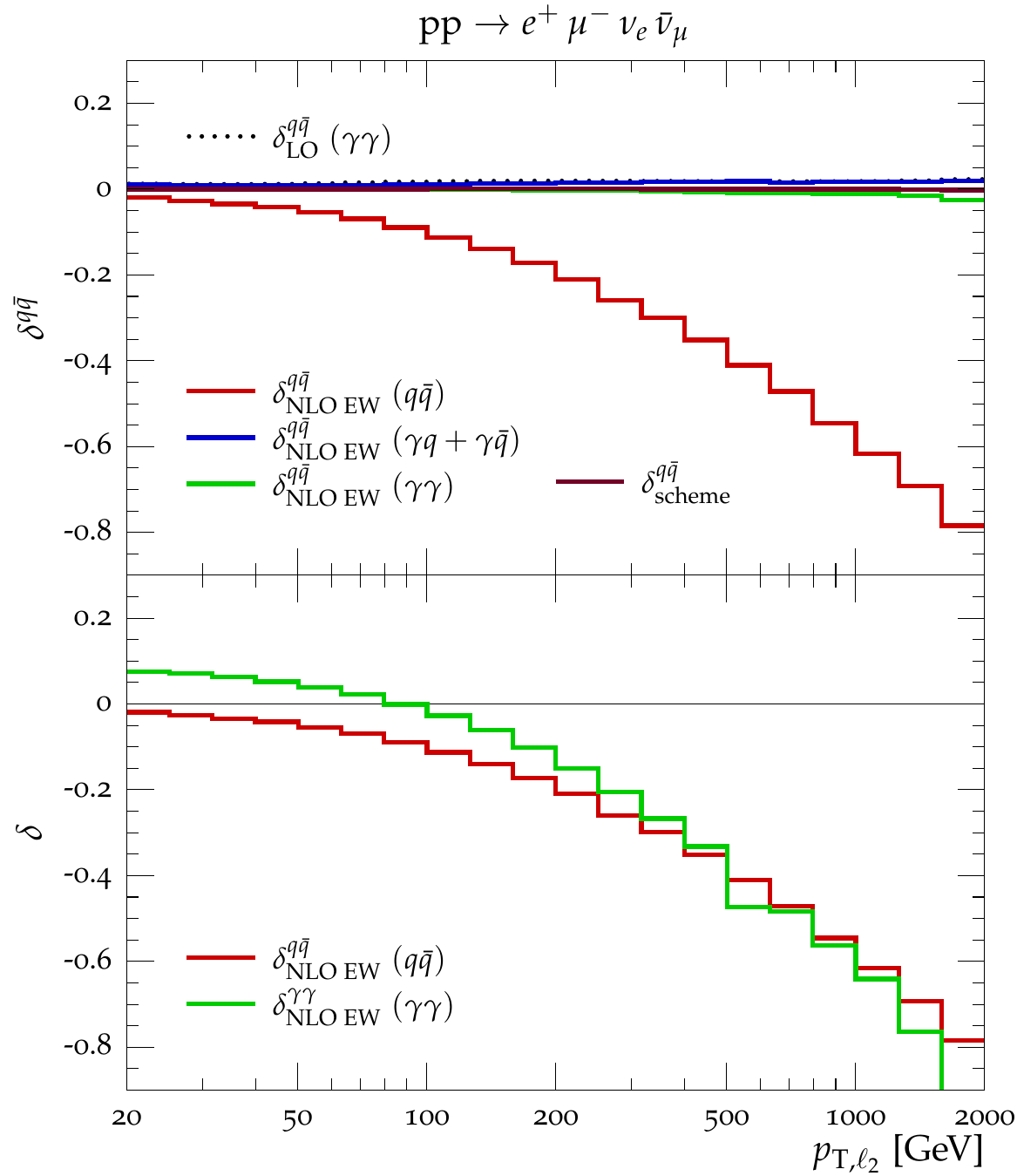}
  \caption{
    Relative corrections in the transverse momentum of the leading and 
    subleading lepton, $p_{\rT,\ell_1}$ and $p_{\rT,\ell_2}$, 
    for $pp\to\llnndf$ at 13\,\TeV. In the upper panel we show 
    the relative corrections to the \LO $\Pq\Pqbar$ channel, induced by 
    the \LO $\aa$ channel ($\delta_{\LO}^{\Pq\Pqbar}$ $(\aa)$),
    and the \NLO \EW $\Pq\Pqbar$-, $\Pq\Pa$/$\Pqbar\Pa$- and $\aa$-induced 
    processes ($\delta_{\NLO\;\EW}^{\Pq\Pqbar}$ $(\Pq\Pqbar)$, 
    $\delta_{\NLO\;\EW}^{\Pq\Pqbar}$ $(\Pq\Pa+\Pqbar\Pa)$, 
    $\delta_{\NLO\;\EW}^{\Pq\Pqbar}$ $(\aa)$, respectively). 
    We further also show the relative size of the scheme conversion 
    term of \refeq{eq:multcombschemedepB} with respect to the \LO 
    $\Pq\Pqbar$ channel ($\delta_\text{scheme}^{\Pq\Pqbar}$).
    The lower panel shows the $\Pq\Pqbar$- and $\aa$-induced \NLO \EW 
    corrections relative to the \LO cross section in the $\Pq\Pqbar$ 
    and $\aa$ channel, respectively. At large transverse momenta, this 
    corresponds to the channels' respective electroweak Sudakov 
    corrections.
    \label{fig:emvv_pTl1_pTl2_delta}
    \vspace*{5mm}
  }
  \centering
  \includegraphics[width=\relplotwidth\textwidth]{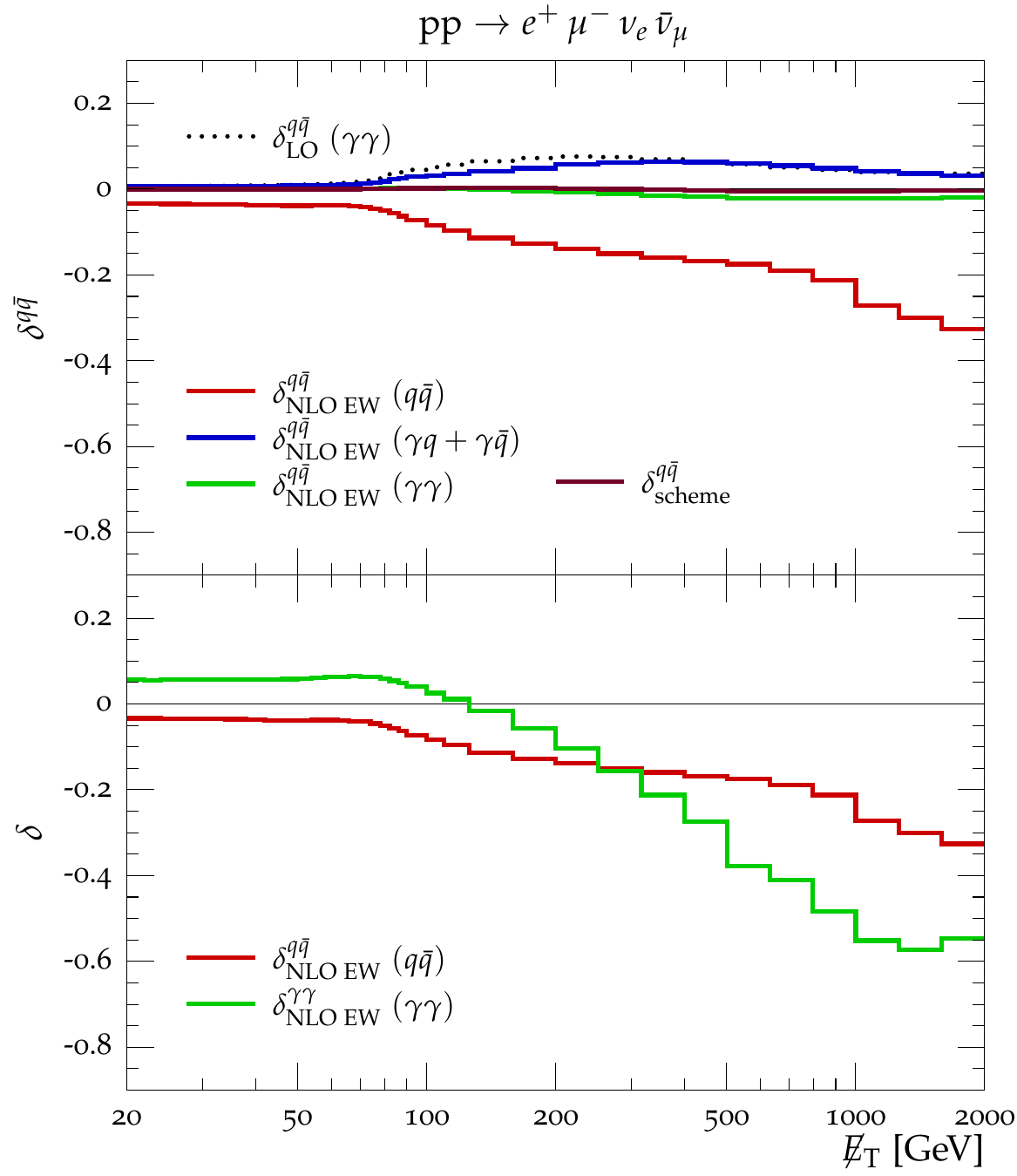}
  \qquad 
  \includegraphics[width=\relplotwidth\textwidth]{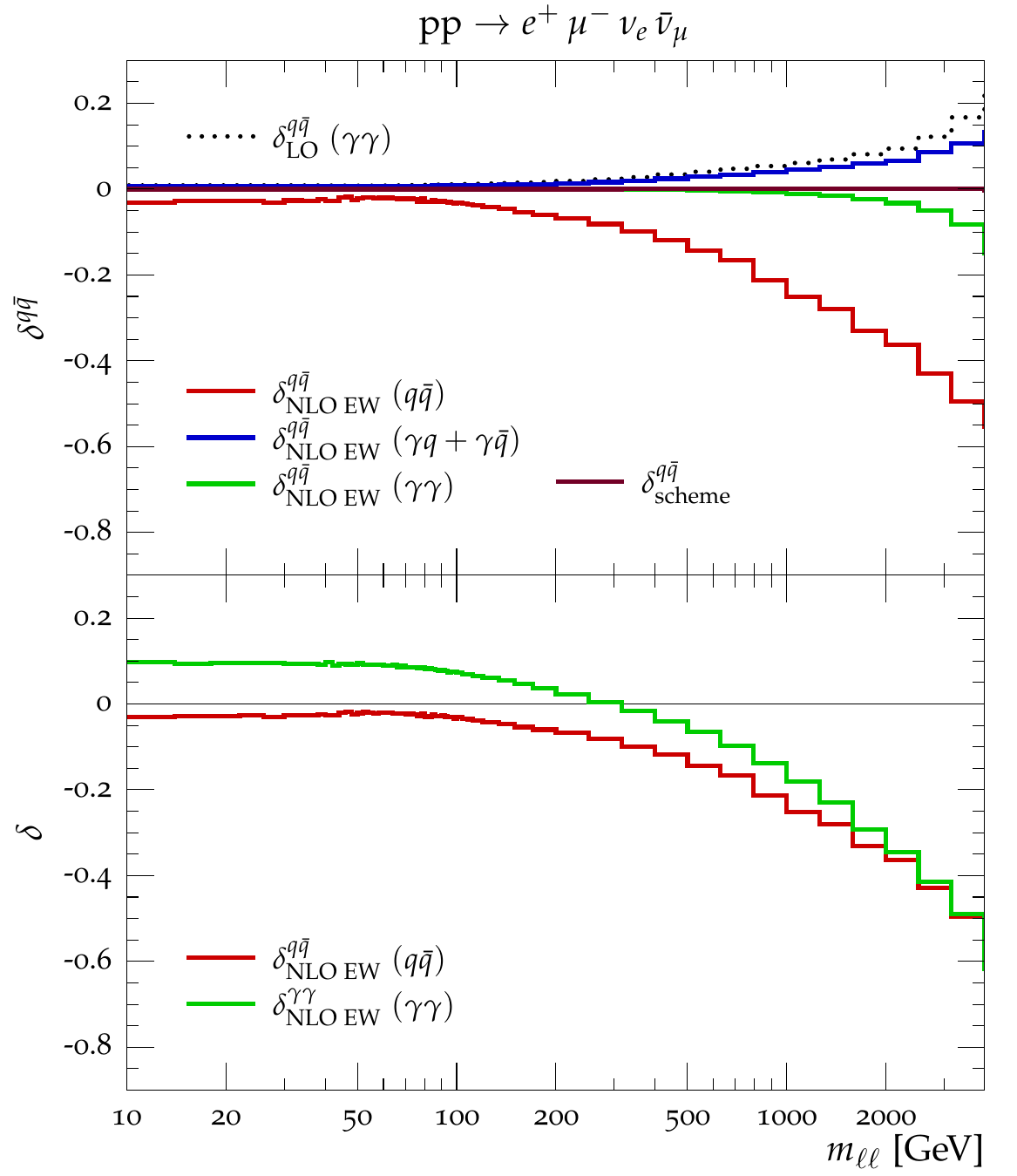}
  \caption{
    Relative corrections in the missing transverse momentum, \missingET, 
    and the invariant mass of the $\Pe^+\Pmu^-$  pair, $m_{\ell\ell}$, 
    for $pp\to\llnndf$ at 13\,\TeV. Details as in 
    \reffi{fig:emvv_pTl1_pTl2_delta}.
    \label{fig:emvv_MET_mll_delta}
  }
\end{figure*}

\begin{figure*}[t!]
  \centering
  \includegraphics[width=\relplotwidth\textwidth]{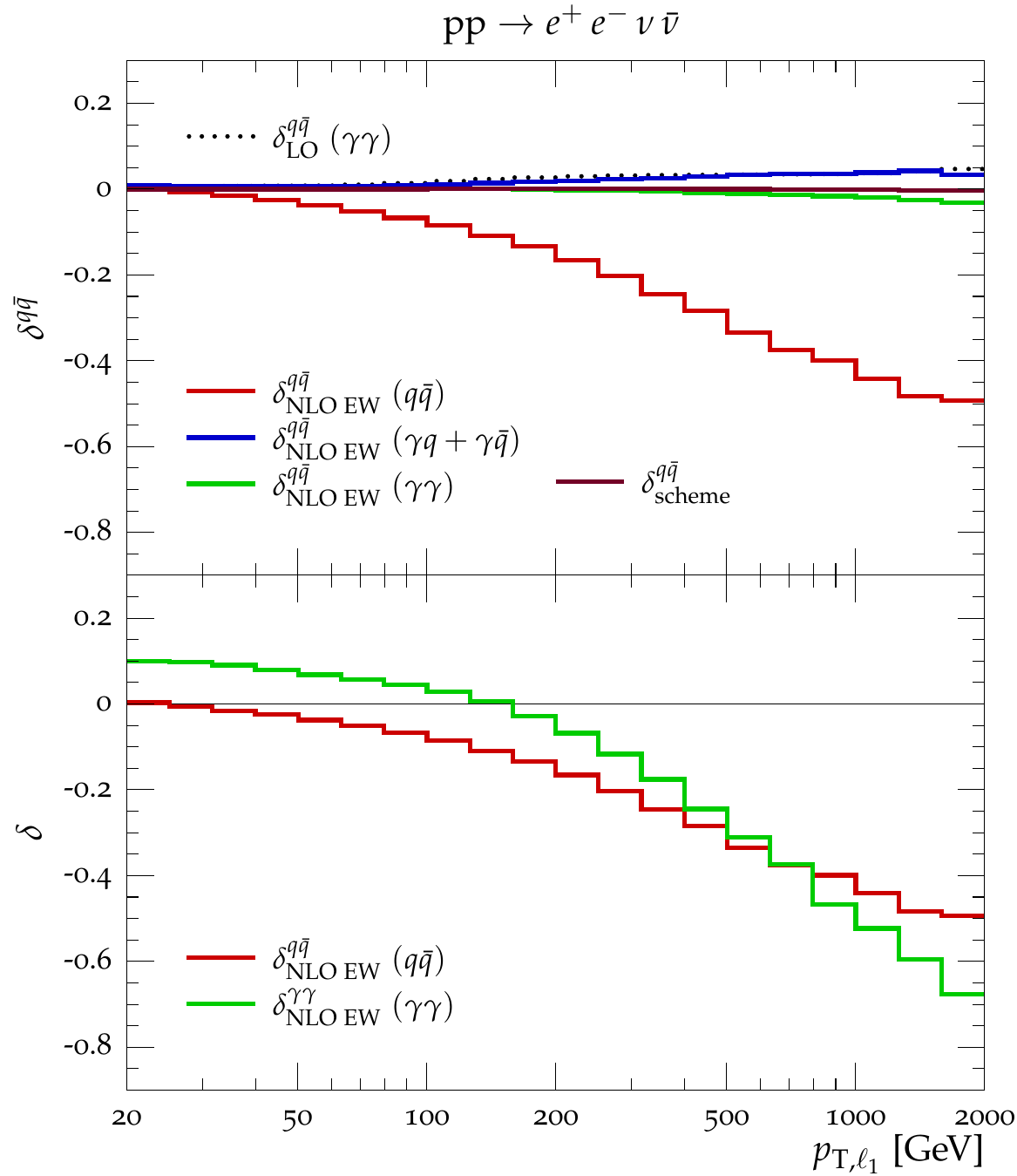}
  \qquad 
  \includegraphics[width=\relplotwidth\textwidth]{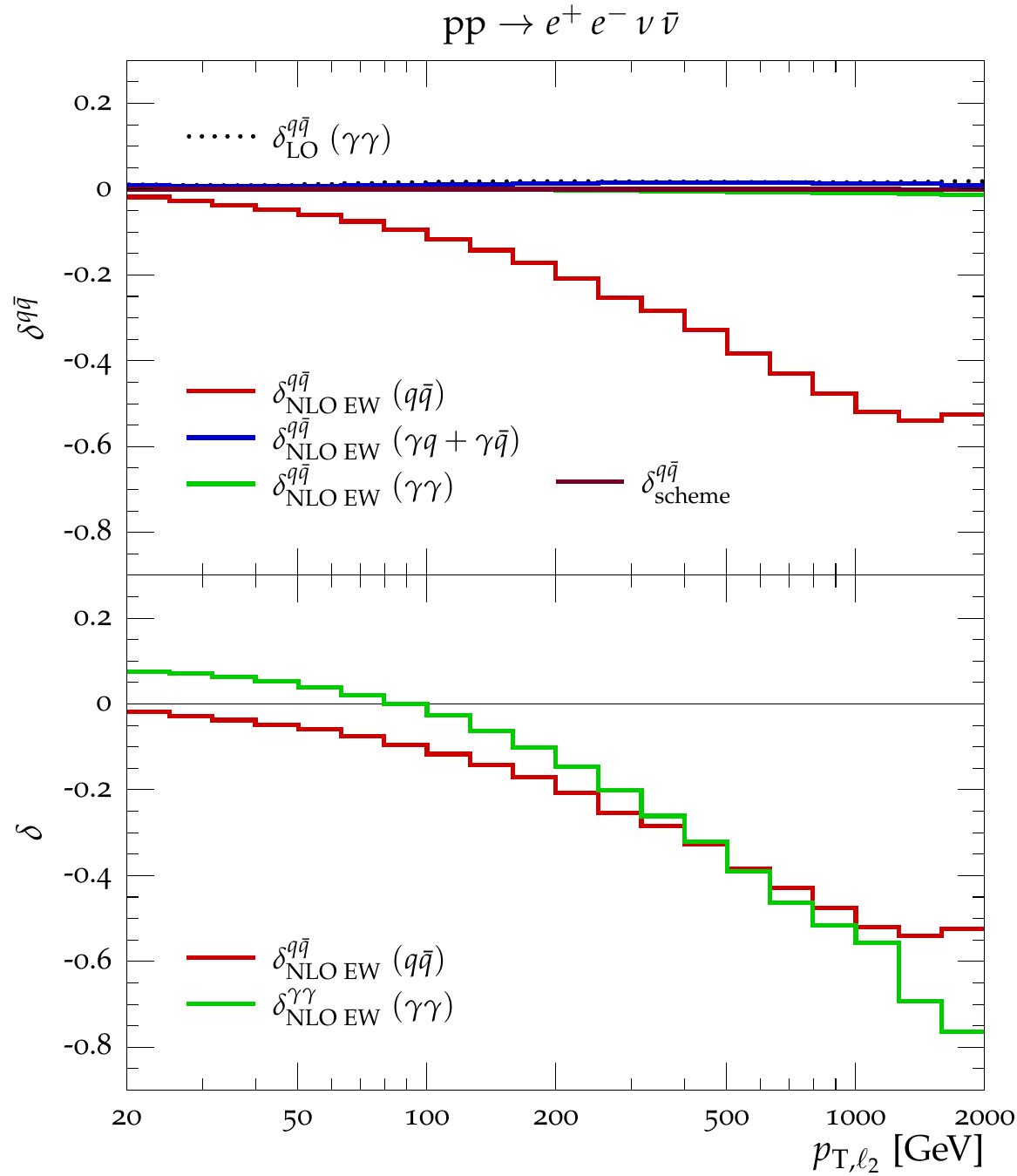}
  \caption{
    Relative corrections in the transverse momentum of the leading and 
    subleading lepton, $p_{\rT,\ell_1}$ and $p_{\rT,\ell_2}$, 
    for $pp\to\llnnsf$ at 13\,\TeV. Details as in 
    \reffi{fig:emvv_pTl1_pTl2_delta}.
    \label{fig:eevv_pTl1_pTl2_delta}
    \vspace*{5mm}
  }
  \centering
  \includegraphics[width=\relplotwidth\textwidth]{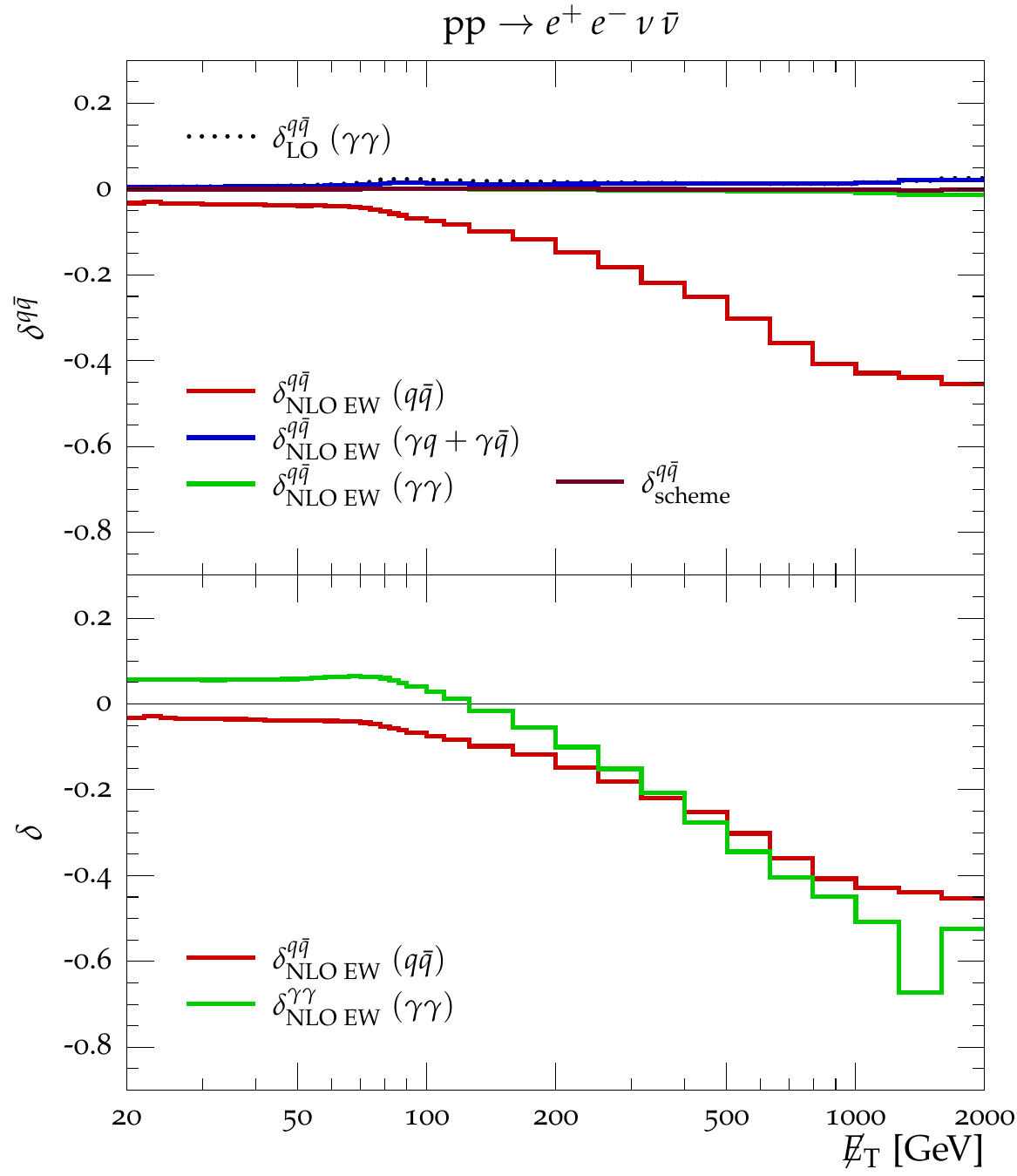}
  \qquad 
  \includegraphics[width=\relplotwidth\textwidth]{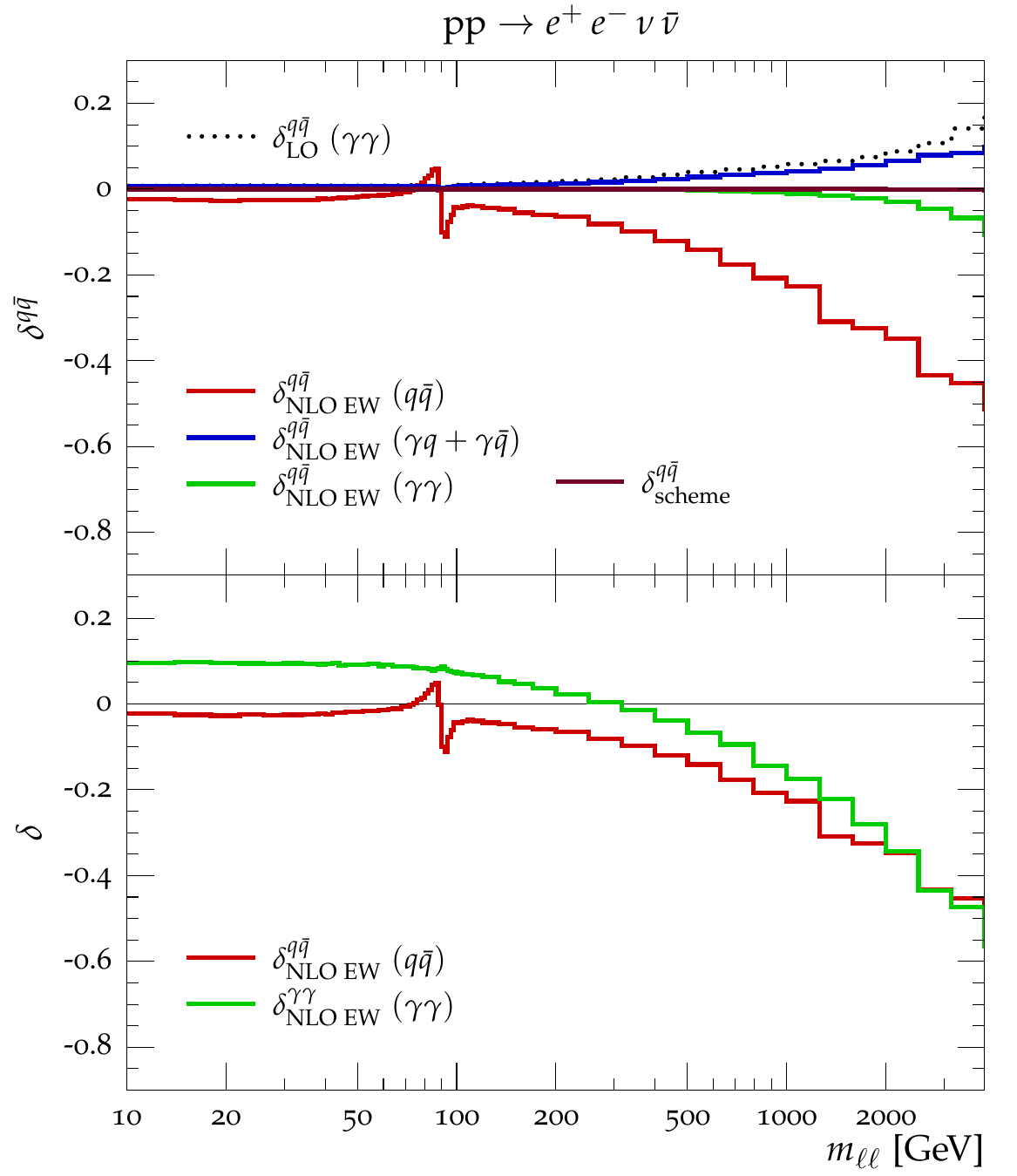}
  \caption{
    Relative corrections in the missing transverse momentum, \missingET, 
    and the invariant mass of the $\Pe^+\Pmu^-$  pair, $m_{\ell\ell}$, 
    for $pp\to\llnnsf$ at 13\,\TeV. Details as in 
    \reffi{fig:emvv_pTl1_pTl2_delta}.
    \label{fig:eevv_MET_mll_delta}
  }
\end{figure*}

\reffis{fig:emvv_pTl1_pTl2_delta}{fig:emvv_MET_mll_delta} and 
\ref{fig:eevv_pTl1_pTl2_delta}--\ref{fig:eevv_MET_mll_delta} detail 
the relative electroweak correction induced by each parton luminosity 
for the DF and SF $pp\to 2\ell 2\nu$ production processes, respectively. 
Here, only the \CTQED PDF is used for the photon density. 
As described in \refse{se:anatomy}, the $pp\to 2\ell2\nu$ 
production process at \LO receives contributions from the $\Pq\Pqbar$ 
and $\aa$ channels, while at \NLO \EW also the $\Pq\Pa$ and 
$\Pqbar\Pa$ channels arise. 
It needs to be noted that the relative contributions from different 
parton luminosities are factorisation scale dependent. 
In each figure, the upper panel shows the relative correction of the 
$\aa$-induced production process at \LO in addition to the relative 
size of the \NLO \EW corrections in the $\Pq\Pqbar$-, 
$\Pa\Pq$/$\Pa\Pqbar$- and $\aa$-induced channels 
relative to the \LO $\Pq\Pqbar$-induced process. 
In addition it quantifies the size of the scheme conversion 
term~\refeq{eq:multcombschemedepB}
that originates when relating our definition of the multiplicative 
combination of \NLO \QCD and \EW correction to another definition 
based upon individual corrections to both \LO production channels. 
This scheme dependence is of relative $\ord(\alphaS\alpha)$ and contributes generally 
0.5\,\textperthousand, rising to 5\,\textperthousand{} in extreme regions.
The lower panel compares the size of both electroweak Sudakov-like 
corrections to their respective Born process. 

Naturally, the \LO $\aa$-induced correction is small but positive 
throughout.
The \NLO \EW corrections are dominated by the $\Pq\Pqbar$-channel 
exhibiting the usual Sudakov suppression at large transverse momenta.
The distribution in the missing transverse momentum in the 
DF case provides an exception, the origins of which and its 
specific characteristics have been discussed in detail in 
\refse{se:dfresults}. 

The $\aa$-induced \NLO \EW corrections are detailed both in the upper 
panel, showing their relative size in comparison to the \LO 
$\Pq\Pqbar$ channel, and the lower panel, showing their relative 
size in comparison to the \LO $\aa$ channel. 
While they contribute only small amounts to the total \NLO \EW 
correction, the comparison against the \LO $\aa$ channel 
clearly exhibits their Sudakov-like behaviour as transverse momenta 
are increasing. 
Despite similar shapes, the size of this Sudakov-type correction 
is found to be slightly 
larger in the $\aa$ channel than in the $\Pq\Pqbar$ channel. 

The \NLO \EW $\Pa\Pq$- and $\Pa\Pqbar$-induced corrections that 
are associated with both \LO processes show a different behaviour. 
At this order, no one-loop diagrams contribute and, thus, the 
Sudakov-type behaviour is absent. Instead, the corrections 
are positive and of a similar magnitude as the \LO $\aa$ channel. 
Please note, since these two channels exhibit a final state quark 
or anti-quark, their precise magnitudes are strongly dependent 
on the chosen form of the jet veto. 
Choosing a tighter veto, e.g.\ by applying a strict veto against 
any jet activity above 30\,GeV, decreases their contribution, 
while loosening it, e.g.\ by not vetoing jets altogether, increases 
it.

In conclusion, while the \NLO \EW correction in the $\aa$ 
channel is dominated by \EW Sudakov logarithms whose magnitude 
in the TeV region balances the additional power of $\alpha$ to 
arrive at a result of the same magnitude as the \LO $\aa$-induced 
contribution, the \NLO \EW correction in the $\Pa\Pq$ and 
$\Pa\Pqbar$ channels uses the replacement of one of its PDF by 
a quark or anti-quark PDF (relative to the \LO $\aa$-induced process) 
to cancel the additional power in $\alpha$. 
Thus, when summing all contributions that depend on the 
photon density in the proton, \LO $\aa$ and \NLO \EW $\Pa\Pq$-, 
$\Pa\Pqbar$- and $\aa$ channels, there are sizeable cancellations 
between the different contributions to the cross sections at 
\NLO \EW accuracy.

\section{Cross section tables}
\label{app:xsecs}

This last section compiles, for reference, a list of 
cross sections and corrections with different phase-space cuts applied. \reftas{tab:xs:df}{tab:xs:sf:sum} 
detail the cross sections for DF and SF $2\ell 2\nu$ production, 
while \reftas{tab:xs:sf:wwzz}{tab:xs:sf:zz} show the contribution 
from the \SFWWZZ{} and one of the two \SFZZ{} channels making up the 
SF signature. We list cross sections and corrections for the inclusive 
fiducial phase space as well as three more exclusive phase-space 
regions focussing on various high-$\pT$ scenarios. In each case, 
the \LO cross section serves as a reference to define the 
\NLO \QCD, \NLO \EW, \NLO \QCDpEW and \NLO \QCDtEW corrections, 
computed in our default setup using the \CTQED PDFs, cf.\ \refse{se:setup}. 
The latter we consider our best prediction for each particle selection.

\begin{table}[t!]
  \begin{center}
        \begin{tabular}{l||C||C|C|C}
      $\boldsymbol{pp\to e^+\mu^-\nu_e\bar{\nu}_\mu}$
        & inclusive& $p_{\rT,\ell_1}>500\,\GeV$& $\missingET>500\,\GeV$& $m_{\ell\ell}>1\,\TeV$ \\\hline\hline
      $\sigma^\text{\LO}$ [fb] \vP
        & $\phantom{.-}\xs{299}{8\%}{6\%}$
        & $\phantom{-}\hspace*{-\unitcharwidth}\xs{0.079}{6\%}{7\%}$
        & $\phantom{-}\hspace*{-\unitcharwidth}\xs{0.017}{7\%}{8\%}$
        & $\phantom{-}\hspace*{-\unitcharwidth}\xs{0.149}{6\%}{7\%}$
         \\\hline\hline
      $\sigma^\text{\NLO}_\text{\QCD} / \sigma^\text{\LO}$ \vP
        & $\phantom{-}\hspace*{\unitcharwidth}\xs{1.04}{5\%\hspace*{\unitcharwidth}}{7\%\hspace*{\unitcharwidth}}$
        & $\phantom{-}\hspace*{\unitcharwidth}\xs{1.34}{9\%\hspace*{\unitcharwidth}}{11\%}$
        & $\phantom{-}\hspace*{\unitcharwidth}\xs{1.41}{10\%}{13\%}$
        & $\phantom{-}\hspace*{\unitcharwidth}\xs{1.06}{5\%\hspace*{\unitcharwidth}}{5\%\hspace*{\unitcharwidth}}$
         \\\hline
      $\sigma^\text{\NLO}_\text{\EW} / \sigma^\text{\LO}$ \vP
        & $\phantom{-}\hspace*{\unitcharwidth}\xs{0.97}{7\%\hspace*{\unitcharwidth}}{6\%\hspace*{\unitcharwidth}}$
        & $\phantom{-}\hspace*{\unitcharwidth}\xs{0.71}{5\%\hspace*{\unitcharwidth}}{6\%\hspace*{\unitcharwidth}}$
        & $\phantom{-}\hspace*{\unitcharwidth}\xs{0.85}{6\%\hspace*{\unitcharwidth}}{7\%\hspace*{\unitcharwidth}}$
        & $\phantom{-}\hspace*{\unitcharwidth}\xs{0.79}{6\%\hspace*{\unitcharwidth}}{7\%\hspace*{\unitcharwidth}}$
         \\\hline
      $\sigma^\text{\NLO}_\text{\QCDpEW} / \sigma^\text{\LO}$ \vP
        & $\phantom{-}\hspace*{\unitcharwidth}\xs{1.01}{5\%\hspace*{\unitcharwidth}}{7\%\hspace*{\unitcharwidth}}$
        & $\phantom{-}\hspace*{\unitcharwidth}\xs{1.05}{8\%\hspace*{\unitcharwidth}}{10\%}$
        & $\phantom{-}\hspace*{\unitcharwidth}\xs{1.27}{10\%}{12\%}$
        & $\phantom{-}\hspace*{\unitcharwidth}\xs{0.85}{5\%\hspace*{\unitcharwidth}}{5\%\hspace*{\unitcharwidth}}$
         \\\hline
      $\sigma^\text{\NLO}_\text{\QCDtEW} / \sigma^\text{\LO}$ \vP
        & $\phantom{-}\hspace*{\unitcharwidth}\xs{1.01}{5\%\hspace*{\unitcharwidth}}{7\%\hspace*{\unitcharwidth}}$
        & $\phantom{-}\hspace*{\unitcharwidth}\xs{0.95}{6\%\hspace*{\unitcharwidth}}{8\%\hspace*{\unitcharwidth}}$
        & $\phantom{-}\hspace*{\unitcharwidth}\xs{1.21}{9\%\hspace*{\unitcharwidth}}{11\%}$
        & $\phantom{-}\hspace*{\unitcharwidth}\xs{0.83}{4\%\hspace*{\unitcharwidth}}{4\%\hspace*{\unitcharwidth}}$
         \\\hline\hline
      $\delta_\text{\,\tiny no \aPDF}^\text{\,\raisebox{1pt}{\tiny \LO}}$ \vP
        & $\hspace*{\unitcharwidth}-1\,\%$ 
        & $\hspace*{\unitcharwidth}-4\,\%$ 
        & $\hspace*{\unitcharwidth}-5\,\%$ 
        & $\hspace*{\unitcharwidth}-6\,\%$ 
         \\\hline
      $\delta_\text{\,\tiny\LUXQED}^\text{\,\raisebox{1pt}{\tiny \LO}}$ \vP
        & $\phantom{-}\hspace*{\unitcharwidth}0\,\%$ 
        & $\hspace*{\unitcharwidth}-0\,\%$ 
        & $\hspace*{\unitcharwidth}-0\,\%$ 
        & $\hspace*{\unitcharwidth}-0\,\%$ 
         \\\hline
      $\delta_\text{\,\tiny\NNPDFQED}^\text{\,\raisebox{1pt}{\tiny \LO}}$ \vP
        & $\hspace*{\unitcharwidth}-0\,\%$ 
        & $\phantom{-}\hspace*{\unitcharwidth}8\,\%$ 
        & $\phantom{-}12\,\%$ 
        & $\phantom{-}\hspace*{\unitcharwidth}6\,\%$ 
         \\\hline\hline
      $\delta_\text{\,\tiny no \aPDF}^\text{\,\raisebox{1pt}{\tiny \NLO \QCDtEW}}$ \vP
        & $\hspace*{\unitcharwidth}-2\,\%$ 
        & $\hspace*{\unitcharwidth}-9\,\%$ 
        & $\hspace*{\unitcharwidth}-9\,\%$ 
        & $-12\,\%$ 
         \\\hline
      $\delta_\text{\,\tiny\LUXQED}^\text{\,\raisebox{1pt}{\tiny \NLO \QCDtEW}}$ \vP
        & $\hspace*{\unitcharwidth}-1\,\%$ 
        & $\hspace*{\unitcharwidth}-1\,\%$ 
        & $\hspace*{\unitcharwidth}-0\,\%$ 
        & $\hspace*{\unitcharwidth}-3\,\%$ 
         \\\hline
      $\delta_\text{\,\tiny\NNPDFQED}^\text{\,\raisebox{1pt}{\tiny \NLO \QCDtEW}}$ \vP
        & $\hspace*{\unitcharwidth}-1\,\%$ 
        & $\phantom{-}10\,\%$ 
        & $\phantom{-}13\,\%$ 
        & $\phantom{-}\hspace*{\unitcharwidth}5\,\%$ 
         \\\lasthline
    \end{tabular}

  \end{center}
  \caption{
    Cross-sections for $pp\to\llnndf$ at 13\,\TeV with \CTQED PDFs
    and fiducial cuts of~\refta{tab:cuts} (1$^\mathrm{st}$ column)
    plus one additional cut on $p_{\rT,\ell_1}$ 
    (2$^\mathrm{nd}$ column), $\missingET$  
    (3$^\mathrm{rd}$ column), or $m_{\ell\ell}$ 
    (4$^\mathrm{th}$ column).
    The top row lists \LO cross sections, while 
    the following four rows give the relative change induced 
    by the \NLO \QCD, \EW, \QCDpEW and \QCDtEW corrections. 
    The sub- and superscripts give their respective relative 
    uncertainties determined through customary $\mur$ and $\muf$ 
    variations, while keeping the reference LO cross section fixed in the ratios.
    The impact of alternative 
    descriptions of the photon density are explored by 
    neglecting it entirely (no \aPDF) or using the 
    densities provided by the \LUXQED and \NNPDFQED sets.
    For quarks and gluons always the central PDF 
    set \CTQED is chosen.
    \label{tab:xs:df}
  }
  \vspace*{4mm}
  \begin{center}
        \begin{tabular}{l||C||C|C|C}
      $\boldsymbol{pp\to e^+e^-\nu\bar{\nu}}$
        & inclusive& $p_{\rT,\ell_1}>500\,\GeV$& $\missingET>500\,\GeV$& $m_{\ell\ell}>1\,\TeV$ \\\hline\hline
      $\sigma^\text{\LO}$ [fb] \vP
        & $\phantom{.-}\xs{368}{7\%}{6\%}$
        & $\phantom{-}\hspace*{-\unitcharwidth}\xs{0.108}{6\%}{7\%}$
        & $\phantom{-}\hspace*{-\unitcharwidth}\xs{0.074}{6\%}{7\%}$
        & $\phantom{-}\hspace*{-\unitcharwidth}\xs{0.158}{6\%}{7\%}$
         \\\hline\hline
      $\sigma^\text{\NLO}_\text{\QCD} / \sigma^\text{\LO}$ \vP
        & $\phantom{-}\hspace*{\unitcharwidth}\xs{1.04}{5\%\hspace*{\unitcharwidth}}{7\%\hspace*{\unitcharwidth}}$
        & $\phantom{-}\hspace*{\unitcharwidth}\xs{1.32}{9\%\hspace*{\unitcharwidth}}{11\%}$
        & $\phantom{-}\hspace*{\unitcharwidth}\xs{1.30}{9\%\hspace*{\unitcharwidth}}{10\%}$
        & $\phantom{-}\hspace*{\unitcharwidth}\xs{1.06}{5\%\hspace*{\unitcharwidth}}{6\%\hspace*{\unitcharwidth}}$
         \\\hline
      $\sigma^\text{\NLO}_\text{\EW} / \sigma^\text{\LO}$ \vP
        & $\phantom{-}\hspace*{\unitcharwidth}\xs{0.97}{7\%\hspace*{\unitcharwidth}}{5\%\hspace*{\unitcharwidth}}$
        & $\phantom{-}\hspace*{\unitcharwidth}\xs{0.68}{5\%\hspace*{\unitcharwidth}}{5\%\hspace*{\unitcharwidth}}$
        & $\phantom{-}\hspace*{\unitcharwidth}\xs{0.68}{4\%\hspace*{\unitcharwidth}}{5\%\hspace*{\unitcharwidth}}$
        & $\phantom{-}\hspace*{\unitcharwidth}\xs{0.78}{6\%\hspace*{\unitcharwidth}}{7\%\hspace*{\unitcharwidth}}$
         \\\hline
      $\sigma^\text{\NLO}_\text{\QCDpEW} / \sigma^\text{\LO}$ \vP
        & $\phantom{-}\hspace*{\unitcharwidth}\xs{1.00}{5\%\hspace*{\unitcharwidth}}{7\%\hspace*{\unitcharwidth}}$
        & $\phantom{-}\hspace*{\unitcharwidth}\xs{1.00}{7\%\hspace*{\unitcharwidth}}{9\%\hspace*{\unitcharwidth}}$
        & $\phantom{-}\hspace*{\unitcharwidth}\xs{0.98}{7\%\hspace*{\unitcharwidth}}{8\%\hspace*{\unitcharwidth}}$
        & $\phantom{-}\hspace*{\unitcharwidth}\xs{0.84}{5\%\hspace*{\unitcharwidth}}{5\%\hspace*{\unitcharwidth}}$
         \\\hline
      $\sigma^\text{\NLO}_\text{\QCDtEW} / \sigma^\text{\LO}$ \vP
        & $\phantom{-}\hspace*{\unitcharwidth}\xs{1.00}{5\%\hspace*{\unitcharwidth}}{6\%\hspace*{\unitcharwidth}}$
        & $\phantom{-}\hspace*{\unitcharwidth}\xs{0.90}{6\%\hspace*{\unitcharwidth}}{7\%\hspace*{\unitcharwidth}}$
        & $\phantom{-}\hspace*{\unitcharwidth}\xs{0.89}{6\%\hspace*{\unitcharwidth}}{7\%\hspace*{\unitcharwidth}}$
        & $\phantom{-}\hspace*{\unitcharwidth}\xs{0.83}{4\%\hspace*{\unitcharwidth}}{4\%\hspace*{\unitcharwidth}}$
         \\\hline\hline
      $\delta_\text{\,\tiny no \aPDF}^\text{\,\raisebox{1pt}{\tiny \LO}}$ \vP
        & $\hspace*{\unitcharwidth}-1\,\%$ 
        & $\hspace*{\unitcharwidth}-3\,\%$ 
        & $\hspace*{\unitcharwidth}-1\,\%$ 
        & $\hspace*{\unitcharwidth}-6\,\%$ 
         \\\hline
      $\delta_\text{\,\tiny\LUXQED}^\text{\,\raisebox{1pt}{\tiny \LO}}$ \vP
        & $\phantom{-}\hspace*{\unitcharwidth}0\,\%$ 
        & $\hspace*{\unitcharwidth}-0\,\%$ 
        & $\hspace*{\unitcharwidth}-0\,\%$ 
        & $\hspace*{\unitcharwidth}-0\,\%$ 
         \\\hline
      $\delta_\text{\,\tiny\NNPDFQED}^\text{\,\raisebox{1pt}{\tiny \LO}}$ \vP
        & $\hspace*{\unitcharwidth}-0\,\%$ 
        & $\phantom{-}\hspace*{\unitcharwidth}6\,\%$ 
        & $\phantom{-}\hspace*{\unitcharwidth}3\,\%$ 
        & $\phantom{-}\hspace*{\unitcharwidth}6\,\%$ 
         \\\hline\hline
      $\delta_\text{\,\tiny no \aPDF}^\text{\,\raisebox{1pt}{\tiny \NLO \QCDtEW}}$ \vP
        & $\hspace*{\unitcharwidth}-2\,\%$ 
        & $\hspace*{\unitcharwidth}-7\,\%$ 
        & $\hspace*{\unitcharwidth}-3\,\%$ 
        & $-10\,\%$ 
         \\\hline
      $\delta_\text{\,\tiny\LUXQED}^\text{\,\raisebox{1pt}{\tiny \NLO \QCDtEW}}$ \vP
        & $\hspace*{\unitcharwidth}-1\,\%$ 
        & $\hspace*{\unitcharwidth}-0\,\%$ 
        & $\hspace*{\unitcharwidth}-0\,\%$ 
        & $\hspace*{\unitcharwidth}-2\,\%$ 
         \\\hline
      $\delta_\text{\,\tiny\NNPDFQED}^\text{\,\raisebox{1pt}{\tiny \NLO \QCDtEW}}$ \vP
        & $\hspace*{\unitcharwidth}-1\,\%$ 
        & $\phantom{-}\hspace*{\unitcharwidth}7\,\%$ 
        & $\phantom{-}\hspace*{\unitcharwidth}3\,\%$ 
        & $\phantom{-}\hspace*{\unitcharwidth}4\,\%$ 
         \\\lasthline
    \end{tabular}

  \end{center}
  \caption{
    Cross-sections for $pp\to\llnnsf$ at 13\,\TeV including all neutrino flavours. 
    Higher-order corrections, scale uncertainties and 
    photon-induced contributions are presented as 
    in~\refta{tab:xs:df}.
    \label{tab:xs:sf:sum}
  }
\end{table}

\begin{table}[t!]
  \begin{center}
        \begin{tabular}{l||C||C|C|C}
      $\boldsymbol{pp\to e^+e^-\nu_e\bar{\nu}_e}$
        & inclusive& $p_{\rT,\ell_1}>500\,\GeV$& $\missingET>500\,\GeV$& $m_{\ell\ell}>1\,\TeV$ \\\hline\hline
      $\sigma^\text{\LO}$ [fb] \vP
        & $\phantom{.-}\xs{322}{8\%}{6\%}$
        & $\phantom{-}\hspace*{-\unitcharwidth}\xs{0.089}{6\%}{7\%}$
        & $\phantom{-}\hspace*{-\unitcharwidth}\xs{0.037}{6\%}{7\%}$
        & $\phantom{-}\hspace*{-\unitcharwidth}\xs{0.152}{6\%}{7\%}$
         \\\hline\hline
      $\sigma^\text{\NLO}_\text{\QCD} / \sigma^\text{\LO}$ \vP
        & $\phantom{-}\hspace*{\unitcharwidth}\xs{1.04}{5\%\hspace*{\unitcharwidth}}{7\%\hspace*{\unitcharwidth}}$
        & $\phantom{-}\hspace*{\unitcharwidth}\xs{1.33}{9\%\hspace*{\unitcharwidth}}{11\%}$
        & $\phantom{-}\hspace*{\unitcharwidth}\xs{1.34}{9\%\hspace*{\unitcharwidth}}{11\%}$
        & $\phantom{-}\hspace*{\unitcharwidth}\xs{1.06}{5\%\hspace*{\unitcharwidth}}{5\%\hspace*{\unitcharwidth}}$
         \\\hline
      $\sigma^\text{\NLO}_\text{\EW} / \sigma^\text{\LO}$ \vP
        & $\phantom{-}\hspace*{\unitcharwidth}\xs{0.97}{7\%\hspace*{\unitcharwidth}}{5\%\hspace*{\unitcharwidth}}$
        & $\phantom{-}\hspace*{\unitcharwidth}\xs{0.69}{5\%\hspace*{\unitcharwidth}}{6\%\hspace*{\unitcharwidth}}$
        & $\phantom{-}\hspace*{\unitcharwidth}\xs{0.73}{5\%\hspace*{\unitcharwidth}}{6\%\hspace*{\unitcharwidth}}$
        & $\phantom{-}\hspace*{\unitcharwidth}\xs{0.78}{6\%\hspace*{\unitcharwidth}}{7\%\hspace*{\unitcharwidth}}$
         \\\hline
      $\sigma^\text{\NLO}_\text{\QCDpEW} / \sigma^\text{\LO}$ \vP
        & $\phantom{-}\hspace*{\unitcharwidth}\xs{1.01}{5\%\hspace*{\unitcharwidth}}{7\%\hspace*{\unitcharwidth}}$
        & $\phantom{-}\hspace*{\unitcharwidth}\xs{1.02}{8\%\hspace*{\unitcharwidth}}{9\%\hspace*{\unitcharwidth}}$
        & $\phantom{-}\hspace*{\unitcharwidth}\xs{1.07}{8\%\hspace*{\unitcharwidth}}{10\%}$
        & $\phantom{-}\hspace*{\unitcharwidth}\xs{0.84}{5\%\hspace*{\unitcharwidth}}{5\%\hspace*{\unitcharwidth}}$
         \\\hline
      $\sigma^\text{\NLO}_\text{\QCDtEW} / \sigma^\text{\LO}$ \vP
        & $\phantom{-}\hspace*{\unitcharwidth}\xs{1.01}{5\%\hspace*{\unitcharwidth}}{7\%\hspace*{\unitcharwidth}}$
        & $\phantom{-}\hspace*{\unitcharwidth}\xs{0.92}{6\%\hspace*{\unitcharwidth}}{7\%\hspace*{\unitcharwidth}}$
        & $\phantom{-}\hspace*{\unitcharwidth}\xs{0.98}{7\%\hspace*{\unitcharwidth}}{8\%\hspace*{\unitcharwidth}}$
        & $\phantom{-}\hspace*{\unitcharwidth}\xs{0.83}{4\%\hspace*{\unitcharwidth}}{4\%\hspace*{\unitcharwidth}}$
         \\\hline\hline
      $\delta_\text{\,\tiny no \aPDF}^\text{\,\raisebox{1pt}{\tiny \LO}}$ \vP
        & $\hspace*{\unitcharwidth}-1\,\%$ 
        & $\hspace*{\unitcharwidth}-4\,\%$ 
        & $\hspace*{\unitcharwidth}-2\,\%$ 
        & $\hspace*{\unitcharwidth}-6\,\%$ 
         \\\hline
      $\delta_\text{\,\tiny\LUXQED}^\text{\,\raisebox{1pt}{\tiny \LO}}$ \vP
        & $\phantom{-}\hspace*{\unitcharwidth}0\,\%$ 
        & $\hspace*{\unitcharwidth}-0\,\%$ 
        & $\hspace*{\unitcharwidth}-0\,\%$ 
        & $\hspace*{\unitcharwidth}-0\,\%$ 
         \\\hline
      $\delta_\text{\,\tiny\NNPDFQED}^\text{\,\raisebox{1pt}{\tiny \LO}}$ \vP
        & $\hspace*{\unitcharwidth}-0\,\%$ 
        & $\phantom{-}\hspace*{\unitcharwidth}7\,\%$ 
        & $\phantom{-}\hspace*{\unitcharwidth}6\,\%$ 
        & $\phantom{-}\hspace*{\unitcharwidth}6\,\%$ 
         \\\hline\hline
      $\delta_\text{\,\tiny no \aPDF}^\text{\,\raisebox{1pt}{\tiny \NLO \QCDtEW}}$ \vP
        & $\hspace*{\unitcharwidth}-2\,\%$ 
        & $\hspace*{\unitcharwidth}-8\,\%$ 
        & $\hspace*{\unitcharwidth}-5\,\%$ 
        & $-11\,\%$ 
         \\\hline
      $\delta_\text{\,\tiny\LUXQED}^\text{\,\raisebox{1pt}{\tiny \NLO \QCDtEW}}$ \vP
        & $\hspace*{\unitcharwidth}-1\,\%$ 
        & $\hspace*{\unitcharwidth}-0\,\%$ 
        & $\hspace*{\unitcharwidth}-0\,\%$ 
        & $\hspace*{\unitcharwidth}-2\,\%$ 
         \\\hline
      $\delta_\text{\,\tiny\NNPDFQED}^\text{\,\raisebox{1pt}{\tiny \NLO \QCDtEW}}$ \vP
        & $\hspace*{\unitcharwidth}-1\,\%$ 
        & $\phantom{-}\hspace*{\unitcharwidth}8\,\%$ 
        & $\phantom{-}\hspace*{\unitcharwidth}6\,\%$ 
        & $\phantom{-}\hspace*{\unitcharwidth}4\,\%$ 
         \\\lasthline
    \end{tabular}

  \end{center}
  \caption[]{
    Cross-sections for $pp\to\llnnsfwwzz$ at 13\,\TeV. 
    Higher-order corrections, scale uncertainties and 
    photon-induced contributions are presented as 
    in~\refta{tab:xs:df}.
    \label{tab:xs:sf:wwzz}
  }
  \vspace*{4mm}
  \begin{center}
        \begin{tabular}{l||C||C|C|C}
      $\boldsymbol{pp\to e^+e^-\nu_\mu\bar{\nu}_\mu}$
        & inclusive& $p_{\rT,\ell_1}>500\,\GeV$& $\missingET>500\,\GeV$& $m_{\ell\ell}>1\,\TeV$ \\\hline\hline
      $\sigma^\text{\LO}$ [fb] \vP
        & $\phantom{.-}\xs{23.0}{6\%}{5\%}$
        & $\phantom{-}\hspace*{-\unitcharwidth}\xs{0.0093}{7\%}{8\%}$
        & $\phantom{-}\hspace*{-\unitcharwidth}\xs{0.0187}{6\%}{7\%}$
        & $\phantom{-}\hspace*{-\unitcharwidth}\xs{0.0032}{7\%}{7\%}$
         \\\hline\hline
      $\sigma^\text{\NLO}_\text{\QCD} / \sigma^\text{\LO}$ \vP
        & $\phantom{-}\hspace*{\unitcharwidth}\xs{1.01}{5\%\hspace*{\unitcharwidth}}{6\%\hspace*{\unitcharwidth}}$
        & $\phantom{-}\hspace*{\unitcharwidth}\xs{1.25}{8\%\hspace*{\unitcharwidth}}{9\%\hspace*{\unitcharwidth}}$
        & $\phantom{-}\hspace*{\unitcharwidth}\xs{1.26}{8\%\hspace*{\unitcharwidth}}{9\%\hspace*{\unitcharwidth}}$
        & $\phantom{-}\hspace*{\unitcharwidth}\xs{1.09}{5\%\hspace*{\unitcharwidth}}{6\%\hspace*{\unitcharwidth}}$
         \\\hline
      $\sigma^\text{\NLO}_\text{\EW} / \sigma^\text{\LO}$ \vP
        & $\phantom{-}\hspace*{\unitcharwidth}\xs{0.95}{6\%\hspace*{\unitcharwidth}}{5\%\hspace*{\unitcharwidth}}$
        & $\phantom{-}\hspace*{\unitcharwidth}\xs{0.65}{4\%\hspace*{\unitcharwidth}}{5\%\hspace*{\unitcharwidth}}$
        & $\phantom{-}\hspace*{\unitcharwidth}\xs{0.63}{4\%\hspace*{\unitcharwidth}}{4\%\hspace*{\unitcharwidth}}$
        & $\phantom{-}\hspace*{\unitcharwidth}\xs{0.81}{5\%\hspace*{\unitcharwidth}}{6\%\hspace*{\unitcharwidth}}$
         \\\hline
      $\sigma^\text{\NLO}_\text{\QCDpEW} / \sigma^\text{\LO}$ \vP
        & $\phantom{-}\hspace*{\unitcharwidth}\xs{0.96}{5\%\hspace*{\unitcharwidth}}{6\%\hspace*{\unitcharwidth}}$
        & $\phantom{-}\hspace*{\unitcharwidth}\xs{0.90}{6\%\hspace*{\unitcharwidth}}{7\%\hspace*{\unitcharwidth}}$
        & $\phantom{-}\hspace*{\unitcharwidth}\xs{0.89}{6\%\hspace*{\unitcharwidth}}{7\%\hspace*{\unitcharwidth}}$
        & $\phantom{-}\hspace*{\unitcharwidth}\xs{0.90}{5\%\hspace*{\unitcharwidth}}{5\%\hspace*{\unitcharwidth}}$
         \\\hline
      $\sigma^\text{\NLO}_\text{\QCDtEW} / \sigma^\text{\LO}$ \vP
        & $\phantom{-}\hspace*{\unitcharwidth}\xs{0.96}{5\%\hspace*{\unitcharwidth}}{6\%\hspace*{\unitcharwidth}}$
        & $\phantom{-}\hspace*{\unitcharwidth}\xs{0.82}{5\%\hspace*{\unitcharwidth}}{6\%\hspace*{\unitcharwidth}}$
        & $\phantom{-}\hspace*{\unitcharwidth}\xs{0.80}{5\%\hspace*{\unitcharwidth}}{6\%\hspace*{\unitcharwidth}}$
        & $\phantom{-}\hspace*{\unitcharwidth}\xs{0.89}{4\%\hspace*{\unitcharwidth}}{5\%\hspace*{\unitcharwidth}}$
         \\\hline\hline
      $\delta_\text{\,\tiny no \aPDF}^\text{\,\raisebox{1pt}{\tiny \LO}}$ \vP
        & $\hspace*{\unitcharwidth}-0.0\,\%$ 
        & $\hspace*{\unitcharwidth}-0.4\,\%$ 
        & $\hspace*{\unitcharwidth}-0.1\,\%$ 
        & $\hspace*{\unitcharwidth}-1.4\,\%$ 
         \\\hline
      $\delta_\text{\,\tiny\LUXQED}^\text{\,\raisebox{1pt}{\tiny \LO}}$ \vP
        & $\phantom{-}\hspace*{\unitcharwidth}0.0\,\%$ 
        & $\hspace*{\unitcharwidth}-0.0\,\%$ 
        & $\hspace*{\unitcharwidth}-0.0\,\%$ 
        & $\hspace*{\unitcharwidth}-0.0\,\%$ 
         \\\hline
      $\delta_\text{\,\tiny\NNPDFQED}^\text{\,\raisebox{1pt}{\tiny \LO}}$ \vP
        & $\hspace*{\unitcharwidth}-0.0\,\%$ 
        & $\phantom{-}\hspace*{\unitcharwidth}0.5\,\%$ 
        & $\phantom{-}\hspace*{\unitcharwidth}0.1\,\%$ 
        & $\phantom{-}\hspace*{\unitcharwidth}1.6\,\%$ 
         \\\hline\hline
      $\delta_\text{\,\tiny no \aPDF}^\text{\,\raisebox{1pt}{\tiny \NLO \QCDtEW}}$ \vP
        & $\hspace*{\unitcharwidth}-0.0\,\%$ 
        & $\hspace*{\unitcharwidth}-0.6\,\%$ 
        & $\hspace*{\unitcharwidth}-0.1\,\%$ 
        & $\hspace*{\unitcharwidth}-1.7\,\%$ 
         \\\hline
      $\delta_\text{\,\tiny\LUXQED}^\text{\,\raisebox{1pt}{\tiny \NLO \QCDtEW}}$ \vP
        & $\hspace*{\unitcharwidth}-0.0\,\%$ 
        & $\hspace*{\unitcharwidth}-0.0\,\%$ 
        & $\hspace*{\unitcharwidth}-0.0\,\%$ 
        & $\hspace*{\unitcharwidth}-0.3\,\%$ 
         \\\hline
      $\delta_\text{\,\tiny\NNPDFQED}^\text{\,\raisebox{1pt}{\tiny \NLO \QCDtEW}}$ \vP
        & $\hspace*{\unitcharwidth}-0.0\,\%$ 
        & $\phantom{-}\hspace*{\unitcharwidth}0.6\,\%$ 
        & $\phantom{-}\hspace*{\unitcharwidth}0.1\,\%$ 
        & $\phantom{-}\hspace*{\unitcharwidth}1.4\,\%$ 
         \\\lasthline
    \end{tabular}

  \end{center}
  \caption[]{
    Cross-sections for $pp\to\llnnsfzzone$ at 13\,\TeV. 
    Higher-order corrections, scale uncertainties and 
    photon-induced contributions are presented as 
    in~\refta{tab:xs:df}.
    \label{tab:xs:sf:zz}
  }
\end{table}

\end{appendix}

\clearpage

\bibliographystyle{JHEP}
\bibliography{2l2n}

\begin{thebibliography}{10}
\bibitem{munich}
{\sc Munich} is the abbreviation of ``MUlti-chaNnel Integrator at Swiss (CH)
  precision''---an automated parton level NLO generator by S.~Kallweit. In
  preparation.
\end{thebibliography}

\end{document}